\documentclass[final]{elsarticle}
\usepackage{lipsum}
\makeatletter
\def\ps@pprintTitle{%
	\let\@oddhead\@empty
	\let\@evenhead\@empty
	\def\@oddfoot{}%
	\let\@evenfoot\@oddfoot}
\makeatother
\usepackage{lineno,hyperref}
\usepackage{caption}
\usepackage{subcaption}
\usepackage{siunitx}
\usepackage{tabularx}
\usepackage{amsmath}
\usepackage{xcolor}
\usepackage{comment}
\usepackage{soul}
\usepackage{setspace}
\usepackage[capitalize]{cleveref}
\modulolinenumbers[0]
\graphicspath{{graphics/}}
\usepackage{caption}
\journal{Powder Technology}

\usepackage{numcompress}

\begin{document}
\captionsetup[figure]{labelfont={bf},labelformat={default},labelsep=period,name={Fig.}}
\renewcommand*{\figureautorefname}{Fig.}
\begin{frontmatter}

\title{Effect of rough wall on drag, lift, and torque on   an ellipsoidal particle  in a linear shear flow}


\author{Atul Manikrao Bhagat}

\author{Partha Sarathi Goswami\corref{mycorrespondingauthor}}
\cortext[mycorrespondingauthor]{Corresponding author. Tel.: +91 22-2576 7230; fax: +91 22-2572 6895}
\ead{psg@iitb.ac.in}

\address{Chemical Engineering Department, Indian Institute of Technology Bombay, Mumbai - 400076 (India)}

\begin{abstract}
The present study provides a detailed description of the forces on an ellipsoidal particle in the vicinity of the rough wall.  Three-dimensional numerical simulations are performed using body-fitted mesh to estimate the drag, lift, and torque coefficients. A large number of simulations are conducted 
over a range of parameters such as shear Reynolds number ($10 \le Re_s \le 100$), orientation angle ($\ang{0}\le\theta\le\ang{180}$), and wall-particle separation distance ($0.1\le\delta\le2.0$) to get a comprehensive description of variation of the above coefficients. Using the simulation results, we develop the correlations for the drag and lift coefficients to describe the effect of rough wall, inclination angles, and particle Reynolds numbers on the hydrodynamic coefficients. The proposed correlations can be used for two phase flow simulation using Eulerian-Lagrangian framework. 

\end{abstract}


\end{frontmatter}


\section{Introduction}
The flow of suspensions containing nonspherical particles is ubiquitous in natural, industrial, and biomedical processes.  Industrial processes like, fluidization \cite{krug2016}, pneumatic transport of solids  \cite{mills2003,kleinstreuer2013}, processing in  paper industry \cite{tadros2011,ogawa1990,Lundell2011}, and wastewater treatments \cite{CUI2007921} deal with the flow of nonspherical particles. Raw materials used in all the processes deviate a lot from the ideal spherical structure. Transport and deposition of nonspherical particles in biological flows are also important issues to be addressed \cite{kleinstreuer2013}. Therefore, understanding the forces and dynamics of the nonspherical particles helps to design and develop these processes. Most of the earlier studies deal with the spherical particles \cite{LeeIJMF2017,zeng2005,ZengPOF2009} where symmetry plays an important role to simplify the problem unless the particles are placed near the solid surfaces. But, nonspherical particles increase the degrees of freedom based on their orientation with respect to the ambient flows.


An attempt to describe the dynamics of a nonspherical particle in viscous flow initiated almost a century back with the phenomenal work by \citet{jeffery1922}. They described the motion of an ellipsoid in the  creeping flow limit. \citet{BRENNER1963I,BRENNER1964II,BRENNER1964III} investigated the forces on particles with arbitrary shape in the Stokes flow limit using analytical theory and provided the expression for drag and lift forces. Later, an expression for the lift force on a nonspherical particle in shear flow was derived by \citet{harper1968} using asymptotic expansions. A comprehensive discussion on the dynamics of elongated particles for a wide range of flow is reported by \citet{LIN2003}.


A number of investigations that analysed the hydrodynamic forces on the 
particles placed near the wall at very low Reynolds number have been reported \cite{GAVZE1997,pozrikidis2005,LeeBEL2015,kim2016,smith2017,zarghami2018,palmer2020}.
The boundary integral method was used to estimate the force and velocity of nonspherical particles in shear flow in the Stokes flow limit \cite{GAVZE1997}. It was reported that the wall effect is very localized near the wall, and there is a  decrease in the effect of wall on hydrodynamic forces and torque as the non-sphericity of the particle increases. A spectral boundary element method was used by \citet{pozrikidis2005} to describe the motion of spheroidal particles for very low particle Reynolds number. A modification of the Jeffery orbit due to the presence of wall was reported by the author. \citet{kim2016} studied the dynamics of a cylindrical particle near a stationary wall in linear shear flow at different  Stokes numbers. They have reported that there is an increase in drag with the aspect ratio of the particle. However, the torque is independent of wall-normal distance.

During the last decade, a number of notable works to provide correlations to calculate drag and lift  are reported for finite Reynolds numbers \cite{HOLZER2008,zastawny2012,ouchene2016,Sanjeevi2018,frohlich2020,FILLINGHAM2021}. Majority of those are obtained through data fitting and  satisfy the theoretical formulation at the limiting cases. Using the experimental data reported in the literature, \citet{HOLZER2008} reported a correlation to calculate drag coefficient, which includes the effect of particle shape, orientation angle, and particle Reynolds number. Their correlation successfully predicts the drag coefficient with a maximum deviation of $14\%$. Immersed Boundary Method (IBM) was used by \citet{zastawny2012} to estimate drag and lift forces on the nonspherical particles. The parameters used in their correlations are exclusive functions of aspect ratio.  \citet{andersson2019} also used IBM to analyse the forces on ellipsoid with a very high aspect ratio in uniform flow at very low Reynolds number. They outlined the challenges in computation at such a low Reynolds numbers for various inclination angles and computation domains. \citet{ouchene2015,ouchene2016} computed drag, lift, and torque coefficients  for uniform flow over ellipsoidal particle at moderate Reynolds number  using body-fitted mesh  and presented the correlations. 
\citet{sanjeevi2017} and \citet{Sanjeevi2018} analyzed the drag on a nonspherical particle at different inclination angles in an uniform flow and for different Reynolds numbers. Forces on the spherical particle in the vicinity of a rough wall for liner shear flow have been estimated by~\citet{LeeIJMF2017}. \citet{frohlich2020} have also performed a number of simulations using the Cartesian cut cell method for an ellipsoid in an unbounded uniform flow. They have reported that beyond certain aspect ratio ($> 3$), the flow is not influenced much with the aspect ratio.  \citet{zarghami2018} examined the hydrodynamic forces and torque acting on a particle placed near the wall with 2D numerical simulations using Lattice Boltzmann method. A significant variation of forces based on wall separation and orientation angle was observed. The presence of a wall was  found to change the directions of lift and torque acting on the particle \cite{zarghami2018}.  \citet{palmer2020} modeled the partially immersed finite body in viscous layers near a wall in a 2-dimensional scenario. Their  study showed the oscillating motion of particles based on the center of mass of the particle. This motion of body directs the fluid displacement in between wall and particle, which results in non-linear pressure responses of a wall. All of these works were carried out either at Stokes flow regime or at a 2-D scenario. Very recently \citet{FILLINGHAM2021} developed correlations for the drag, lift, and torque coefficients on an ellipsoid particle placed on a smooth wall in a linear shear flow. They carried out a number of simulations at several combinations of parameters such as incident angle (in the plane perpendicular to the plane of shear), Reynolds number of flow, and aspect ratio of particle. The data from simulations were used to formulate the correlations, which  reproduce the results for a spherical particle \cite{ZengPOF2009} as a limiting case. They have reported a maximum deviation of $16\%$ as predicted by the correlation.

In the present study, we analyse the forces acting on an ellipsoidal particle in a linear shear flow. The particle is placed in the vicinity of a rough wall. 3D finite element method with body-fitted meshes is used for the simulations.
The ellipsoidal particle is assumed to be fixed at a predefined distance away from the rough wall with an inclination angle. The objective is to estimate the forces acting on a particle at different distances from the rough wall and at various orientations with flow direction in the plane of shear. The effects of variation of the above parameters and the flow Reynolds number on the drag, lift, and torque coefficients  acting on the  particle are reported in detail. For all parameters investigated here,  corresponding studies are carried out for  the domain with a smooth wall. The results for the rough wall have been compared with the results for the smooth wall. Semi-empirical correlations are reported to predict the drag and lift coefficients for different inclination angles and a wide range of Reynolds numbers.



\section{Numerical method}\label{sec:numerical}
This section outlines the numerical technique used for the simulations. First, we  describe the system configuration followed by the  governing equations. Then, a validation study is presented for the flow conditions reported in the literature. Finally, an assessment for the  flow development and finding particle location to obtain steady forces are presented.
\begin{figure}[htb]
	\centering
	\begin{subfigure}[b]{0.45\textwidth}
		\includegraphics[width=\textwidth]{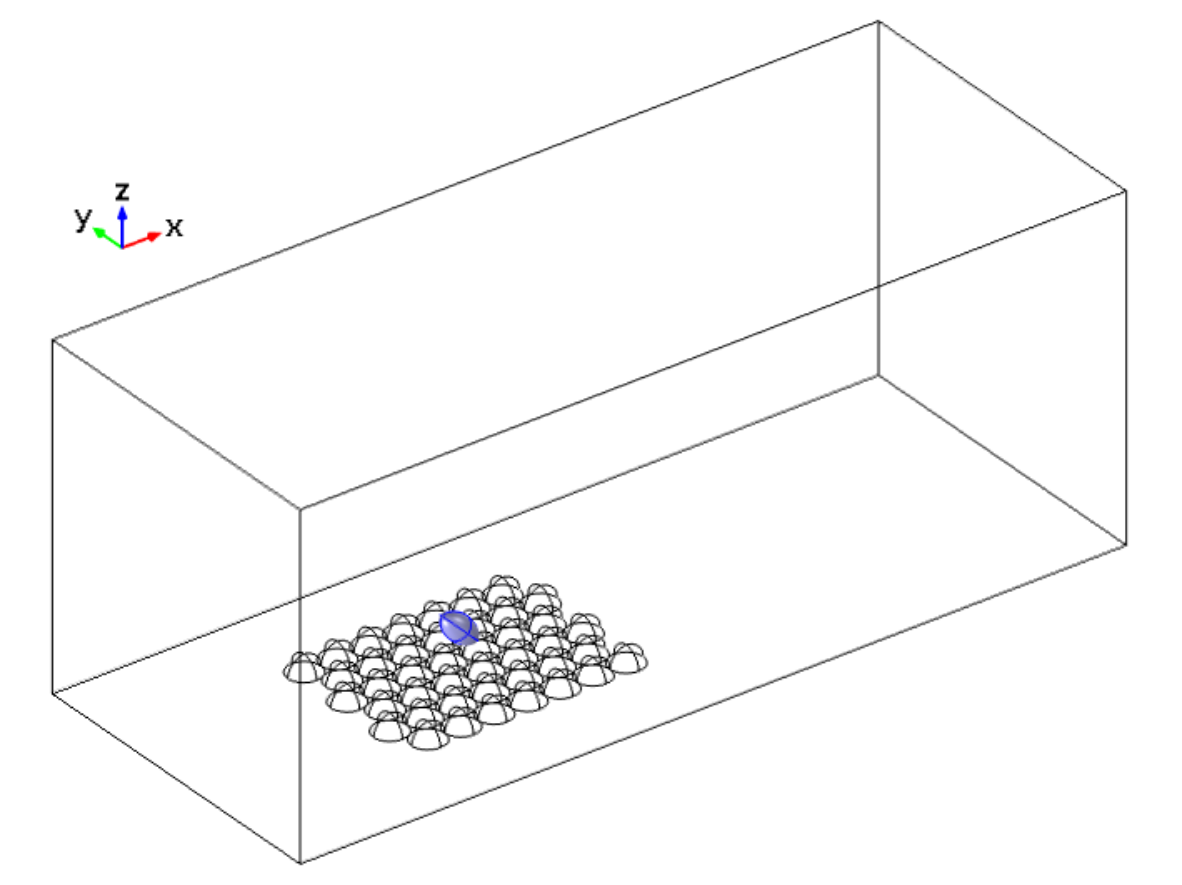}
		\caption{}
		\label{fig:RoughDomain1}
	\end{subfigure}
	\quad
	\begin{subfigure}[b]{0.45\textwidth}
		\includegraphics[width=\textwidth]{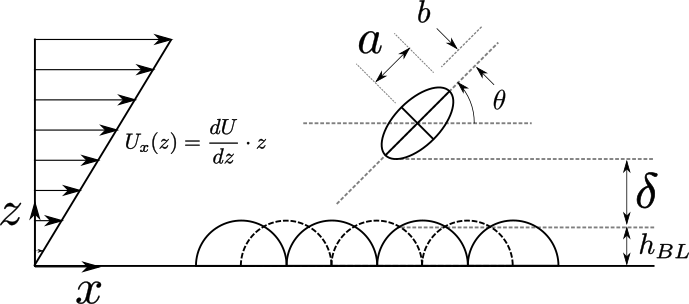} \vspace{1cm}
		\caption{}
		\label{fig:RoughDomain2}
	\end{subfigure}
	\caption[Domain for numerical analysis]{(a) Domain for numerical analysis (3D) (b) Ellipsoid particle in domain}
	\label{fig:RoughDomain12}
\end{figure}

A schematic of the domain, along with coordinate system used, is shown in \autoref{fig:RoughDomain12}. An ellipsoidal particle with semi-major axis $a$, semi-minor axis $b$ (aspect ratio $AR=a/b$), is placed at a location of $L$ from the horizontal base wall. In the present study, all kinds of motion of ellipsoid such as rolling, tumbling, and linear movement are not considered; essentially forces on the stationary particle are analysed. The origin of a coordinate system is located on the base wall. A rough base plane is constructed using hemispherical particles of diameter $d$. These hemispheres are arranged in a triangular array fashion and in contiguity to generate a compact structure. The location of these hemispheres on the base plane is calculated using the method given by \citet{LeeIJMF2017} as,
\begin{equation}
\begin{aligned}
x_{ij} = & \ \begin{cases}
x_o + j & \text{if i is odd} \\
x_o +(j-\frac{1}{2}) & \text{if i is even}
\end{cases} \\
y_{ij} = & \ \ y_o + sin\ \bigg(\frac{\pi}{3}\bigg)(i-1) \\
z_{ij} = & \ \ 0
\end{aligned}
\end{equation}
Here $x$, $y$, and $z$ are scaled with equivalent diameter of particle $d_p$. Hemispheres are arranged in rows ($i$) and columns ($j$). The first row of hemispheres is kept at 5 particle diameters away from the entrance (at $x_0 = 5$). The value of $y_o$ ($\approx-2.165$) is maintained such that there should be equal numbers of rows of hemispheres across ellipsoidal particle in channel cross-section.

Due to the presence of roughness on the wall, the effective clearance between the ellipsoid and 
the base plane is different as compared to a smooth wall. In case of the smooth wall, 
clearance (wall separation) is the distance from base plane to closest point on the ellipsoid in vertical direction. On the other hand, for rough wall, the minimum vertical distance changes with location of particle due to non-uniform surface. In that case, two different base levels are  needed to be defined. First one is the actual wall base where the hemispheres are centered, and the other one is the bed height. Based on two-dimensional configuration the bed height is defined as minimum vertical distance between base level and lowest point on the ellipsoid, when it is seating on the bed. For the present configuration the bed height $h_{BL}$ is taken to be $0.3165d_p$. Further, the clearance of the ellipsoid from the bed height ($\delta$) and orientation angle ($\theta$) are the key parameters  considered in the present study. $\delta$ is scaled with volume equivalent spherical particle diameter. The rotation of ellipsoid defined through orientation angle ($\theta$), is about $y$-axis, and anticlockwise rotation is assumed to be positive.

A linear shear flow is considered to advance towards the rough surface with mean a velocity component in $x$-direction only and no flow in the span-wise direction. The velocity profile  can be expressed as $U_x(z) = G_x z$ with $G_x=dU_x/dz$, a shear rate of approaching flow. Here, suffix $x$, represents inlet flow in $x-$direction. The shear Reynolds number based on the applied shear can be defined as, 
	\begin{equation}
		Re_s  = \frac{G_x d_p^2}{\nu}.
	\end{equation}
	Here, $d_p$ is related to major and minor axes of ellipsoid as $r_p = 2\times(ab^2)^{1/3}$. 
The hydrodynamic coefficients for drag, lift, and torque based on the volume equivalent spherical diameter and ambient fluid velocity ($U_x$) at the center of mass location of ellipsoid($z^\prime = L$) are defined as
\begin{equation}
	C_D = \frac{F_d}{\frac{1}{2}\rho U_x^2\pi\frac{d_p^2}{4}}, C_L = \frac{F_L}{\frac{1}{2}\rho U_x^2\pi\frac{d_p^2}{4}}, C_M = \frac{F_T}{\frac{1}{2}\rho U_x^2\pi\frac{d_p^3}{8}}.
\end{equation}
\subsection{Governing equations}
Flow of incompressible Newtonian fluid in the described domain is governed by the continuity and Navier-Stokes equations as, 
\begin{equation}
\rho\bigg(\frac{\partial \textbf{u}}{\partial t} + \textbf{u}\cdot\nabla\textbf{u}
\bigg) - \nabla\cdot\sigma = 0 \hspace{1cm} \mathrm{on} \ \Omega (\textbf{x}, t)
\end{equation}
and
\begin{equation}
\nabla\textbf{u} = 0 \hspace{1cm} \mathrm{on} \ \Omega (\textbf{x}, t).
\end{equation}
Here, 
$\sigma$ is a stress tensor, and  is expressed as,
\begin{equation}
\sigma = -p\textbf{I} + \textbf{T}
\end{equation}
\begin{equation}
\textbf{T} = 2\mu\varepsilon(\textbf{u})
\end{equation}
and $\varepsilon(\textbf{u})$, the deformation tensor can be expressed as,
\begin{equation}
\varepsilon(\textbf{u}) = \frac{1}{2}\bigg[(\nabla\textbf{u})+(\nabla\textbf{u})^{\mathrm{T}}\bigg].
\end{equation}
 $p$ and $t$ are the pressure and time respectively. The  simulations are done using COMSOL Multiphysics. The no-slip boundary condition is applied on the surface of ellipsoid and on the rough wall. Fully developed linear shear flow profile is imposed at the inlet. 
 Free slip condition with no-flow across it, is set for the top wall. The domain is  discretized using non-uniform tetrahedral grids. An {\it extremely fine} grid size is imposed on the surface of rough wall and ellipsoid, in order to capture the flow gradient correctly. 
 A large number of  simulations have been carried out using different angles of inclination of the particle and the wall normal separation distances. Hydrodynamic forces on the particle are calculated and results obtained are discussed in the next section.

\subsection{Validation}
\begin{figure}[htb]
	\centering
	\begin{subfigure}[b]{0.48\textwidth}
		\includegraphics[width=\textwidth]{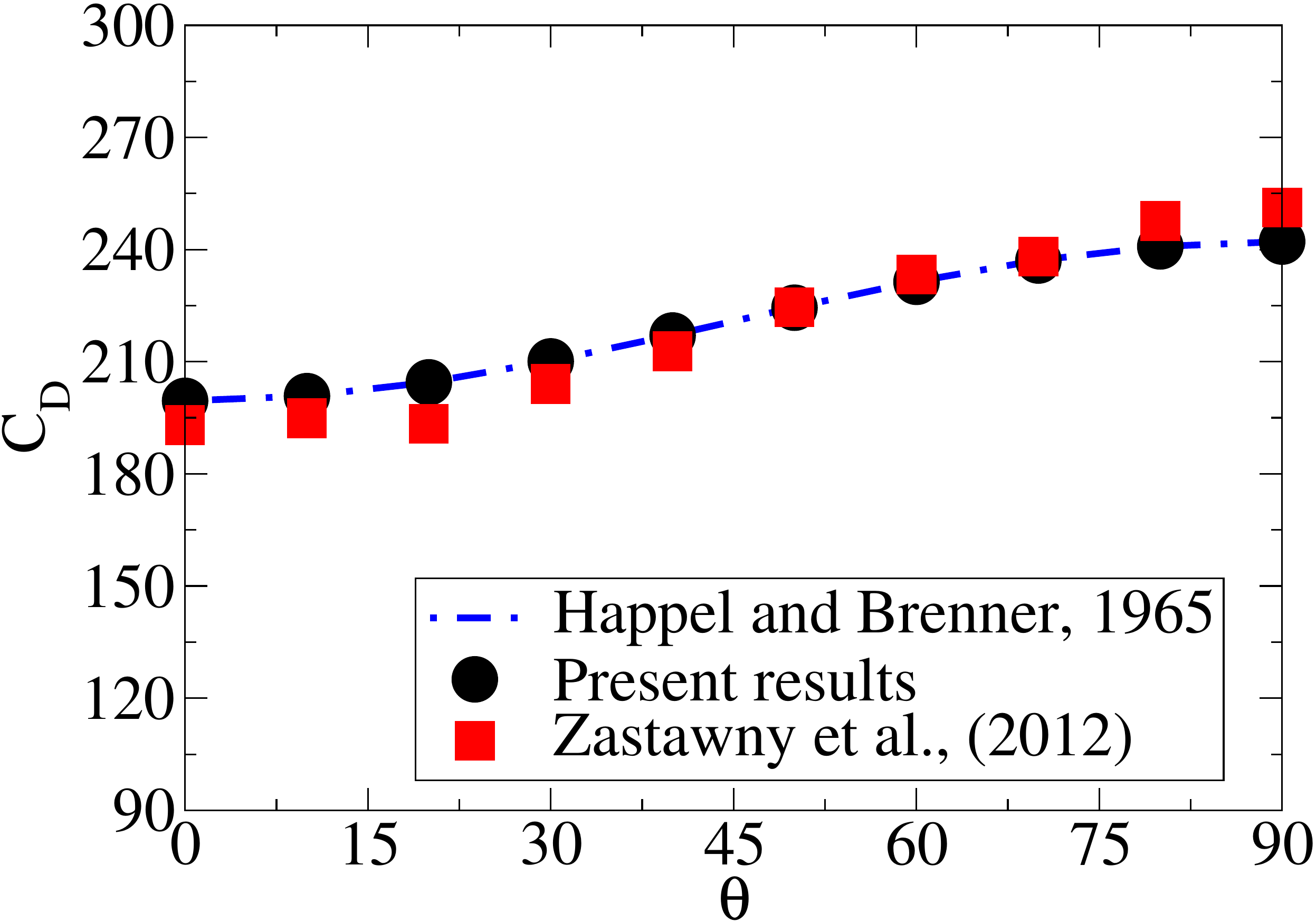}
		\caption{}
		\label{fig:CdVsThetaZastawnyValidation}
	\end{subfigure}
	\quad
	\begin{subfigure}[b]{0.48\textwidth}
		\includegraphics[width=\textwidth]{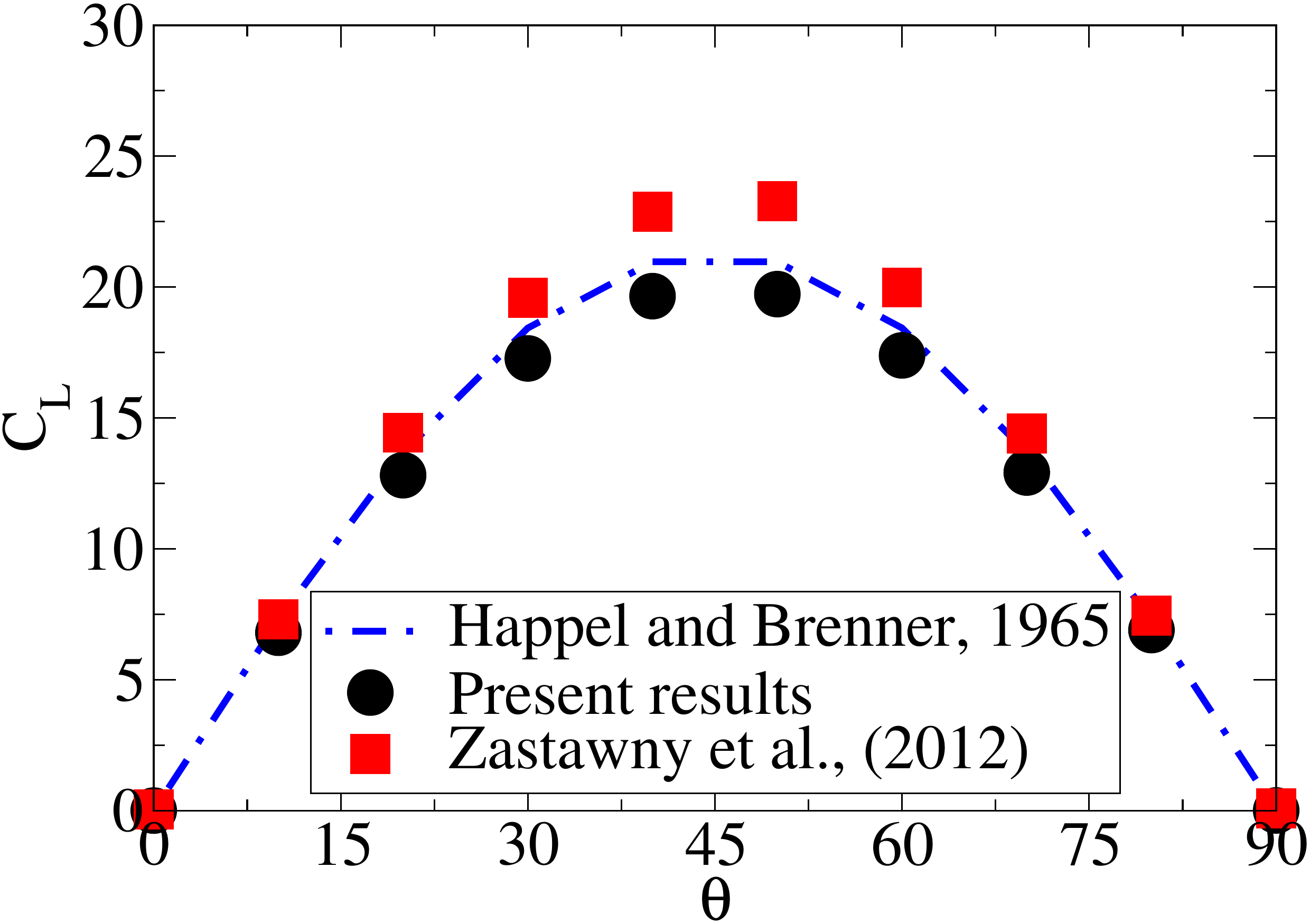}
		\caption{}
		\label{fig:ClVsThetaZastawnyValidation}
	\end{subfigure}
	\caption{Force coefficients for ellipsoid particle in uniform flow, compared with results of \citet{zastawny2012}}
	\label{fig:CdClVsThetaZastawnyValidation}
\end{figure}

\begin{figure}[htb]
	\centering
	\begin{subfigure}[b]{0.48\textwidth}
		\includegraphics[width=\textwidth]{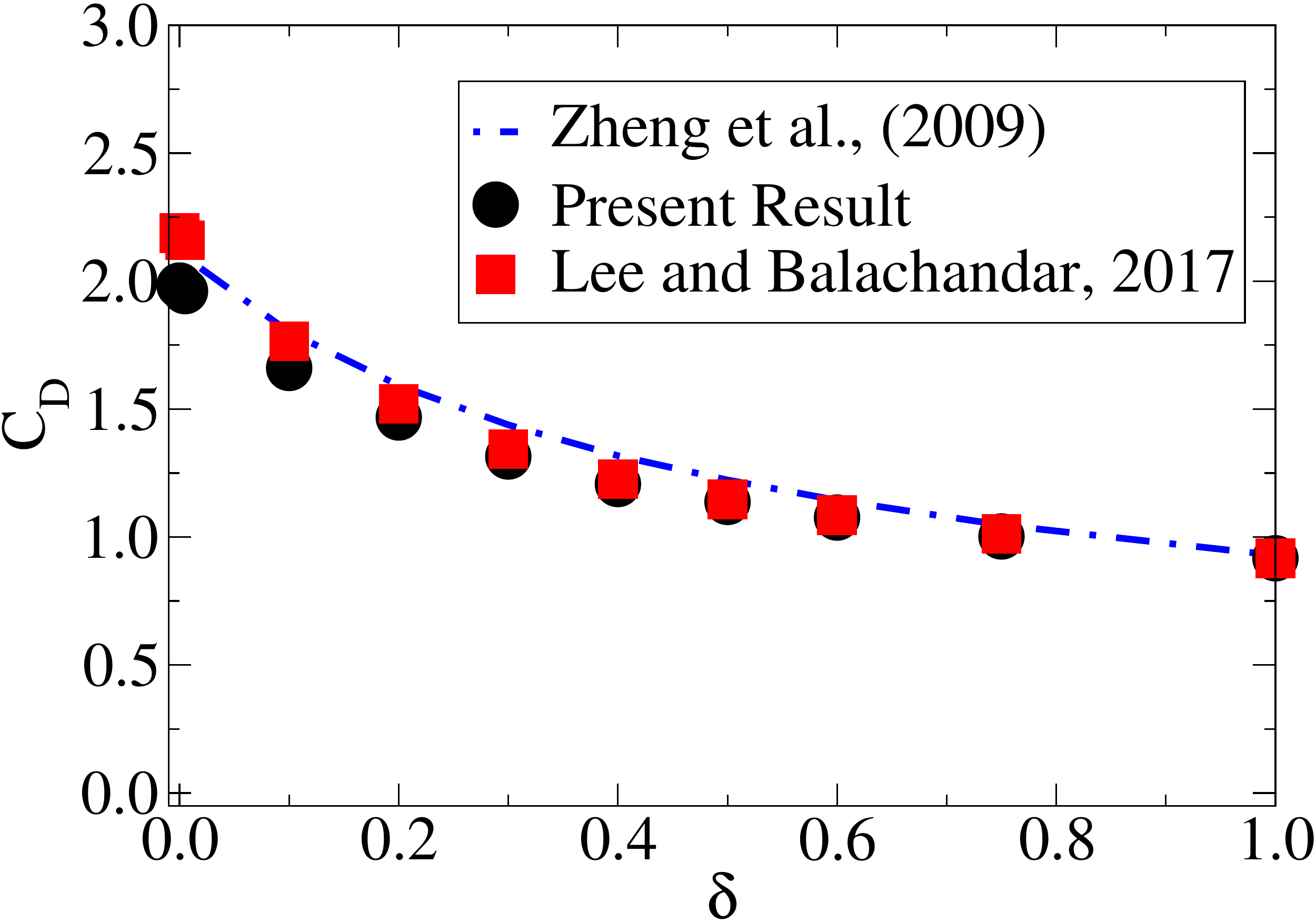}
		\caption{}
		\label{fig:CompRes100CdVsDelta}
	\end{subfigure}
	\quad
	\begin{subfigure}[b]{0.48\textwidth}
		\includegraphics[width=\textwidth]{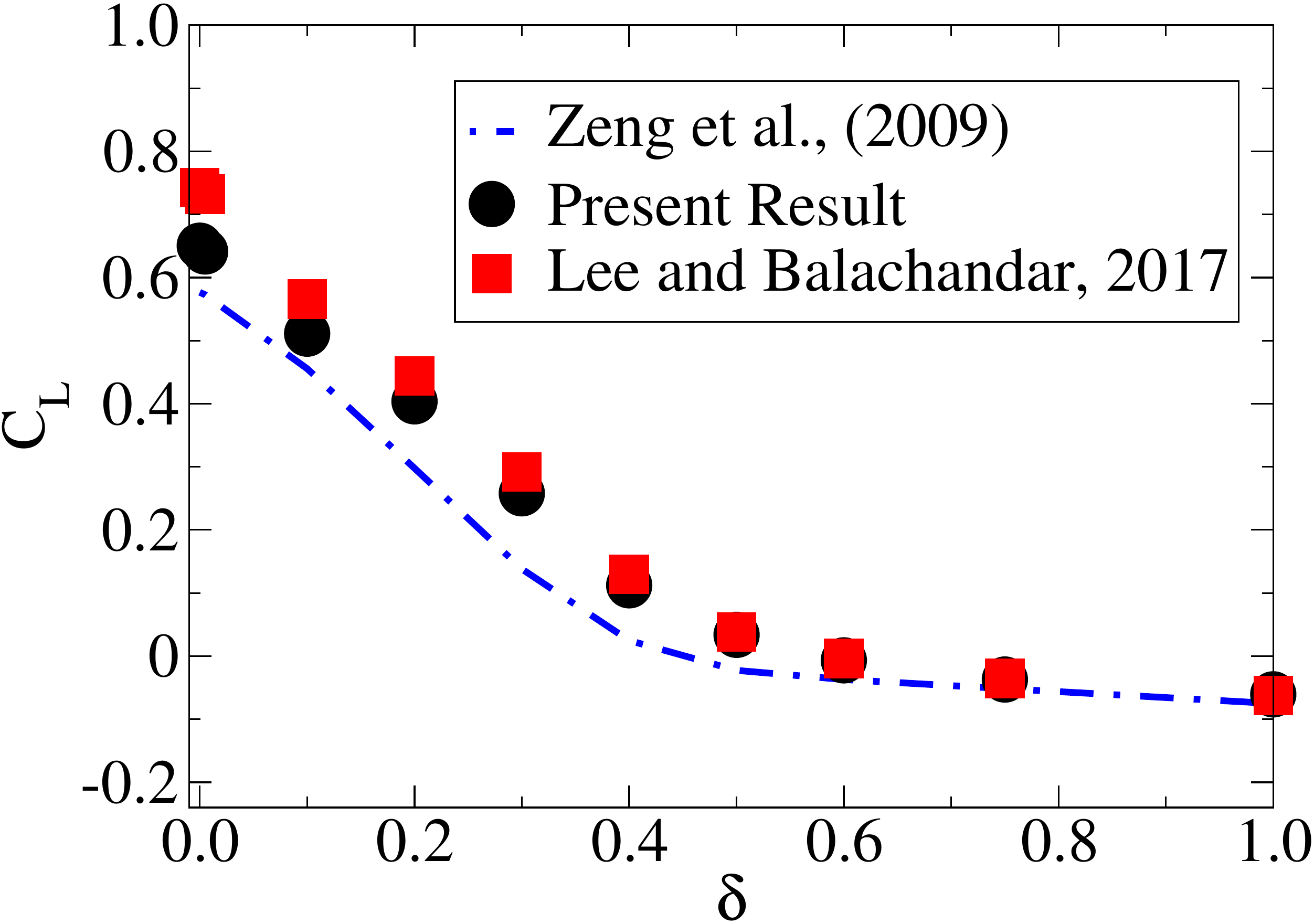}
		\caption{}
		\label{fig:CompRes100ClVsDelta}
	\end{subfigure}
	\caption{Force coefficients for spherical particle in uniform shear flow, compared with results of \citet{LeeIJMF2017}}
	\label{fig:CompRes100CdClVsDelta}
\end{figure}

Before analyzing the results related to forces on an ellipsoid, 
the validation is conducted using the data from other simulations reported in the literature  and with theoretical results, if available. The validations have been  performed for two different cases. In the first case, the drag and the lift coefficients for ellipsoid (aspect ratio AR = 2.5) in uniform flow at low Reynolds number ($Re_p = 0.1$) is computed. The results have  been compared with DNS results of \citet{zastawny2012}, as shown in the  \cref{fig:CdVsThetaZastawnyValidation,fig:ClVsThetaZastawnyValidation}. The figures show a variation of the  coefficients with angle of inclination. Drag and lift coefficients predicted by the  present simulations show a good agreement with that predicted theoretically by \citet{HappelBook1983}. There is an insignificant difference (about $0.3\%$) in drag coefficient as predicted by \citet{zastawny2012} using Immersed Boundary Method (IBM). Our results on lift coefficient, show a deviation of less than $5\%$ from that predicted by \citet{zastawny2012} at an angle of $\theta = \ang{45}$. 
The second set of validations is performed by analysing the flow over a spherical particle near the rough surface. The results (\autoref{fig:CompRes100CdClVsDelta}) are compared with \citet{LeeIJMF2017} who investigated hydrodynamic forces acting on spherical particle in an uniform shear flow over rough wall using IBM for a shear Reynolds number $Re_s \le 100$. \cref{fig:CompRes100CdVsDelta,fig:CompRes100ClVsDelta} show the drag and lift coefficients for particle at different location above the rough wall at $Re_s = 100$. 
It is observed that the results are in good agreement with the results of \citet{LeeIJMF2017}. A small difference is observed when the  particle is placed very close to  the wall. In any of the cases, maximum difference is always less than  $5\%$. 
The results from the  correlation by  \citet{ZengPOF2009} are plotted with similar conditions and found to under predict the value.  It is to be noted that results reported by \citet{ZengPOF2009} are for the case of smooth wall.

\subsection{Flow development analysis}
\begin{figure}[htb]
	\centering
	\begin{subfigure}[b]{0.45\textwidth}
		\includegraphics[width=\textwidth]{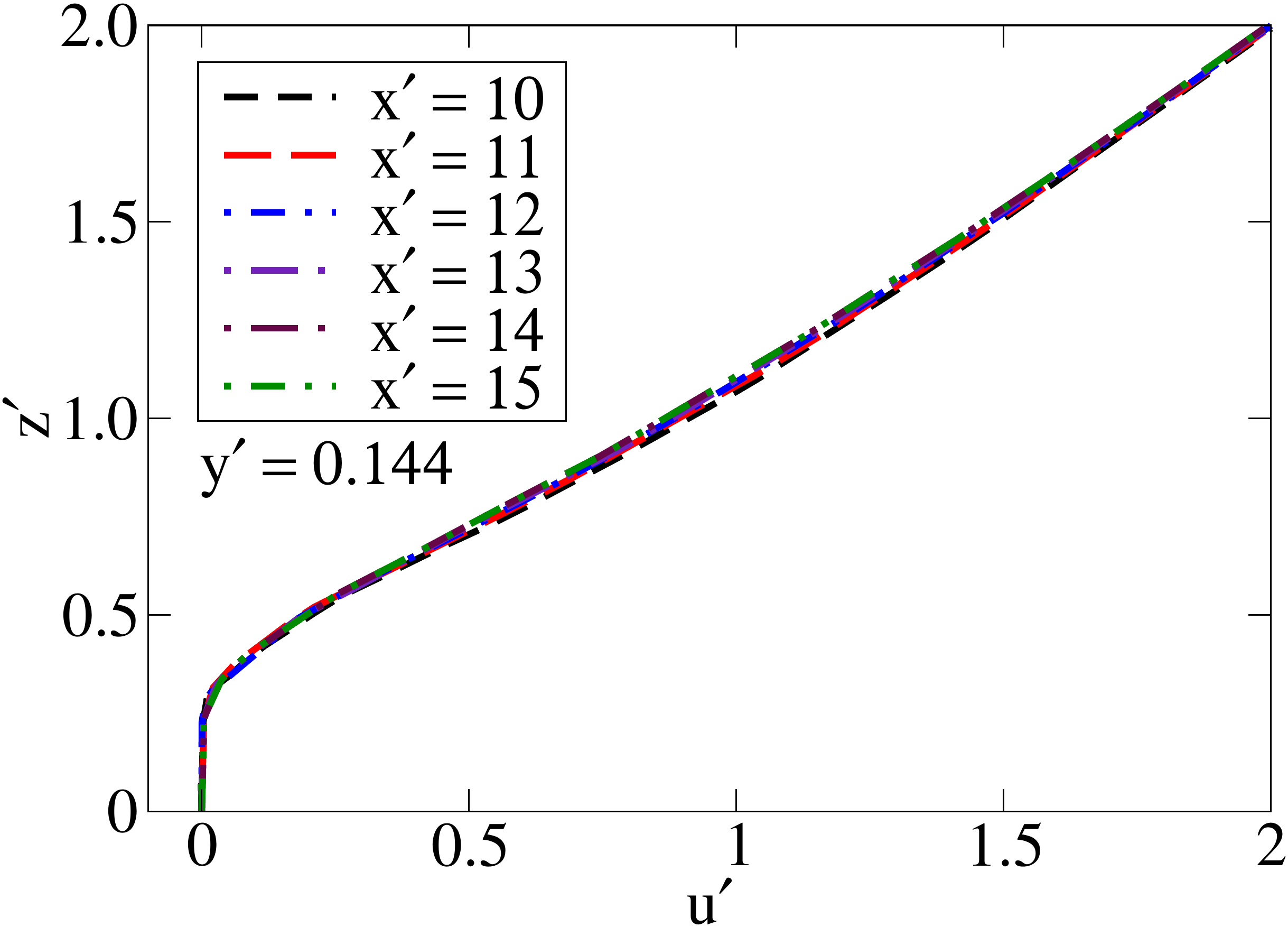}
		\caption{}
		\label{fig:Res50UVsZyPos0p144X}
	\end{subfigure}
	\quad
	\begin{subfigure}[b]{0.45\textwidth}
		\includegraphics[width=\textwidth]{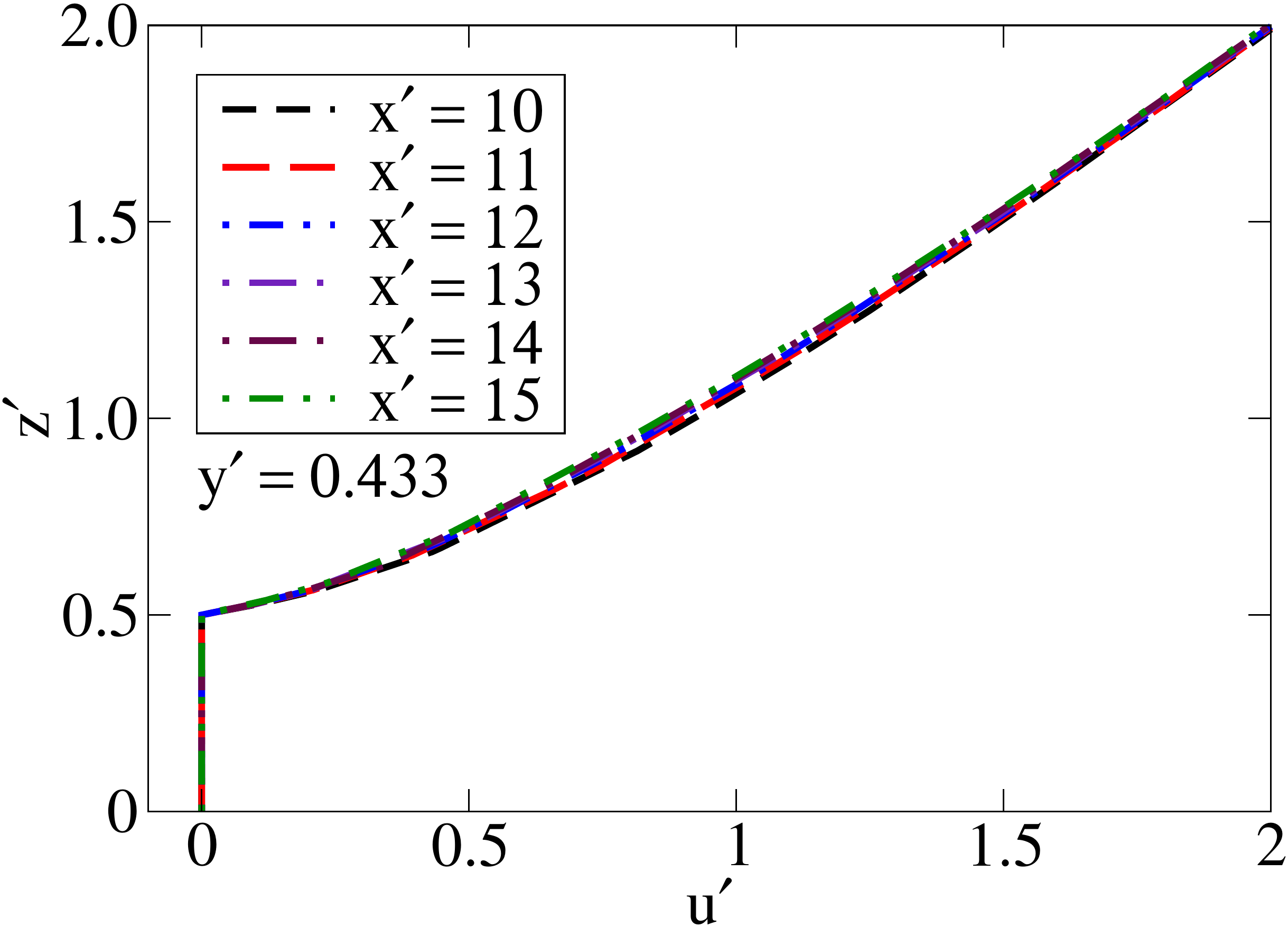}
		\caption{}
		\label{fig:Res50UVsZyPos0p433X}
	\end{subfigure}
	\quad
	\begin{subfigure}[b]{0.45\textwidth}
		\includegraphics[width=\textwidth]{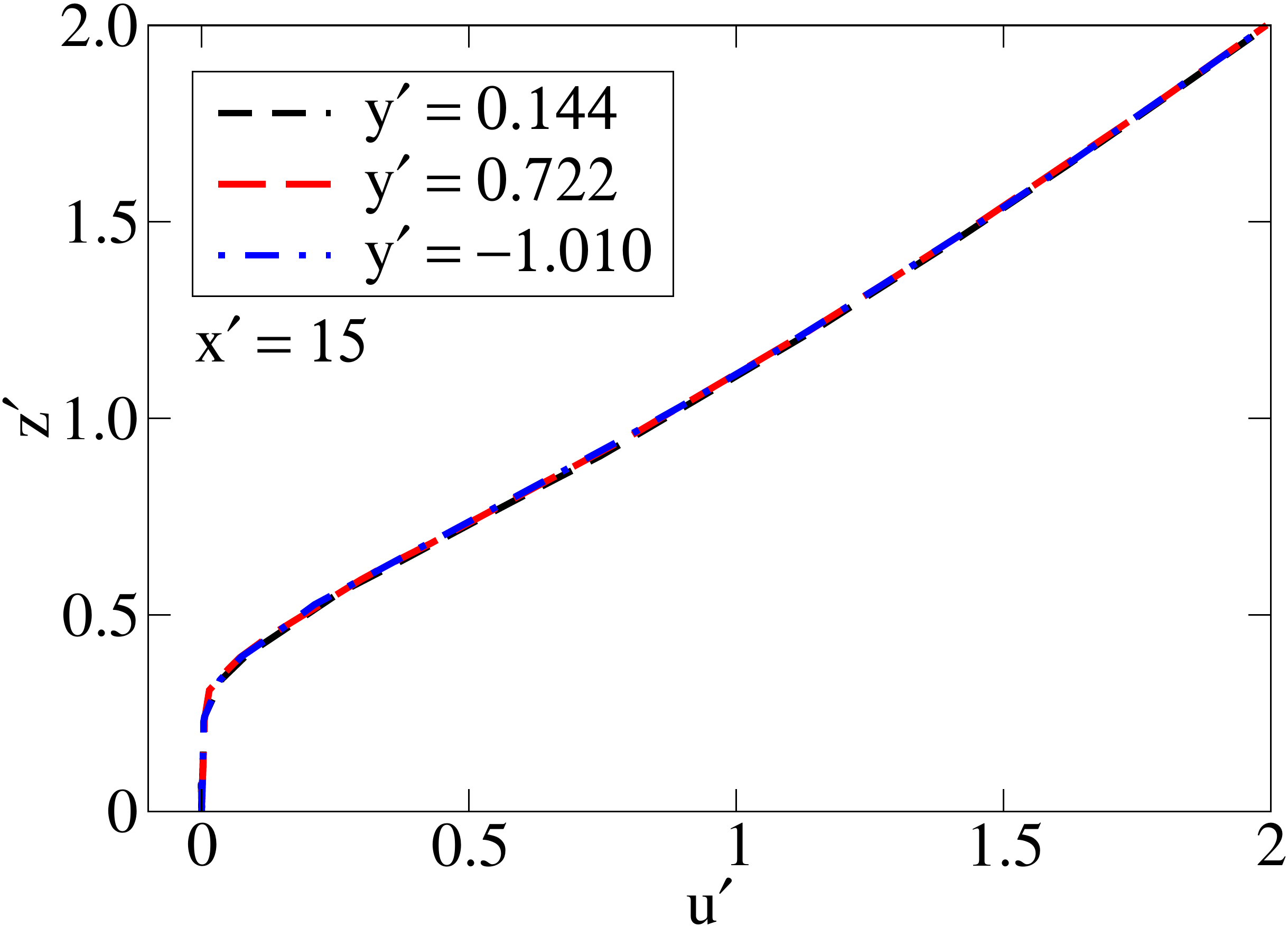}
		\caption{}
		\label{fig:Res50UVsZxPos15Y}
	\end{subfigure}
	\caption{Ambient velocity u variation with all normal direction z for $Re_s = 50$, at (a) $y^\prime$ = 0.144 and several x-locations, (b) $y^\prime$ = 0.433 and various x-locations, (c) $x^\prime$ = 15 and different y-locations}
	\label{fig:Res50UVsZyPosAndxPos}
\end{figure}
A fully resolved 3D simulations with body fitted fine mesh is computationally very much expensive. In the present study, the bottom wall  is made rough using hemispheres of equivalent diameter of the ellipsoid. Such an arrangement also adds up to the overall computational cost. One of the major reason for this is the requirement of fine mesh on these hemispheres as well to capture the boundary layer on the particle. Therefore, to optimize the computational cost, only a small section of the bottom wall is made rough with an ordered arrangement of the hemispheres as shown in \autoref{fig:RoughDomain1}. First, simulations are carried out for flow over rough wall in the absence of  particle. The rough bed is constructed arranging hemisphere in 10 rows and 20 columns.
The shear Reynolds number is varied  between 10 to 100. The flow profile obtained from the simulations are  plotted at different span-wise locations in domain. \cref{fig:Res50UVsZyPos0p144X,fig:Res50UVsZyPos0p433X,fig:Res50UVsZxPos15Y} show ambient flow profiles as a functions of wall normal direction ($z$) at different $x^\prime$ and $y^\prime$ locations over rough bed. It is observed that the flow remains developed for all the cases. \autoref{fig:Res50UVsZyPos0p144X} shows the velocity profile at $y^\prime = 0.144$, which represents the deep cavity (along trough) in to the rough surface. Similarly \autoref{fig:Res50UVsZyPos0p433X} shows velocity profile at $y^\prime = 0.433$ that is the top position (along crest) on surface of hemispheres. In both the \cref{fig:Res50UVsZyPos0p144X,fig:Res50UVsZyPos0p433X}, the velocity profiles are shown  at different stream-wise locations ($x^\prime$ = 10, 11, 12, 13, 14, and 15). The figures depict that the flow becomes developed after $x^\prime$ = 14, that is after $9^{th}$ row of hemispheres. Further investigation are carried out at $x^\prime$ = 15 and $y^\prime$ = 0.144, 0.722, and -1.010, which represent a deepest cavities along $y^\prime$ = 0.144 column. The corresponding profiles are shown in \autoref{fig:Res50UVsZxPos15Y}. Here, velocity profiles for all the considered positions along $y^\prime$ are seen to be merged into a single line, give  the confidence that the velocity is almost identical along that column of hemispheres around $x^\prime$ = 15.
\begin{figure}[htb]
	\centering
	\begin{subfigure}[b]{0.45\textwidth}
		\includegraphics[width=\textwidth]{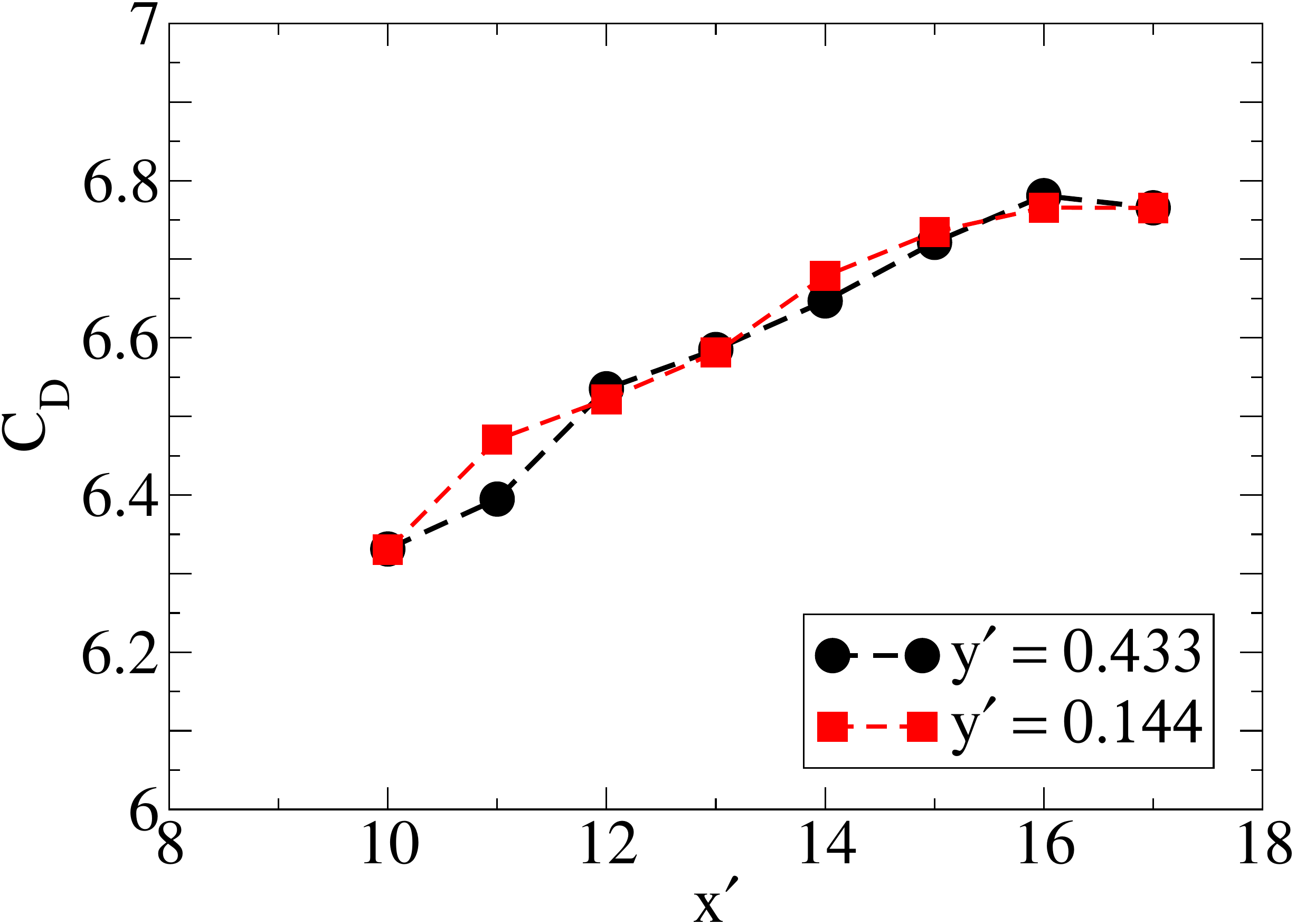}
		\caption{}
		\label{fig:Re10CdVsXyPos144}
	\end{subfigure}
	\quad
	\begin{subfigure}[b]{0.45\textwidth}
		\includegraphics[width=\textwidth]{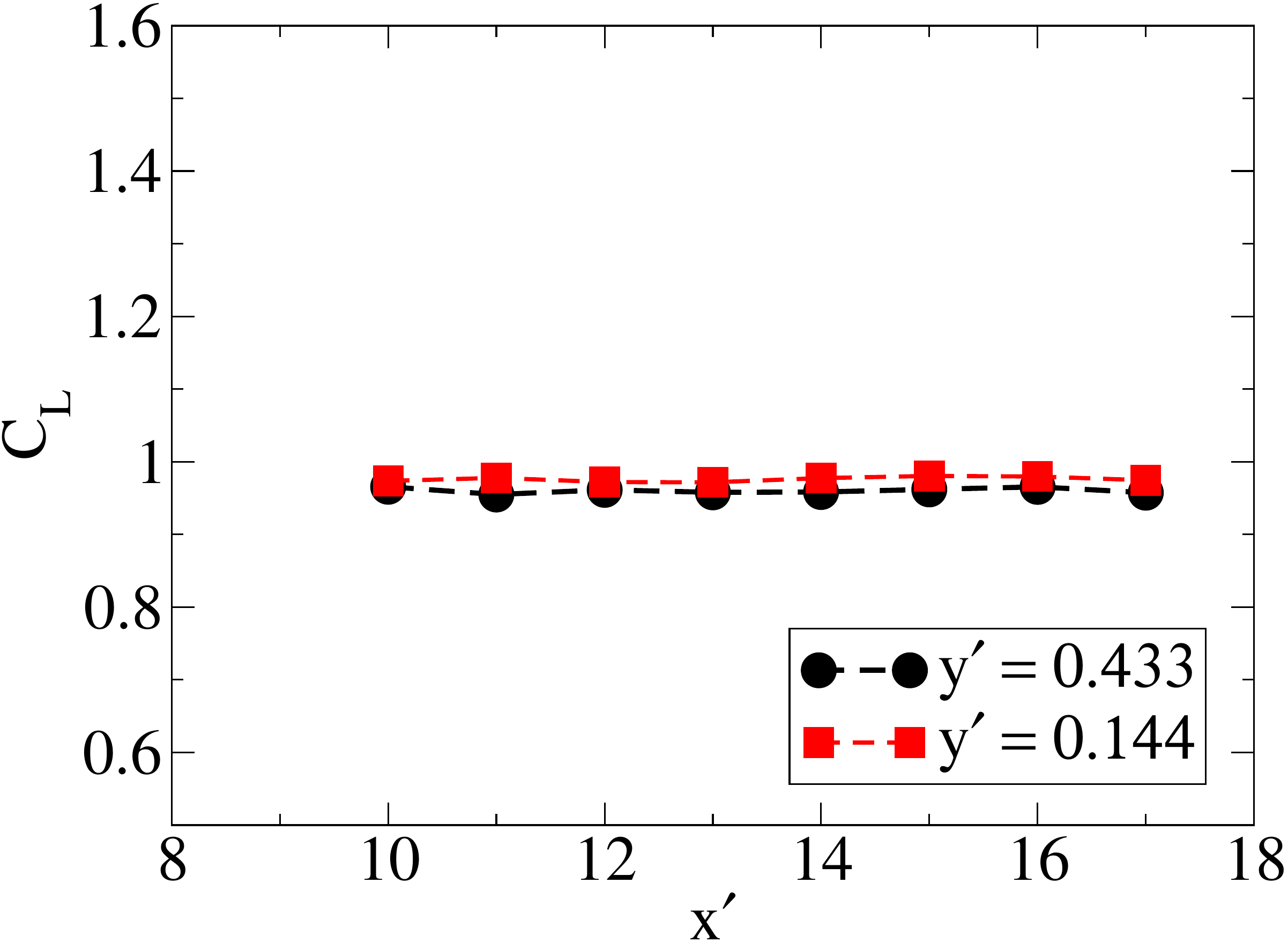}
		\caption{}
		\label{fig:Re10ClVsXyPos144}
	\end{subfigure}
	\caption{(a) Drag and (b) lift coefficients variation over several x-position on rough bed ($\delta = 0.1$) for $y^\prime$ = 0.144, $Re_s = 10$ and AR = 2.}
	\label{fig:CdClVsXyPos0p144Re10}
\end{figure}
\begin{figure}[htb]
	\centering
	\includegraphics[width=0.5\textwidth]{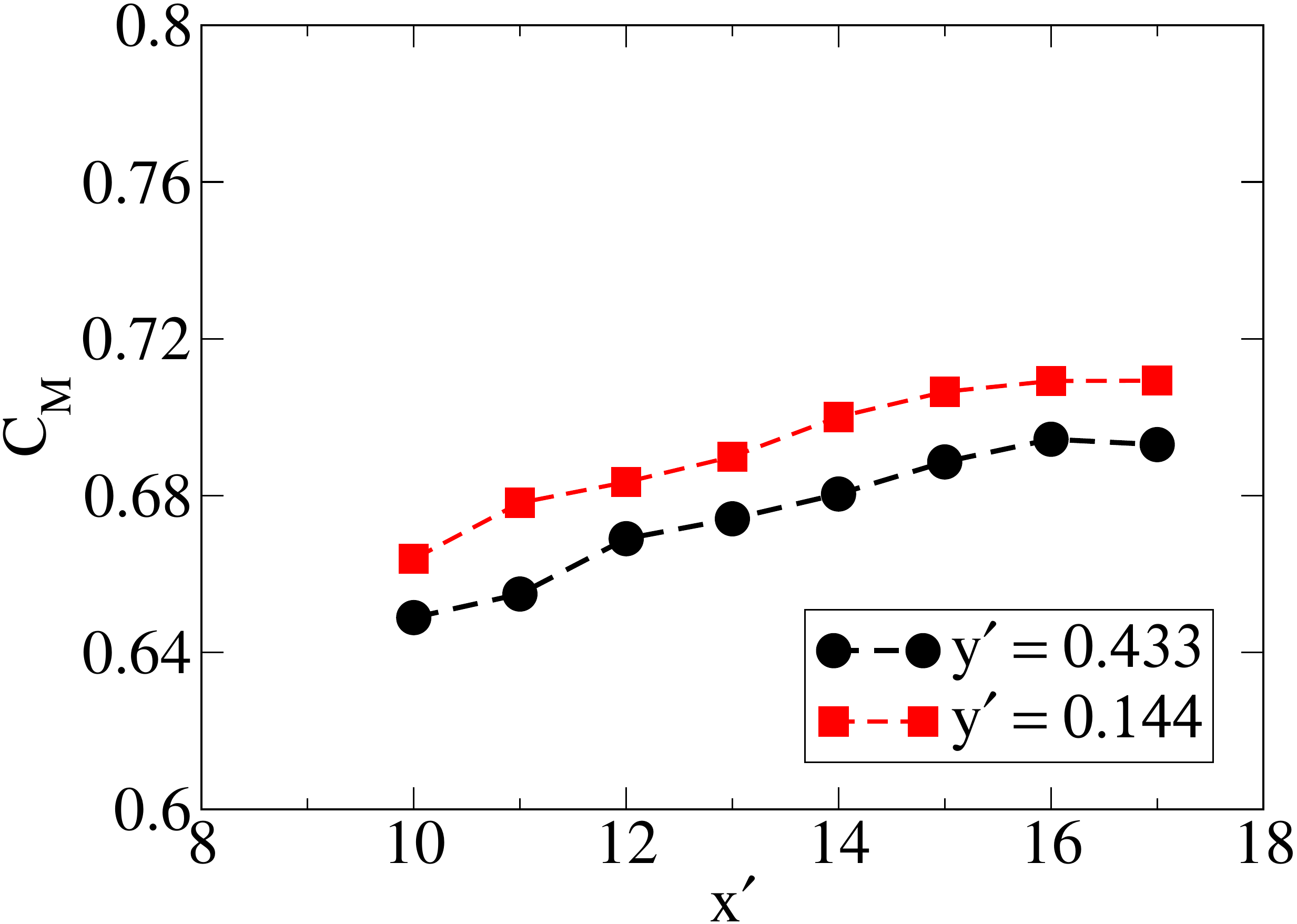}
	\caption{Torque coefficient variation over several x-position on rough bed ($\delta = 0.1$) for $y^\prime$ = 0.144, $Re_s = 10$ and AR = 2.}
	\label{fig:Re10CmVsXyPos144}
\end{figure}
The next set of simulations is conducted to explore the location on the rough bed where the hydrodynamic forces on the particle are not be influenced by the edge effect of the rough bed. Three different cases are simulated based on the the 
Reynolds numbers ($Re_s$) 10, 50, and 100. Since presence of the wall affects the hydrodynamic forces acting on particle \citep{ZengPOF2009,LeeIJMF2017,zarghami2018},  it is important to carry out these test cases  with particle positioned near the wall. Here we place the particle at $\delta = 0.1$ in $z$-direction, which presents the situation  where the particle  just touches the rough bed. The  span-wise position, $y^\prime = 0.144$ and the particle position is varied in $x$-direction ($x^\prime$ = 10, 11, 12, 13, 14, 15, 16, and 17). The drag, lift, and torque coefficients are then calculated  and are shown in \autoref{fig:CdClVsXyPos0p144Re10}. 
The figure shows that the variation of drag coefficient  is only $0.5\%$ for position higher than $x^\prime=15$. In case of the lift coefficient (\autoref{fig:Re10ClVsXyPos144}), the variation is small and it remains almost constant with an approximate change of  $0.5\%$  over entire range of the particle positions.  \autoref{fig:Re10CmVsXyPos144} shows the variation of torque coefficient with particle position in stream-wise direction. An insignificant change is observed for particle position $x^\prime < 15$ and for $x^\prime > 15$ the 
coefficient appears to be constant  having a maximum deviation of $0.2\%$. The above analysis confirms that a stream-wise distance $x^\prime = 15$ or higher and a span-wise location $y^\prime = 0.144$ can be considered to obtain hydrodynamic forces which are free of edge effects of rough bed.

\subsection{Grid independence study}
\begin{figure}[htb]
	\centering
	\begin{subfigure}[b]{0.45\textwidth}
		\includegraphics[width=\textwidth]{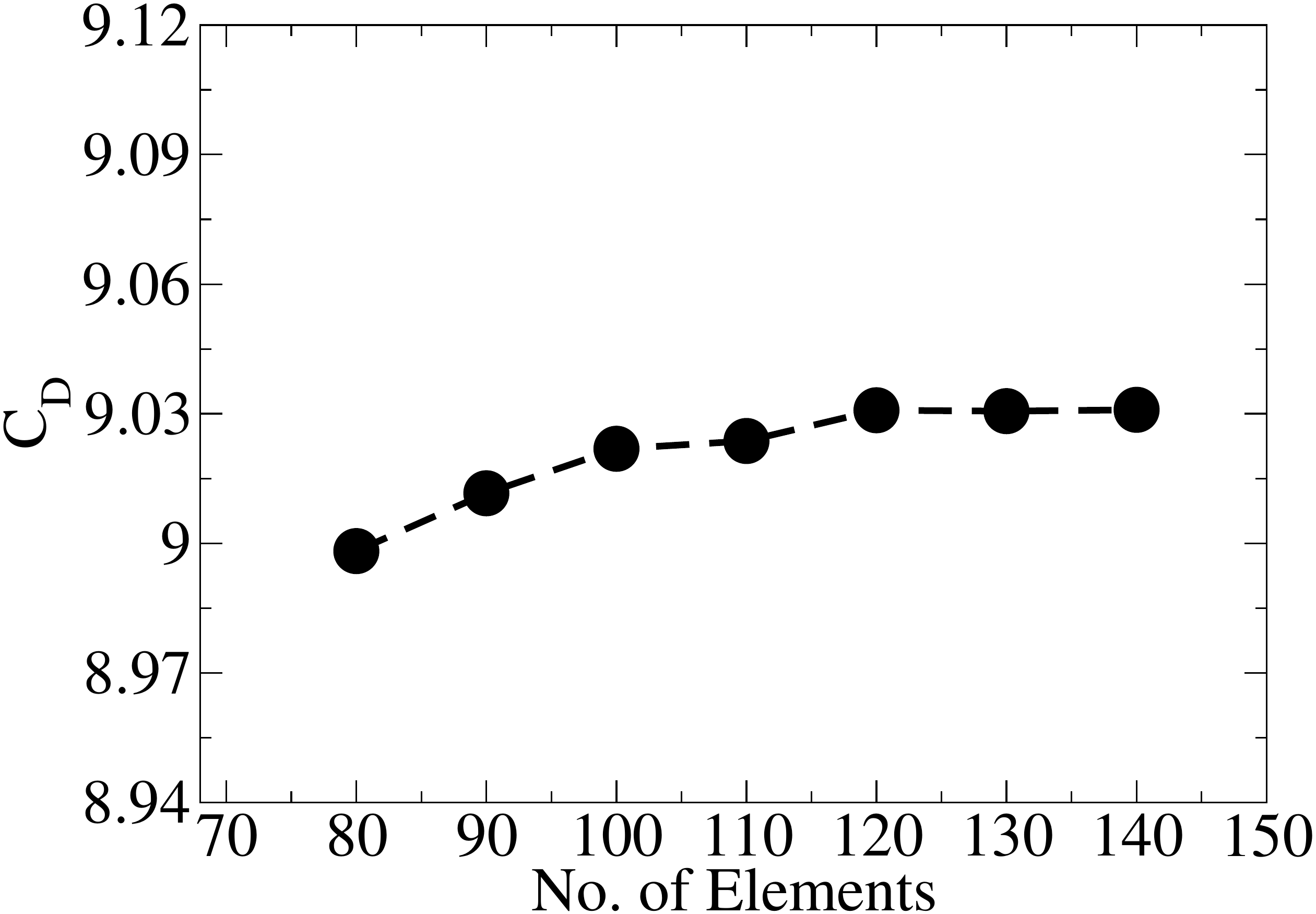}
		\caption{}
		\label{fig:cdVsNumGridPtsAtRes10Delta0p1Theta0}
	\end{subfigure}
	\quad
	\begin{subfigure}[b]{0.45\textwidth}
		\includegraphics[width=\textwidth]{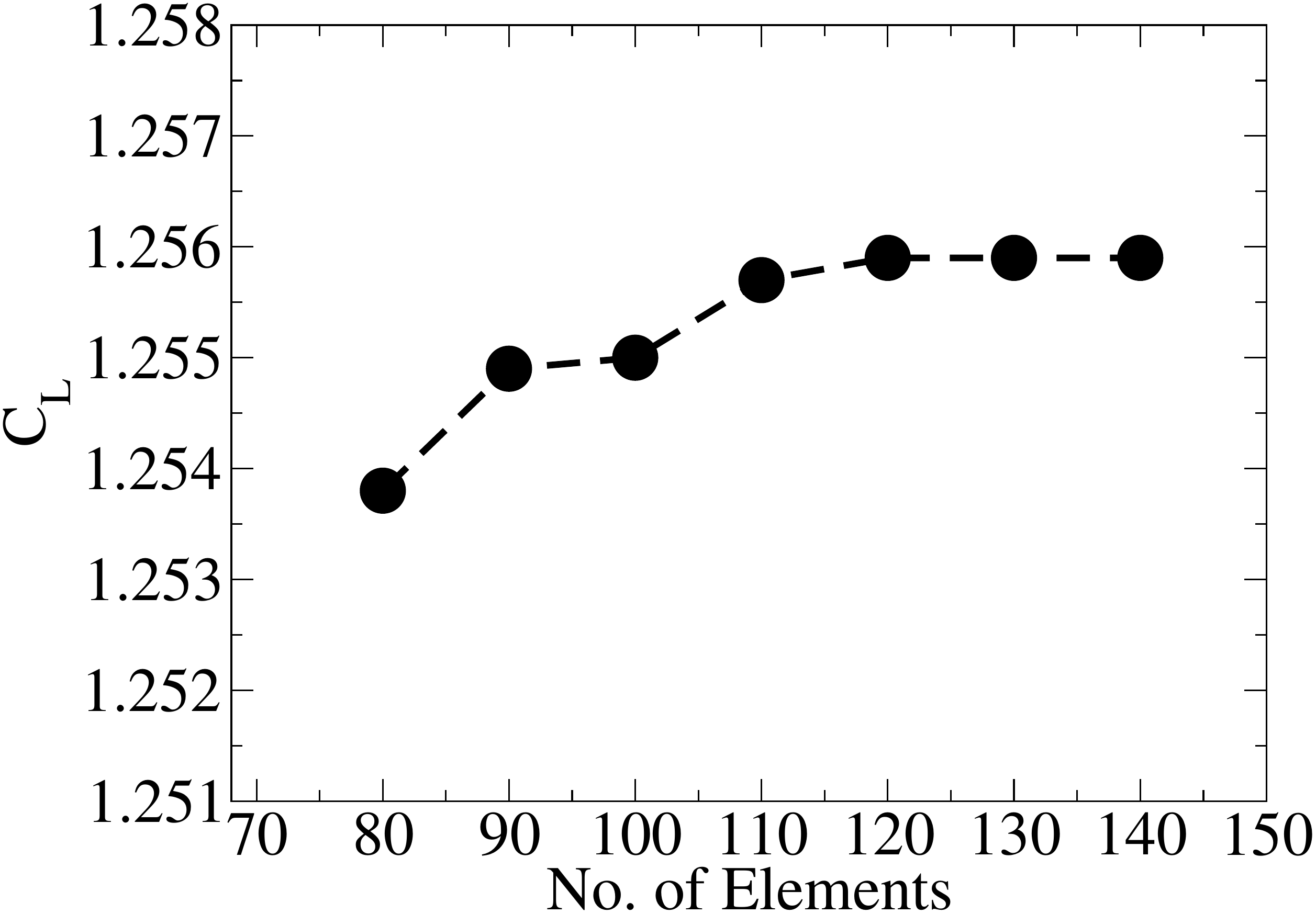}
		\caption{}
		\label{fig:clVsNumGridPtsAtRes10Delta0p1Theta0}
	\end{subfigure}
	\quad
	\begin{subfigure}[b]{0.45\textwidth}
		\includegraphics[width=\textwidth]{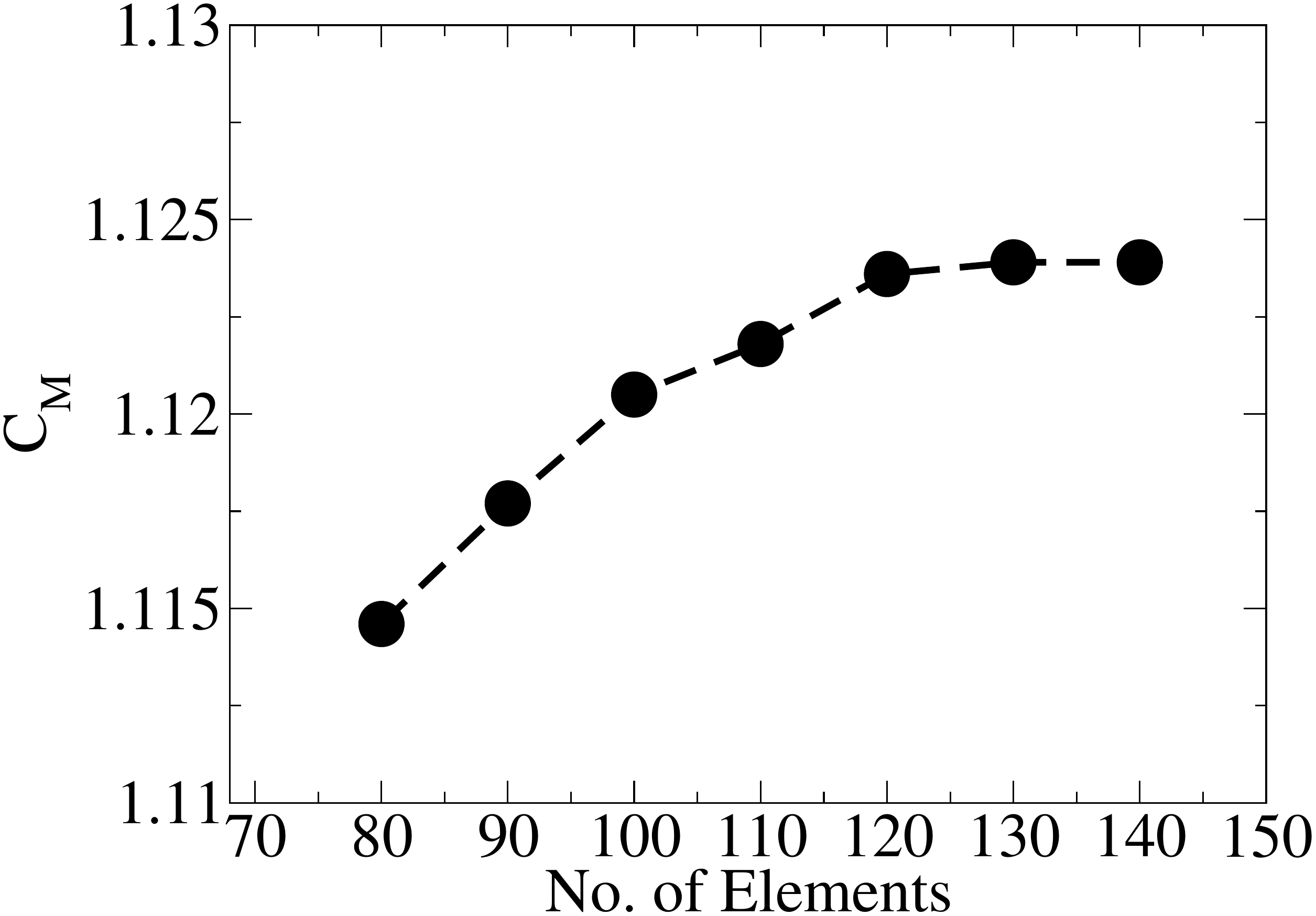}
		\caption{}
		\label{fig:cmVsNumGridPtsAtRes10Delta0p1Theta0}
	\end{subfigure}
	\caption{Hydrodynamic coefficients variation with number of elements on ellipsoid surface at $Re_s = 10$ and $\theta = \ang{0}$}
	\label{fig:cdClCmVsNumGridPtsAtRes10Delta0p1Theta0}
\end{figure}
A proper grid resolution is a must to calculate accurate forces on the particle. Either the CFD results are to be validated with experiments to get an idea of proper resolution or in case of unavailability of the experimental data, grid-independent results are to be produced through different levels of grid refinements. 
There is no standard protocol yet to set-up the required grid resolution to analyse the flow over an object when it is placed near a wall with finite roughness \cite{LeeIJMF2017}. Therefore, a grid independence study for the present work is of foremost importance. Use of an extremely fine mesh throughout the domain will also be computationally very expensive.  Thus, an extremely fine mesh has been  employed on the rough-surface using 3 layers of {\it boundary layer meshing}. The number of elements on the ellipsoid is then systematically varied in order to obtain a converge solution of the flow. The effect of variation of number of mesh-elements 
on the drag, lift, and torque coefficients  is shown in the \autoref{fig:cdClCmVsNumGridPtsAtRes10Delta0p1Theta0}. Although it is observed that there is an insignificant variation ($\approx 1\%$) of the coefficients with the change in  the number of elements considered here, we have considered 120 elements on the ellipsoid to perform all the simulations reported here.


\section{Results and discussions}\label{sec:results}
\subsection{Flow structure around ellipsoid}
\begin{figure}[htb]
	\centering
	\begin{subfigure}[b]{0.22\textwidth}
		\includegraphics[width=\textwidth]{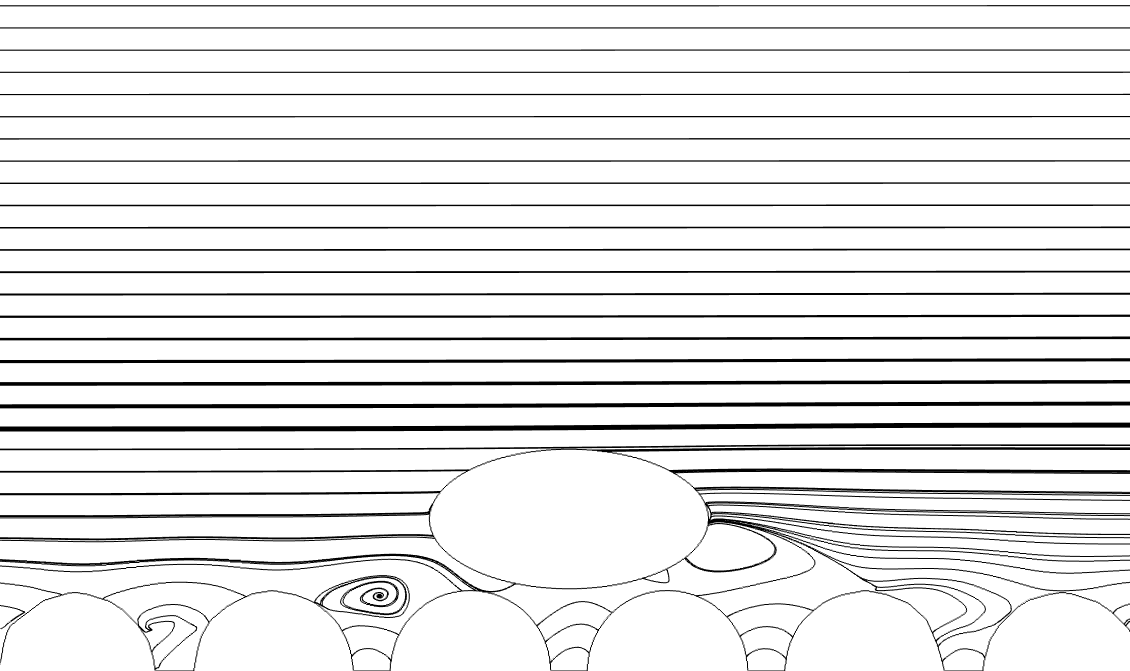}
		\caption{}
		\label{fig:streamlineRe100theta0AR2Delta0p1}
	\end{subfigure}
	\quad
	\begin{subfigure}[b]{0.22\textwidth}
		\includegraphics[width=\textwidth]{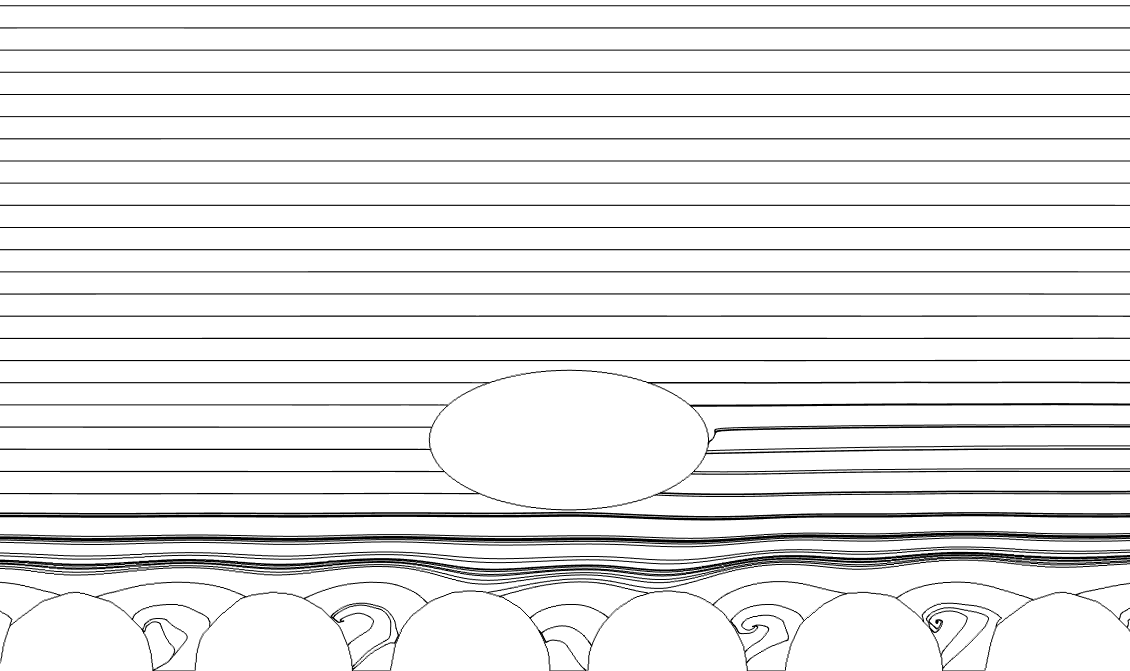}
		\caption{}
		\label{fig:streamlineRe100theta0AR2Delta0p5}
	\end{subfigure}
	\quad
	\begin{subfigure}[b]{0.22\textwidth}
		\includegraphics[width=\textwidth]{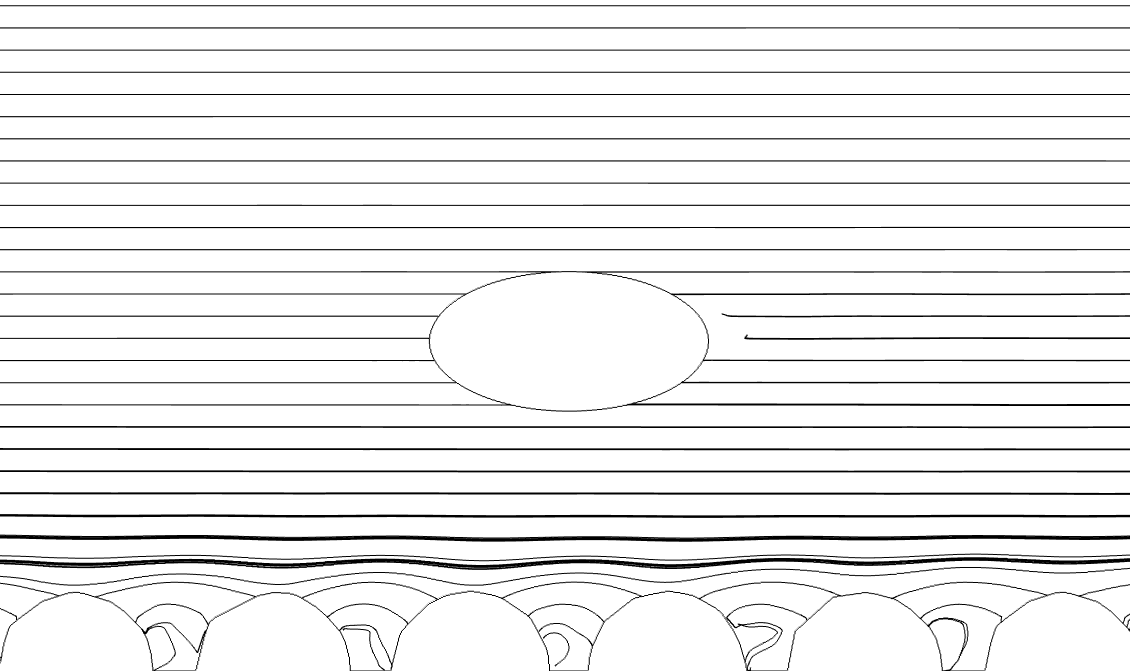}
		\caption{}
		\label{fig:streamlineRe100theta0AR2Delta1p0}
	\end{subfigure}
	\quad
	\begin{subfigure}[b]{0.22\textwidth}
		\includegraphics[width=\textwidth]{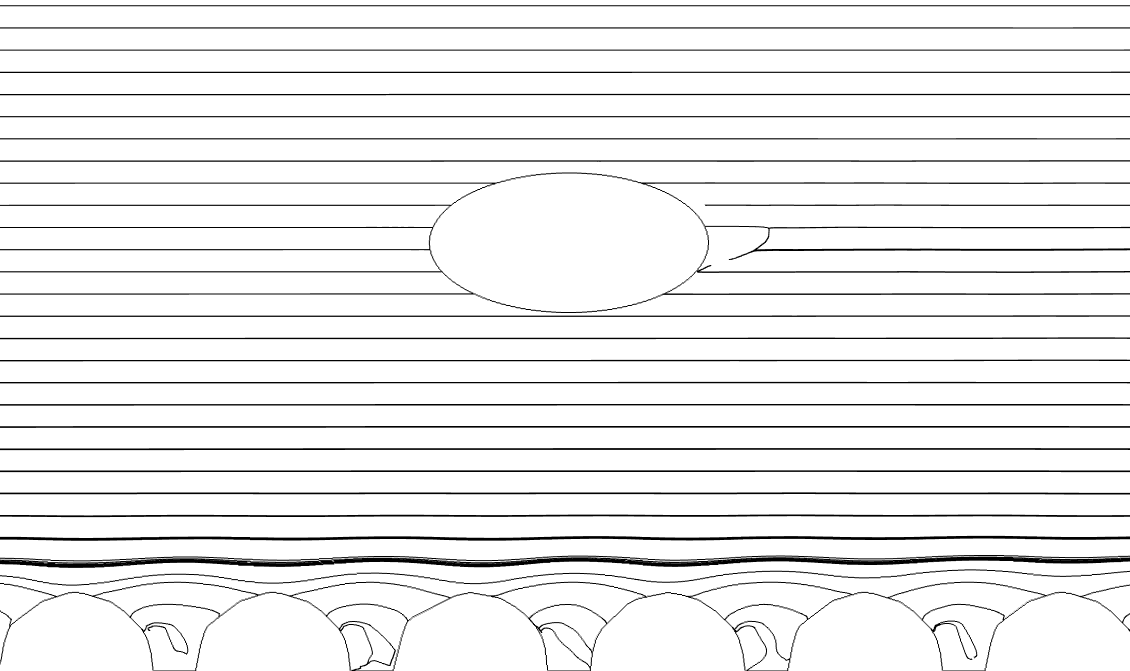}
		\caption{}
		\label{fig:streamlineRe100theta0AR2Delta1p5}
	\end{subfigure}
	\\
	\begin{subfigure}[b]{0.22\textwidth}
		\includegraphics[width=\textwidth]{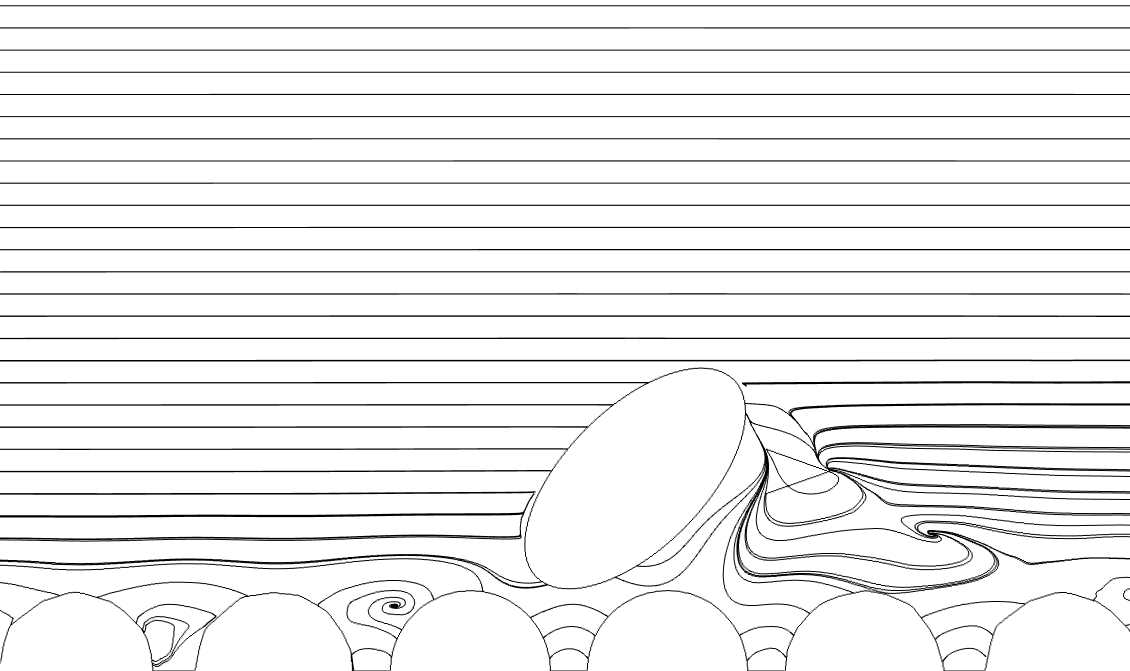}
		\caption{}
		\label{fig:streamlineRe100theta45AR2Delta0p1}
	\end{subfigure}
	\quad
	\begin{subfigure}[b]{0.22\textwidth}
		\includegraphics[width=\textwidth]{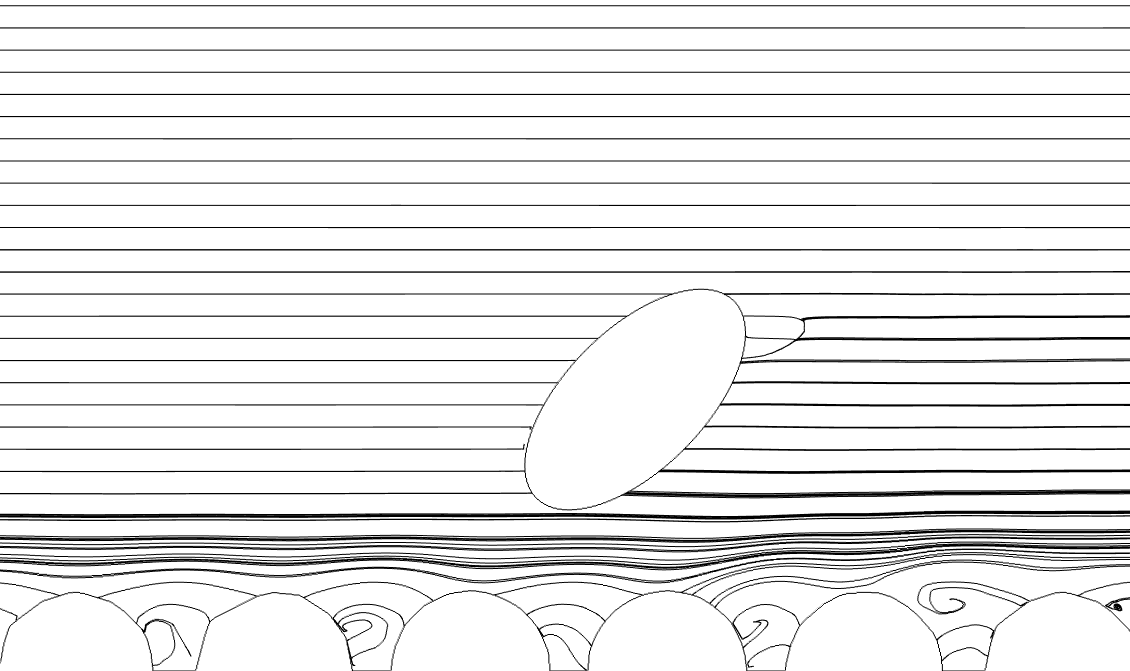}
		\caption{}
		\label{fig:streamlineRe100theta45AR2Delta0p5}
	\end{subfigure}
	\quad
	\begin{subfigure}[b]{0.22\textwidth}
		\includegraphics[width=\textwidth]{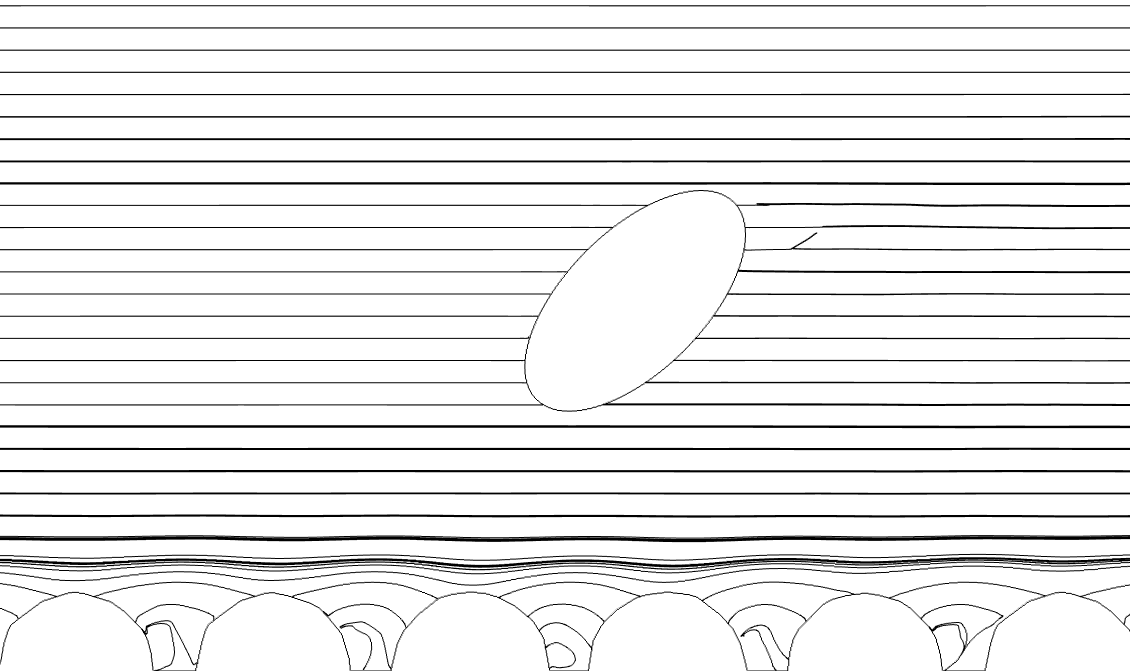}
		\caption{}
		\label{fig:streamlineRe100theta45AR2Delta1p0}
	\end{subfigure}
	\quad
	\begin{subfigure}[b]{0.22\textwidth}
		\includegraphics[width=\textwidth]{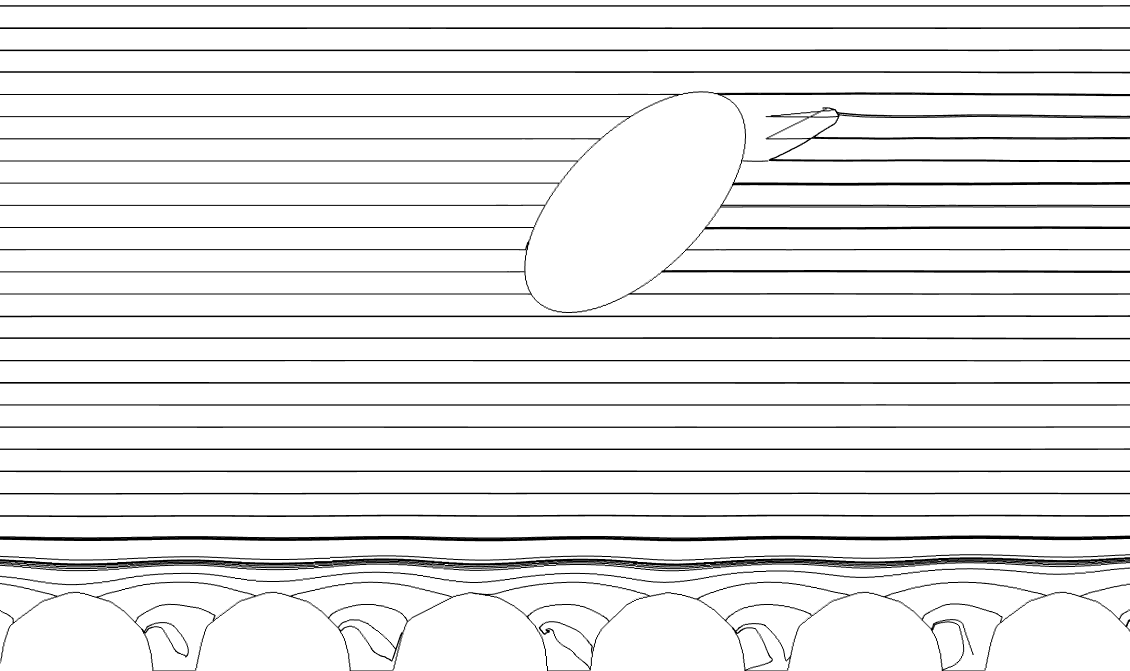}
		\caption{}
		\label{fig:streamlineRe100theta45AR2Delta1p5}
	\end{subfigure}
	\\
	\begin{subfigure}[b]{0.22\textwidth}
		\includegraphics[width=\textwidth]{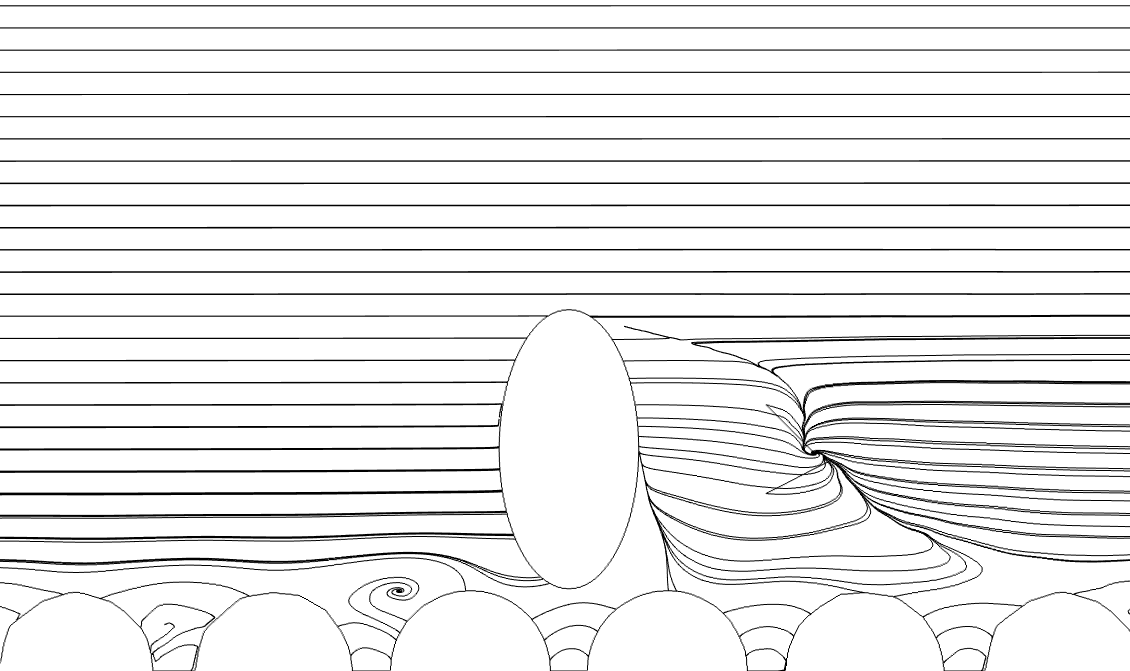}
		\caption{}
		\label{fig:streamlineRe100theta90AR2Delta0p1}
	\end{subfigure}
	\quad
	\begin{subfigure}[b]{0.22\textwidth}
		\includegraphics[width=\textwidth]{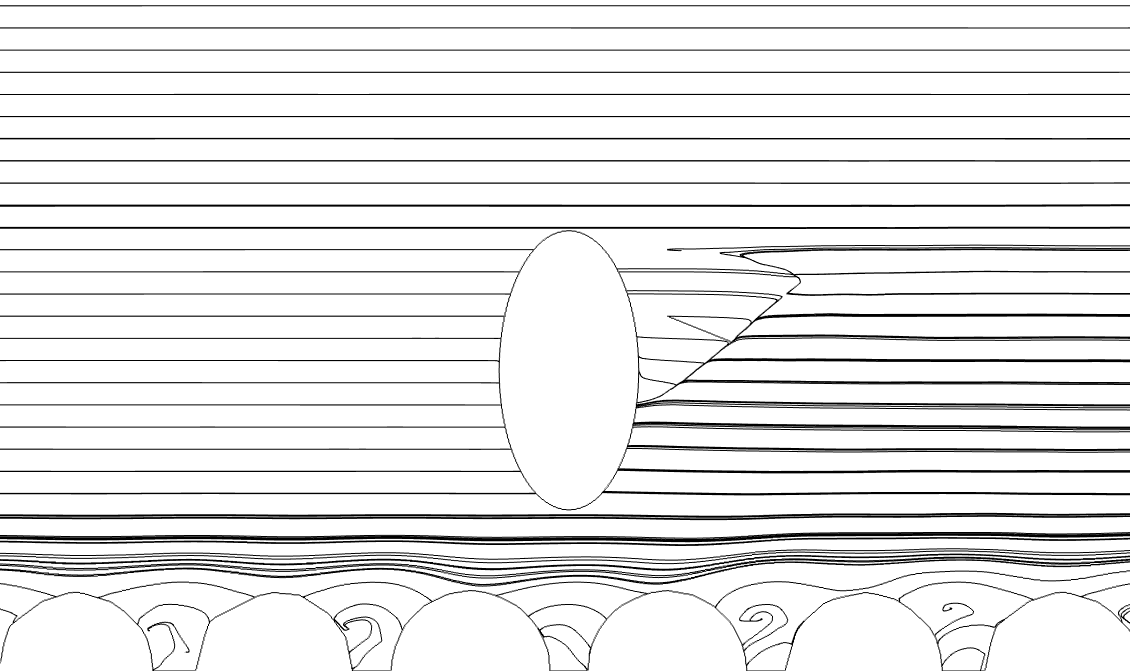}
		\caption{}
		\label{fig:streamlineRe100theta90AR2Delta0p5}
	\end{subfigure}
	\quad
	\begin{subfigure}[b]{0.22\textwidth}
		\includegraphics[width=\textwidth]{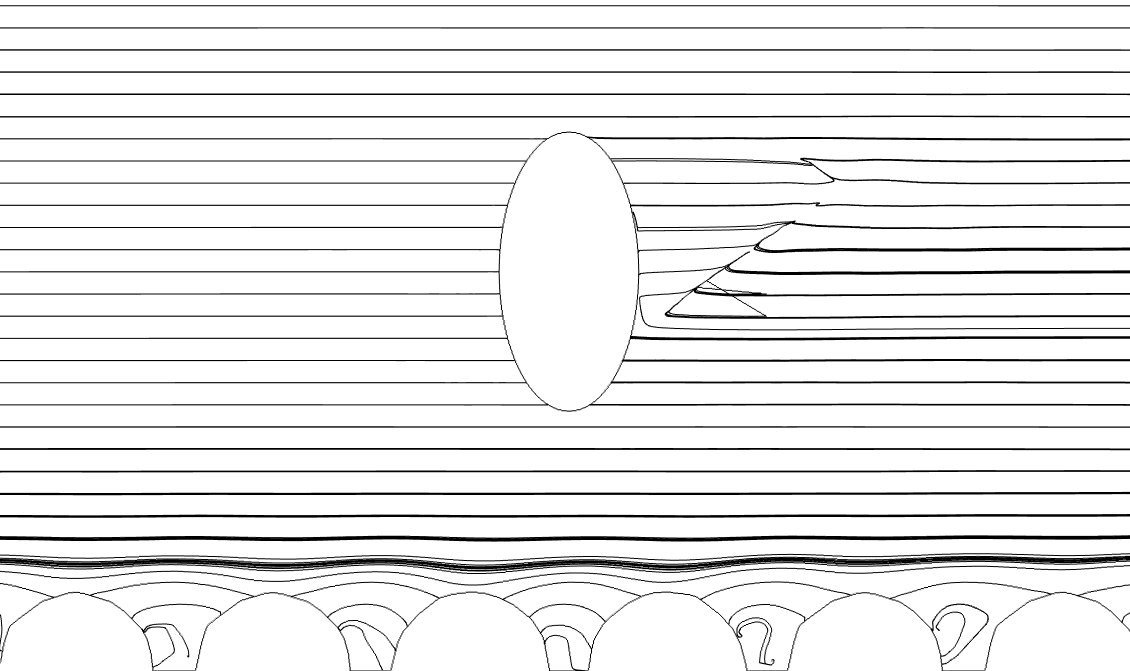}
		\caption{}
		\label{fig:streamlineRe100theta90AR2Delta1p0}
	\end{subfigure}
	\quad
	\begin{subfigure}[b]{0.22\textwidth}
		\includegraphics[width=\textwidth]{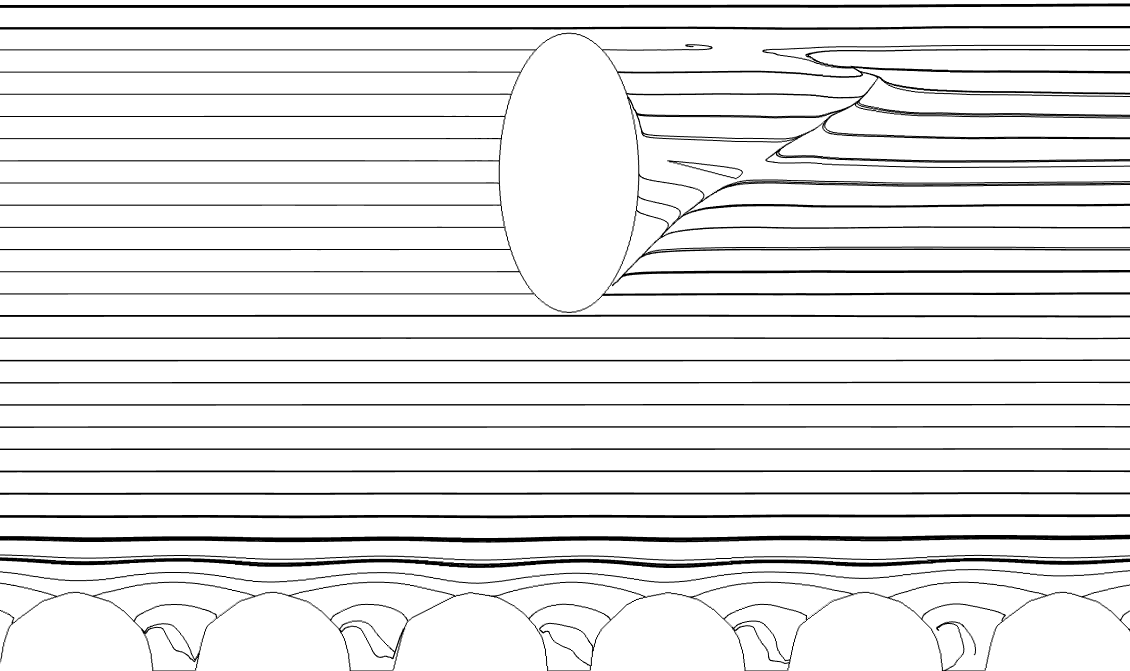}
		\caption{}
		\label{fig:streamlineRe100theta90AR2Delta1p5}
	\end{subfigure}
	\\
	\begin{subfigure}[b]{0.22\textwidth}
		\includegraphics[width=\textwidth]{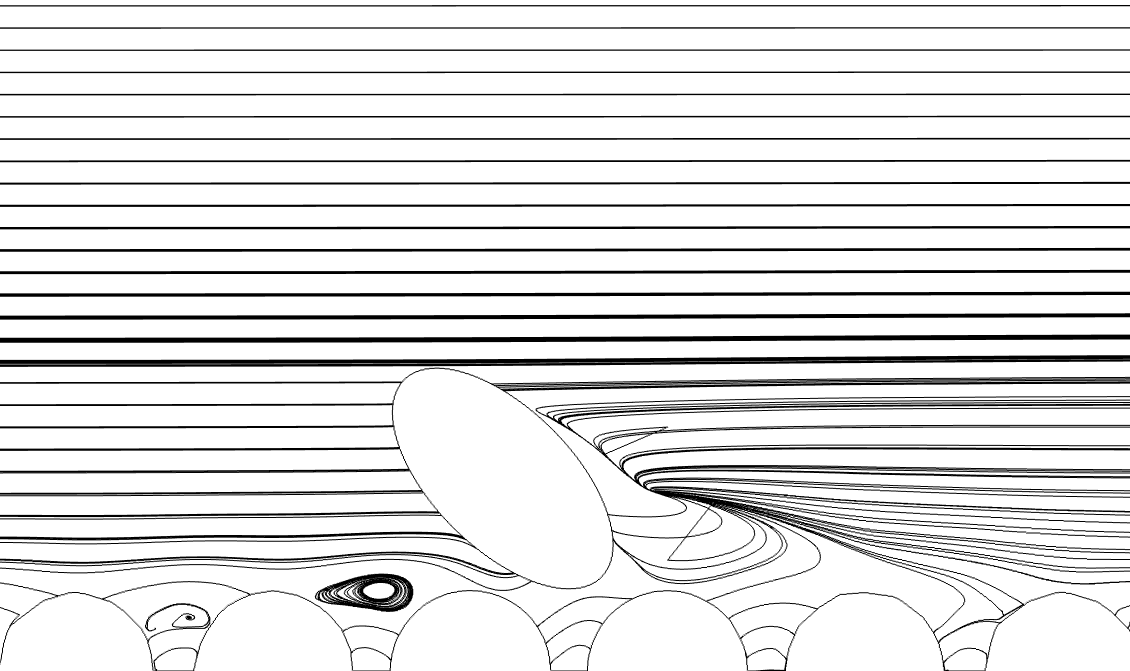}
		\caption{}
		\label{fig:streamlineRe100theta135AR2Delta0p1}
	\end{subfigure}
	\quad
	\begin{subfigure}[b]{0.22\textwidth}
		\includegraphics[width=\textwidth]{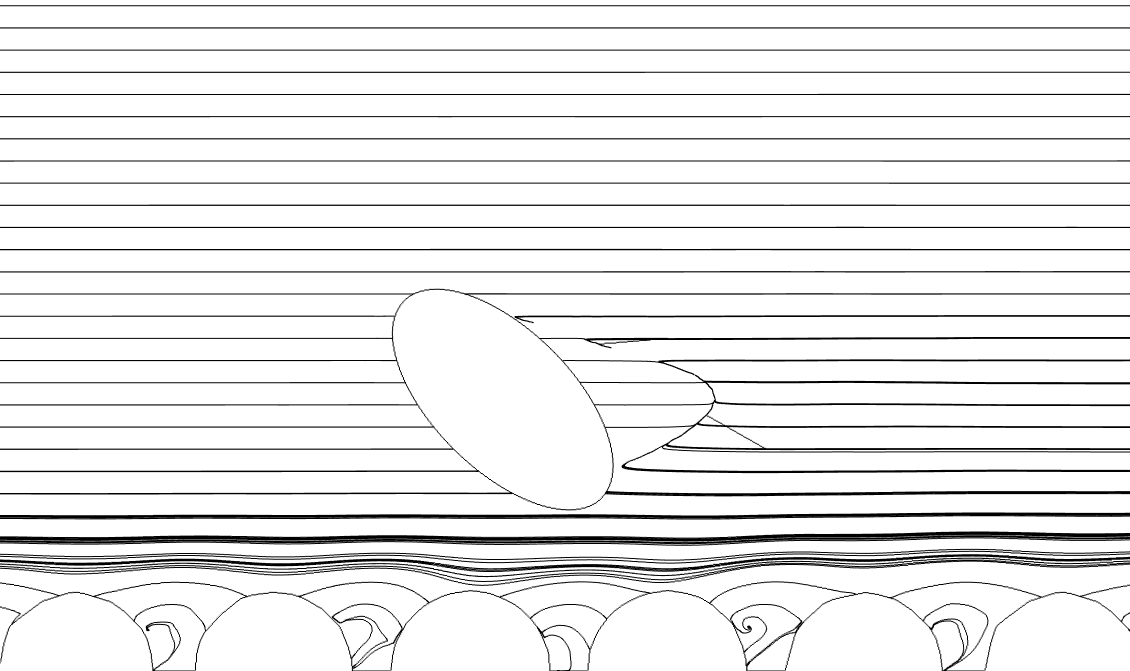}
		\caption{}
		\label{fig:streamlineRe100theta135AR2Delta0p5}
	\end{subfigure}
	\quad
	\begin{subfigure}[b]{0.22\textwidth}
		\includegraphics[width=\textwidth]{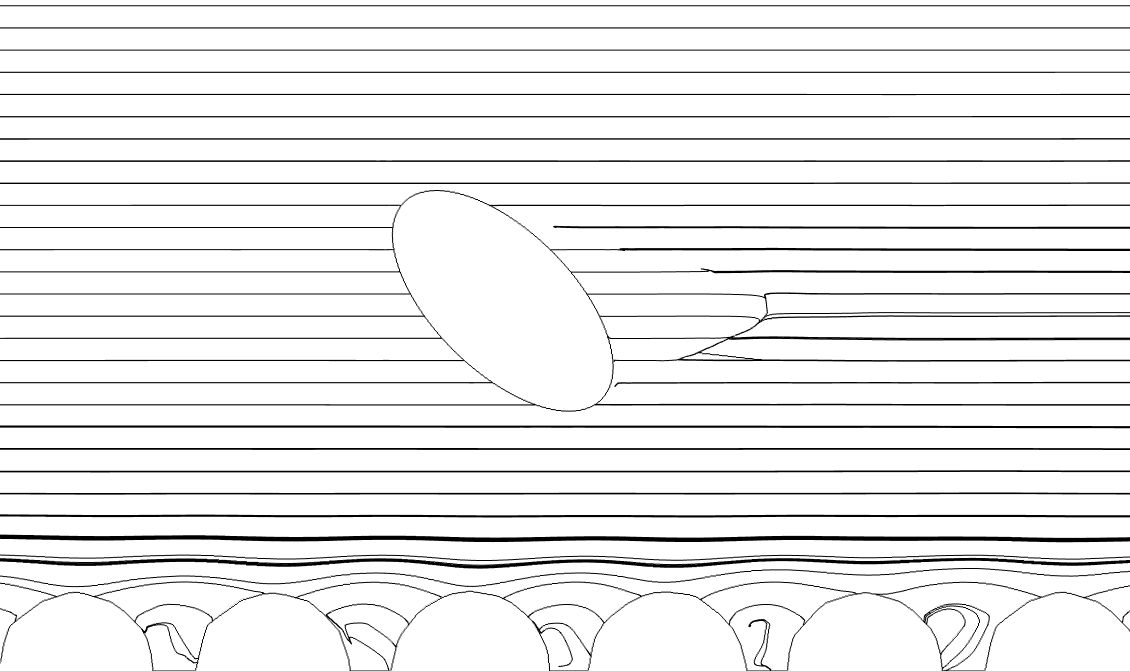}
		\caption{}
		\label{fig:streamlineRe100theta135AR2Delta1p0}
	\end{subfigure}
	\quad
	\begin{subfigure}[b]{0.22\textwidth}
		\includegraphics[width=\textwidth]{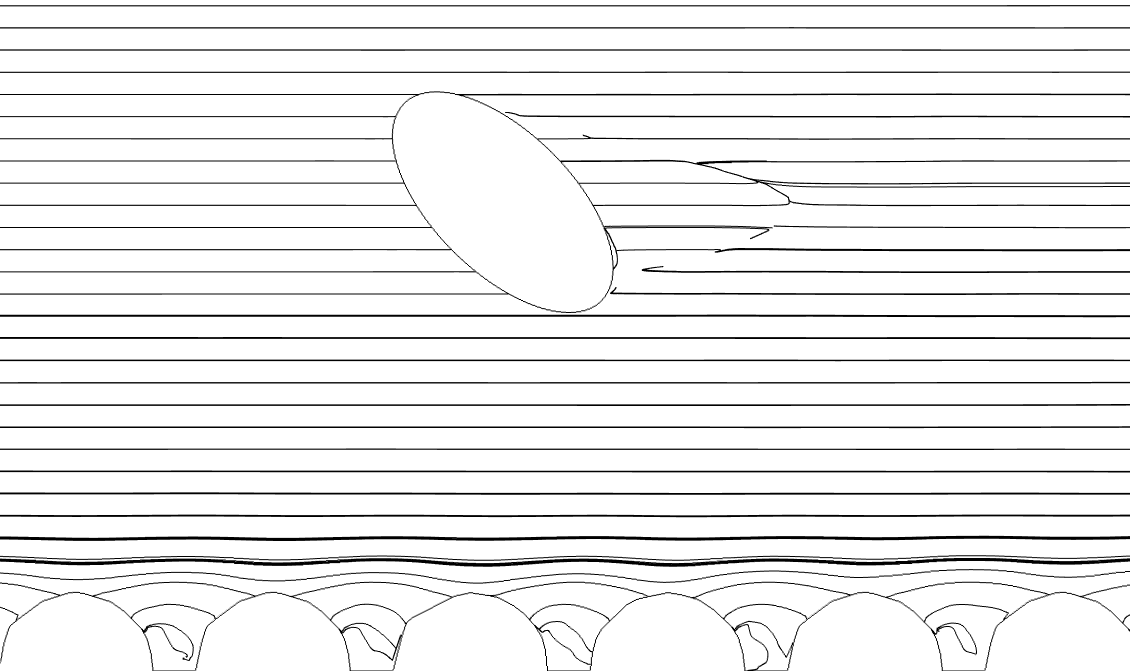}
		\caption{}
		\label{fig:streamlineRe100theta135AR2Delta1p5}
	\end{subfigure}
	\caption{Streamlines across particle at $Re_s = 100$, $\theta =\ang{0}$ (first row), \ang{45} (second row), \ang{90} (third row) and $\theta =\ang{135}$ (fourth row) and columns represents constant $\delta\ (= 0.1, 0.5, 1.0, 1.5)$ from left to right}
	\label{fig:streamlineRe100thetaAllAR2DeltaAll}
\end{figure}

The effects of shear Reynolds number $Re_s$ and wall-normal distance along with orientation angle of the ellipsoid are shown with the roughness and particle induced disturbed flow field. There are significant contribution from the particle-wall separation distance and orientation of the particle to decide the flow characteristics  around the ellipsoid. \autoref{fig:streamlineRe100thetaAllAR2DeltaAll} shows streamlines across the ellipsoid at the plane of center (i.e. $y^\prime$ = 0), at $Re_s = 100$ for  different value of wall-normal distances $\delta = 0.1, 0.5, 1.0, 1.5$ and orientation angle $\theta = \ang{0}, \ang{45}, \ang{90}, $ and $\ang{135}$. 
The streamline plots show that as the $\delta$ increases, the effect of wall diminishes rapidly. 
For $\delta > 0.5$, at all orientation angles, only the effect of ambient shear but not the wall roughness is observed. A similar observation is followed for all simulations carried out at lower $Re_s$'s (not shown here). Appearance of the wake behind the ellipsoid is mainly observed for higher values of $Re_s$'s. The structure of wake is also found to vary with separation from wall and the orientation angle. 
The location  of the wake on either top (\autoref{fig:streamlineRe100theta45AR2Delta1p5}) or bottom (\autoref{fig:streamlineRe100theta135AR2Delta1p5}) part of the ellipsoid depends on the orientation angle to which it is exposed to the background flow. In the case of $\theta = \ang{90}$, when a particle is away from the wall, the asymmetric wake is only due to ambient shear of flow.
	
\begin{figure}[htb]
		\centering
		\begin{subfigure}[b]{0.3\textwidth}
			\includegraphics[width=\textwidth]{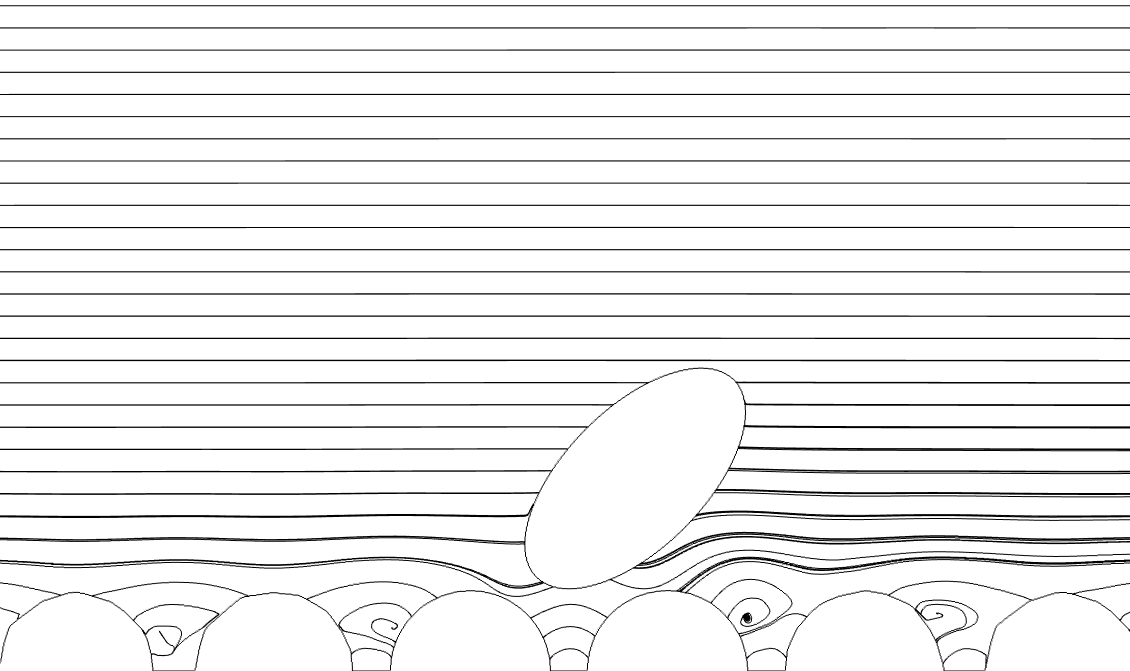}
			\caption{}
			\label{fig:streamlineRe10theta45AR2Delta0p1}
		\end{subfigure}
		\quad
		\begin{subfigure}[b]{0.3\textwidth}
			\includegraphics[width=\textwidth]{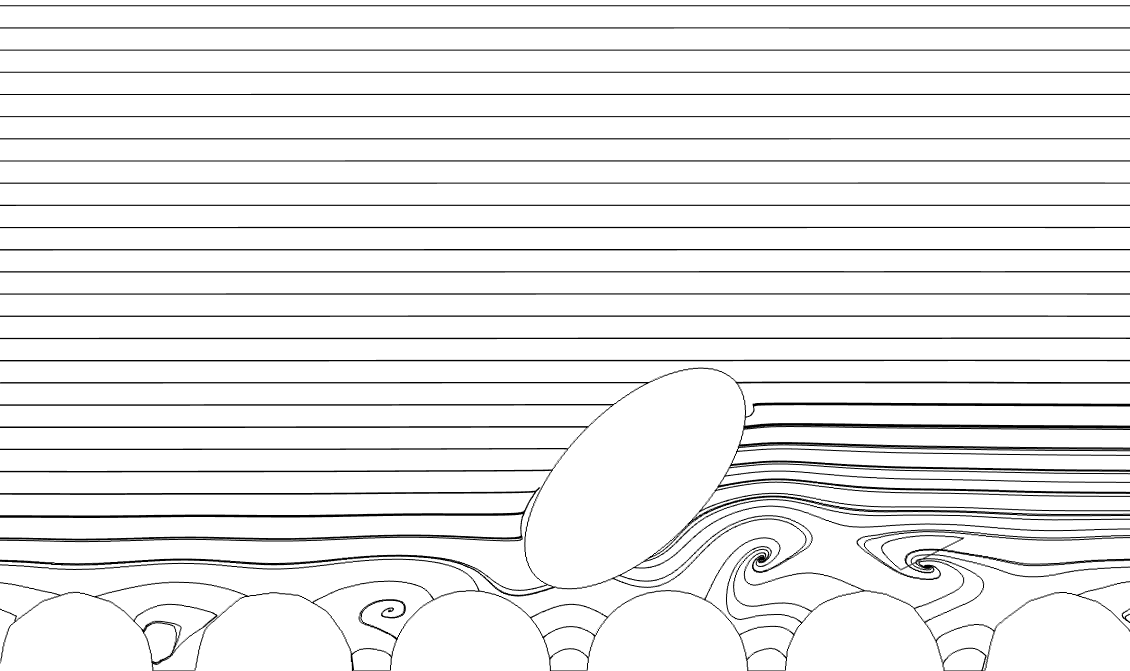}
			\caption{}
			\label{fig:streamlineRe50theta45AR2Delta0p1}
		\end{subfigure}
		\quad
		\begin{subfigure}[b]{0.3\textwidth}
			\includegraphics[width=\textwidth]{streamlineRe100theta45AR2Delta0p1.png}
			\caption{}
			\label{fig:streamlineRe100theta45AR2Delta0p1}
		\end{subfigure}
		\caption{Flow streamlines around the ellipsoid for AR = 2, $\theta$ = $\ang{45}$, at (a) $Re_s = 10$ (b) $Re_s = 50$ (c) $Re_s = 100$}
		\label{fig:streamlineReAlltheta45AR2Delta0p1}
	\end{figure}

The effect of Reynolds number on the  development of wake behind the ellipsoid is visualised through the streamline plots \autoref{fig:streamlineReAlltheta45AR2Delta0p1}. The figure shows the streamlines  for $\delta = 0.1$, $\theta = \ang{45}$, and at $Re_s$ 10, 50, 100. For $\delta = 0.1$, the ellipsoid is sitting in the deepest cavity of rough wall. The flow between ellipsoid and wall is mainly due to channeling through the cavities on the rough wall.  The accelerated flow observed in the gap for $\delta > 0.1$ is absent here, as ellipsoid acts as a barrier to the flow. Another important parameter is orientation angle which decides the location of the  appearance of the  recirculating zones. 
\autoref{fig:streamlineReAlltheta45AR2Delta0p1} shows that at lower $Re_s$, the flow is mostly streamlined over the ellipsoid, and a small recirculating zone is seen in the cavity of the rough wall.  As $Re_s$ increases (to 50), 
two counter-rotating vortices appear near the wall and flow separation happens at the bottom (near wall) part of the ellipsoid.
 Finally, for highest value of $Re_s$ (i.e., 100) considered in the present study, the flow from bottom and top merges in the down stream and separation happens  from the top of the ellipsoid. 
\begin{figure}[htb]
	\centering
	\begin{subfigure}[b]{0.3\textwidth}
		\includegraphics[width=\textwidth]{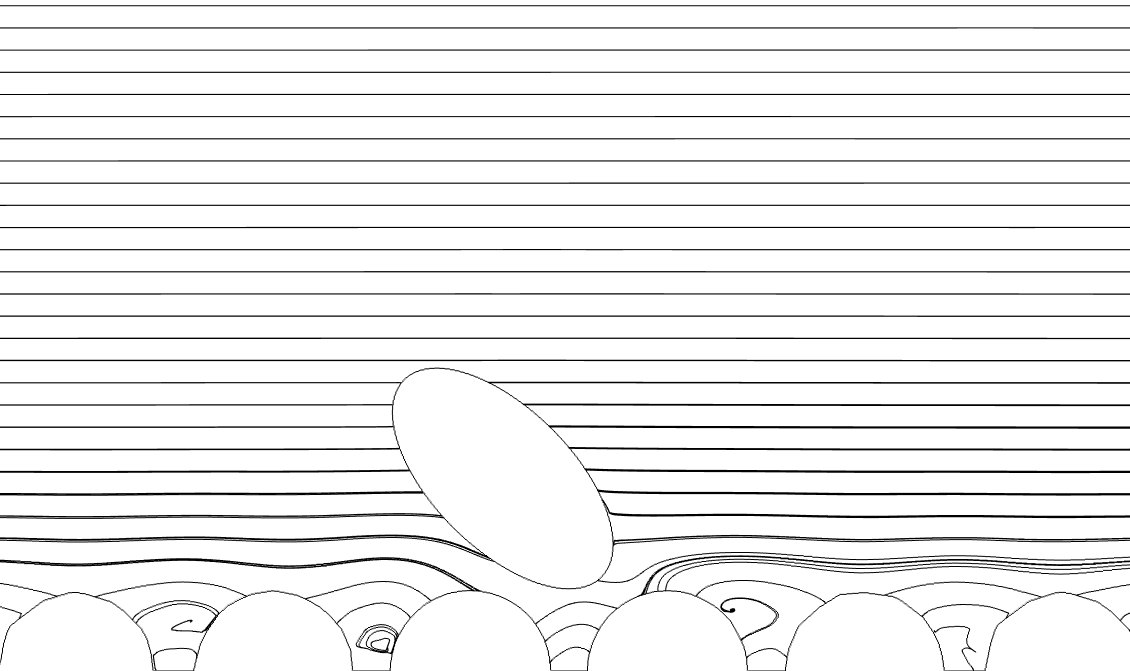}
		\caption{}
		\label{fig:streamlineRe10theta135AR2Delta0p1}
	\end{subfigure}
	\quad
	\begin{subfigure}[b]{0.3\textwidth}
		\includegraphics[width=\textwidth]{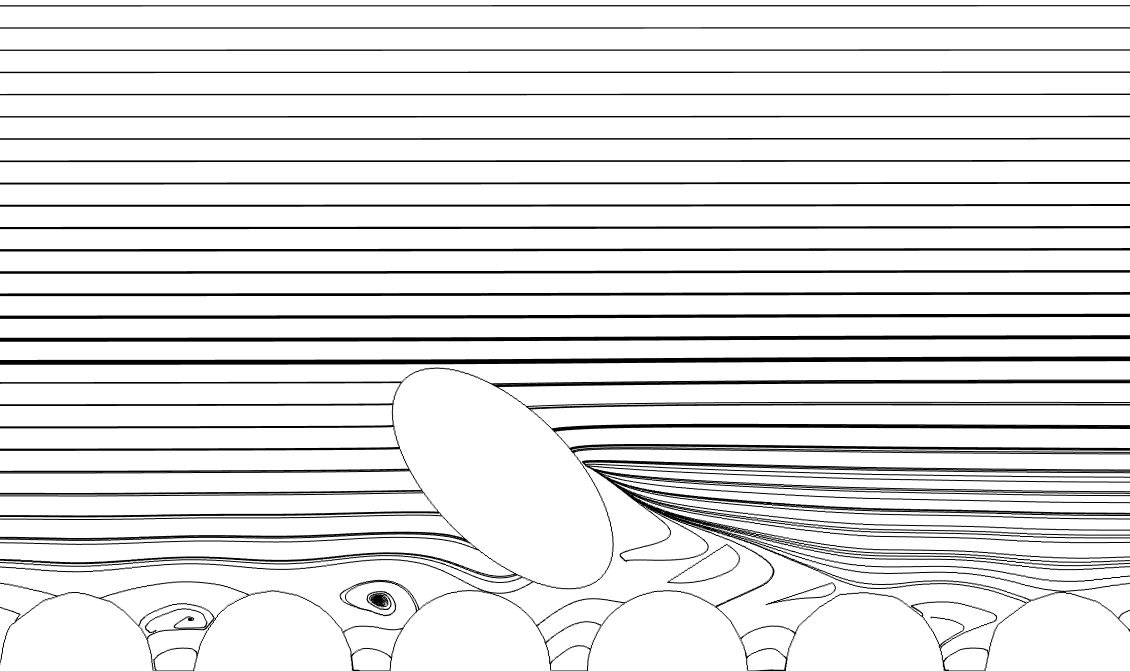}
		\caption{}
		\label{fig:streamlineRe50theta135AR2Delta0p1}
	\end{subfigure}
	\quad
	\begin{subfigure}[b]{0.3\textwidth}
		\includegraphics[width=\textwidth]{streamlineRe100theta135AR2Delta0p1.png}
		\caption{}
		\label{fig:streamlineRe100theta135AR2Delta0p1}
	\end{subfigure}
	\caption{Flow streamlines around the ellipsoid for AR = 2, $\theta$ = $\ang{135}$, at (a) $Re_s = 10$ (b) $Re_s = 50$ (c) $Re_s = 100$}
		\label{fig:streamlineReAlltheta135AR2Delta0p1}
\end{figure}
 It is shown in \autoref{fig:streamlineReAlltheta135AR2Delta0p1} that at $\theta = \ang{135}$ the streamlines are completely different to those which are observed for $\theta = \ang{45}$. 
 At $Re_s = 10$ (i.e. lowest value considered in study), a small vortex appear at upstream rather than downstream as in case of $\theta = \ang{45}$.
 Strength of circulation increases with increase in $Re_s$.
  Further, the detachment of boundary layer at high $Re_s$ also observed at lower half of ellipsoid, distinctly different from the case of $\ang{45}$. This has significance in deciding the lift and torque acting on particle. As we will see in later sections, the lift and torque are found to change the direction based on detachment of the boundary layer, which in turn depends on orientation angle of the ellipsoid with the incident flow. 
\begin{figure}[htb]
	\centering
	\begin{subfigure}[b]{0.48\textwidth}
		\includegraphics[width=\textwidth]{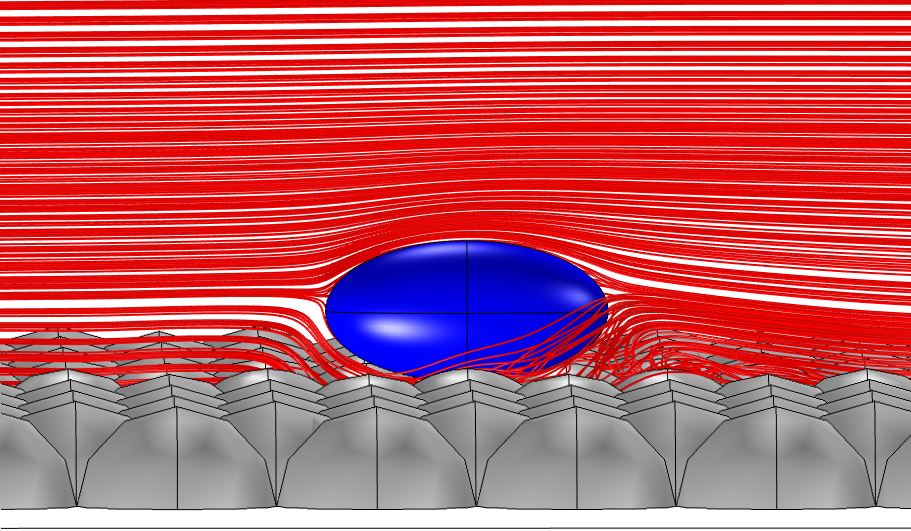}
		\caption{}
		\label{fig:AR2Res50Theta0Delta0p1Side3DStrml}
	\end{subfigure}
	\quad
	\begin{subfigure}[b]{0.48\textwidth}
		\includegraphics[width=\textwidth]{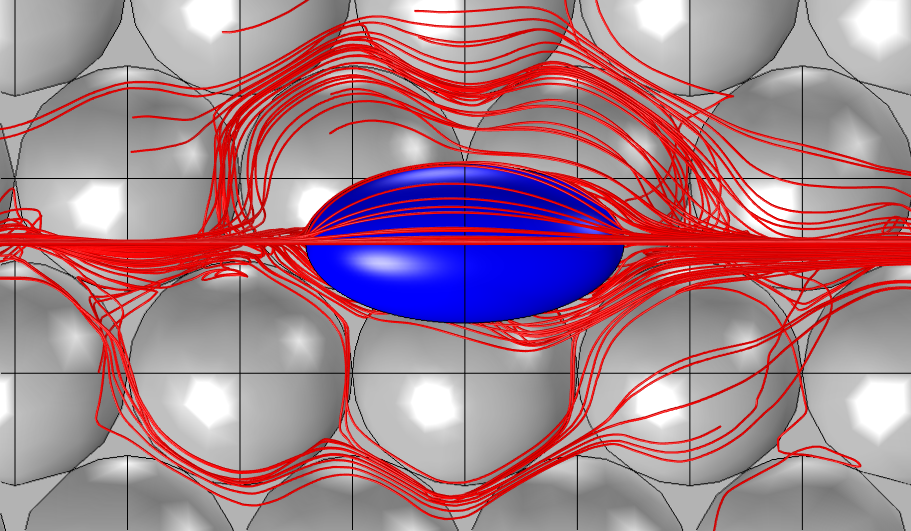}
		\caption{}
		\label{fig:AR2Res50Theta0Delta0p1Top3DStrml}
	\end{subfigure}
	\caption{3D view of flow streamlines around the ellipsoid for AR = 2, Re = 50, $\theta$ = $\ang{0}$, $\delta = 0.1$ from (a) side and (b) top view}
	\label{fig:AR2Res50Theta0Delta0p1SideTop3DStrml}
\end{figure}
\\In the above, all streamlines  are shown in  x-z 2D plane. 
A simultaneous 2-D view of the streamlines in  x-z and x-y  are shown for 
$\delta = 0.1d_p$ in at plane of particle center. \cref{fig:AR2Res50Theta0Delta0p1Side3DStrml,fig:AR2Res50Theta0Delta0p1Top3DStrml}. 
The figures show the side and top view of 3D streamlines for $Re_s = 50$ and $\theta = \ang{0}$. As discussed earlier, the ellipsoid at this location acts as a barrier to flow. The channeling of flow through the rough cavity can be observed. 
 Due to the roughness pattern on a surface and weak flow below ellipsoid, the recirculating vortices appear in deep cavities.

\subsection{Drag coefficient}
\begin{figure}[htb]
	\centering
	\begin{subfigure}[b]{0.45\textwidth}
		\includegraphics[width=\textwidth]{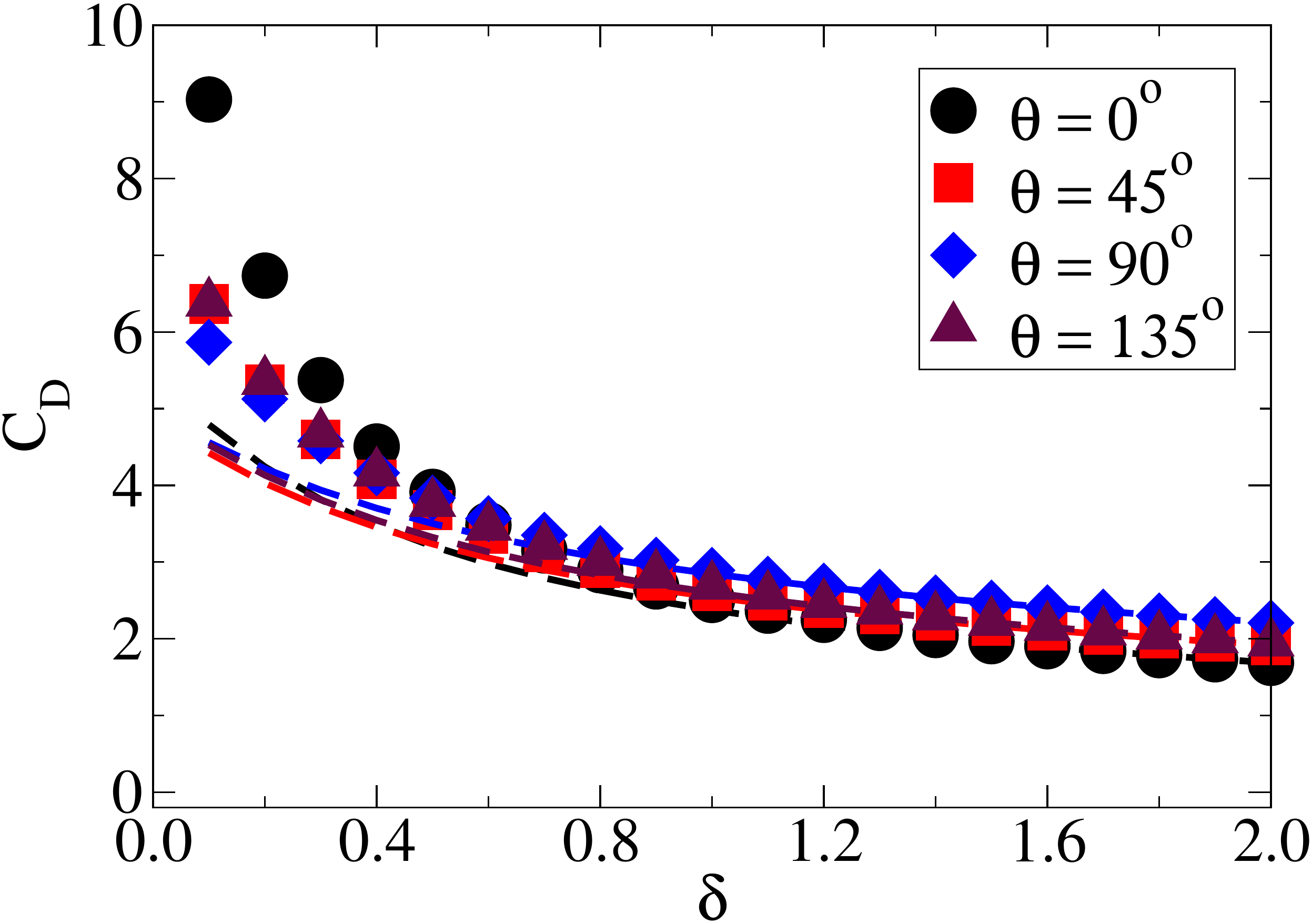}
		\caption{$Re_s = 10$}
		\label{fig:CdReShear10ThetaAll}
	\end{subfigure}
	\quad
	\begin{subfigure}[b]{0.45\textwidth}
		\includegraphics[width=\textwidth]{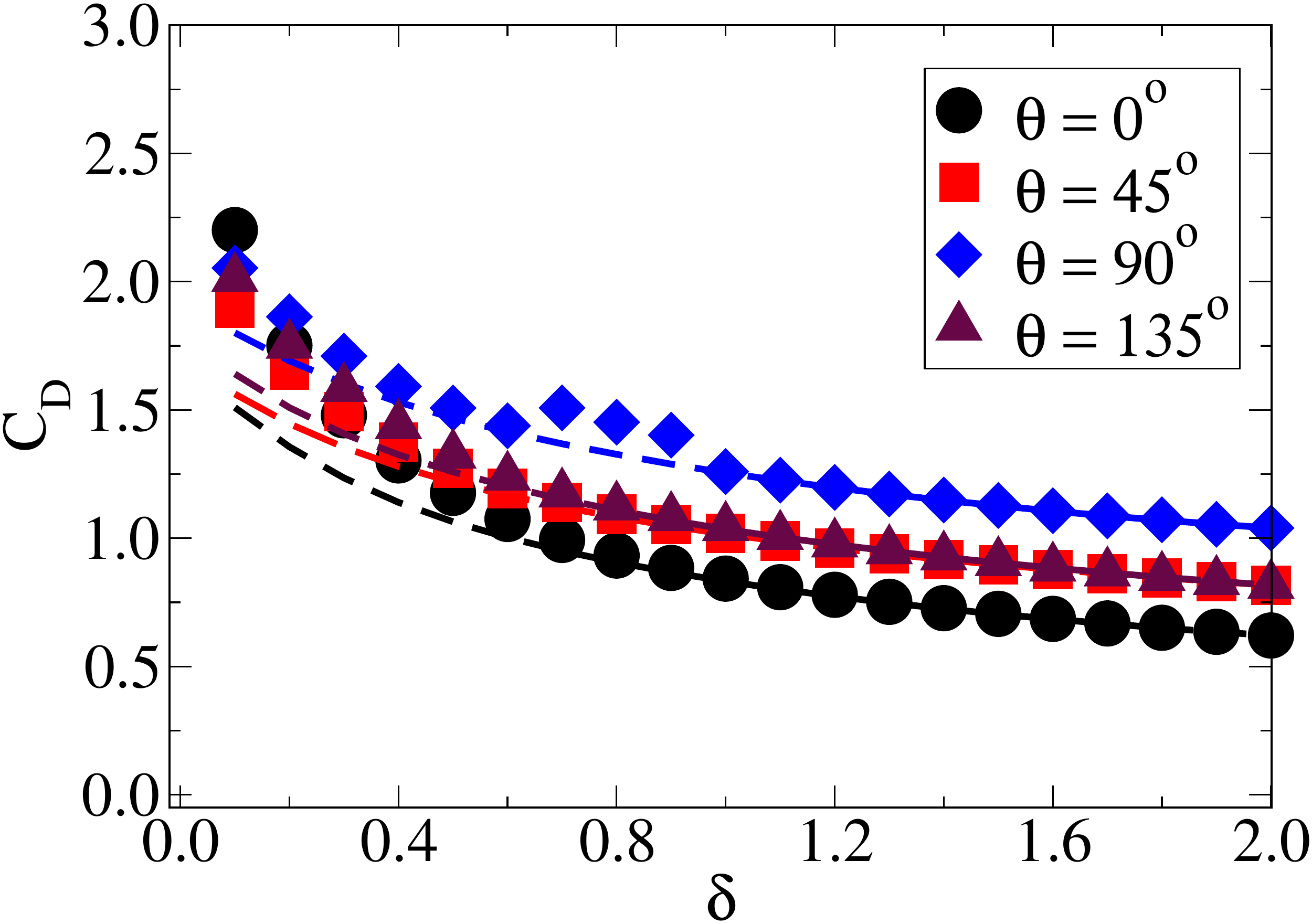}
		\caption{$Re_s = 50$}
		\label{fig:CdReShear50ThetaAll}
	\end{subfigure}
	\quad
	\begin{subfigure}[b]{0.45\textwidth}
		\includegraphics[width=\textwidth]{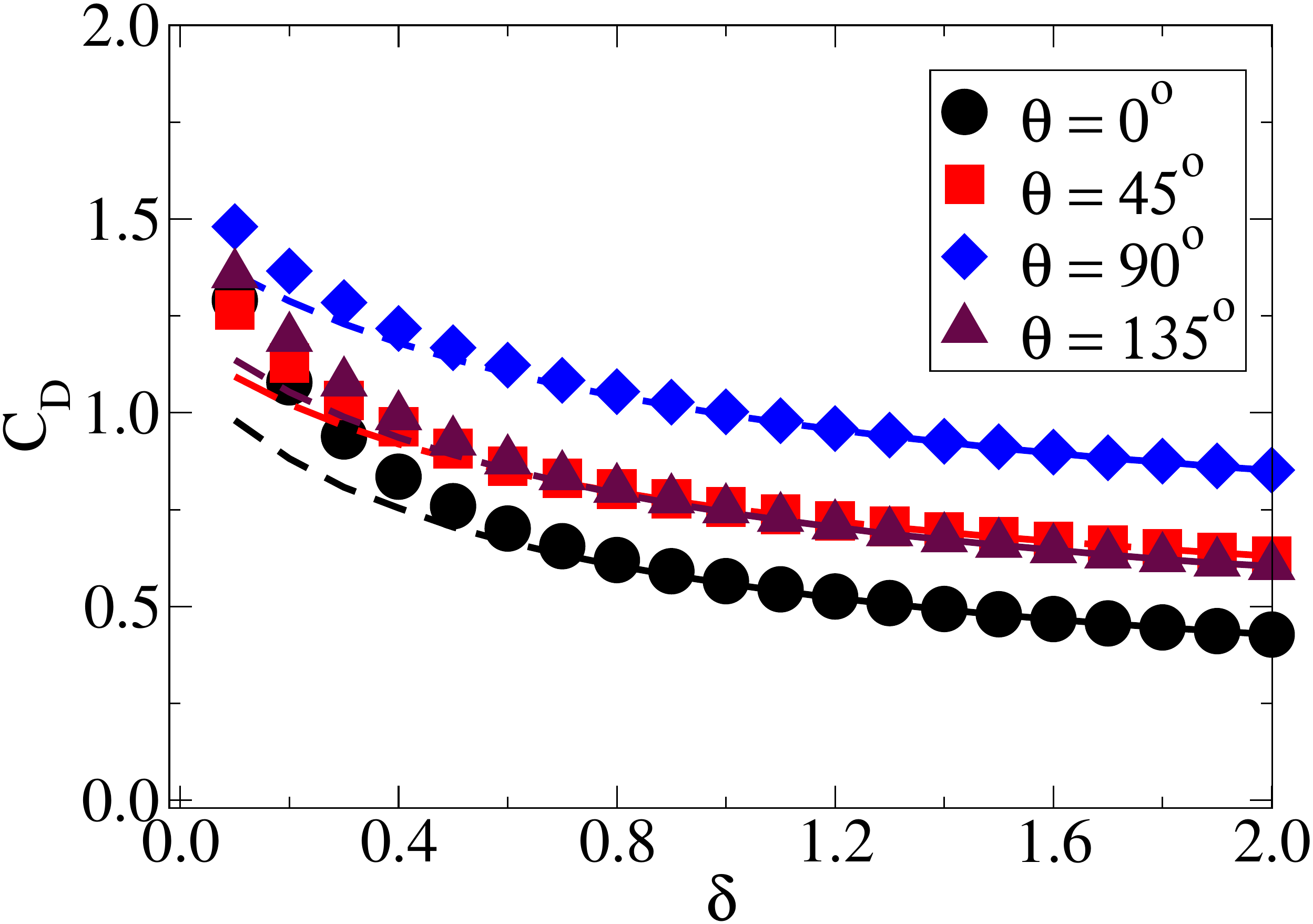}
		\caption{$Re_s = 100$}
		\label{fig:CdReShear100ThetaAll}
	\end{subfigure}
	\caption{Drag coefficient ($C_d$) as a function of wall normal distance $\delta$ at several Reynolds number. A dashed line with same color represent the corresponding values of $C_d$ for smooth wall condition.}
	\label{fig:CdReShearAllThetaAll}
\end{figure}
In this section, we present the drag coefficient as a function of wall-particle separation distance ($\delta$) for different orientation angles ($\theta$) and shear Reynolds numbers ($Re_s$), ranging between 10 and 100. In each of the cases, we have compared the results obtained for a rough wall with that of a smooth wall. Symbols with different shapes show $C_D$ values obtained from the simulations with the rough wall, and dashed lines represent the $C_D$s for a smooth wall. For all Reynolds numbers, drag coefficients are much higher when the particle is placed near the wall (\autoref{fig:CdReShearAllThetaAll}). There is a distinct difference in the effect of roughness on $C_D$  at different Reynolds numbers. At low $Re_s$, if the particle is placed horizontally near the wall ($\delta < 0.5$), $C_D$ for a rough wall is almost two times higher than that for the smooth wall. With an increase in wall-normal distance, the differences between two drag coefficients decrease for all the Reynolds numbers reported here. For $\delta > 1.0$, the effect of roughness becomes significant. Another notable observation at low Reynolds number is the significantly higher $C_D$ values for horizontally aligned ellipsoid compared to the other orientation angles. This difference eventually vanishes at elevated shear Reynolds number. It is worth noting that for higher wall-particle gap ($\delta >1.0$) drag coefficient in case of a vertically aligned particle is higher ($\approx 50 \%$ for $Re_s$ = 100) than the horizontal arrangement, even though the undisturbed velocity interpolated at particle center is higher in case of the  former. At low Reynolds number ($Re_s = 10$), the difference between drag coefficients for different angles of orientation decreases for the separation distance ($\delta$) greater than 0.5. This observation motivates us to investigate the variation of drag coefficient as a function of angle of inclination with approaching fluid stream.

\begin{figure}[htb]
	\centering
	\begin{subfigure}[b]{0.45\textwidth}
		\includegraphics[width=\textwidth]{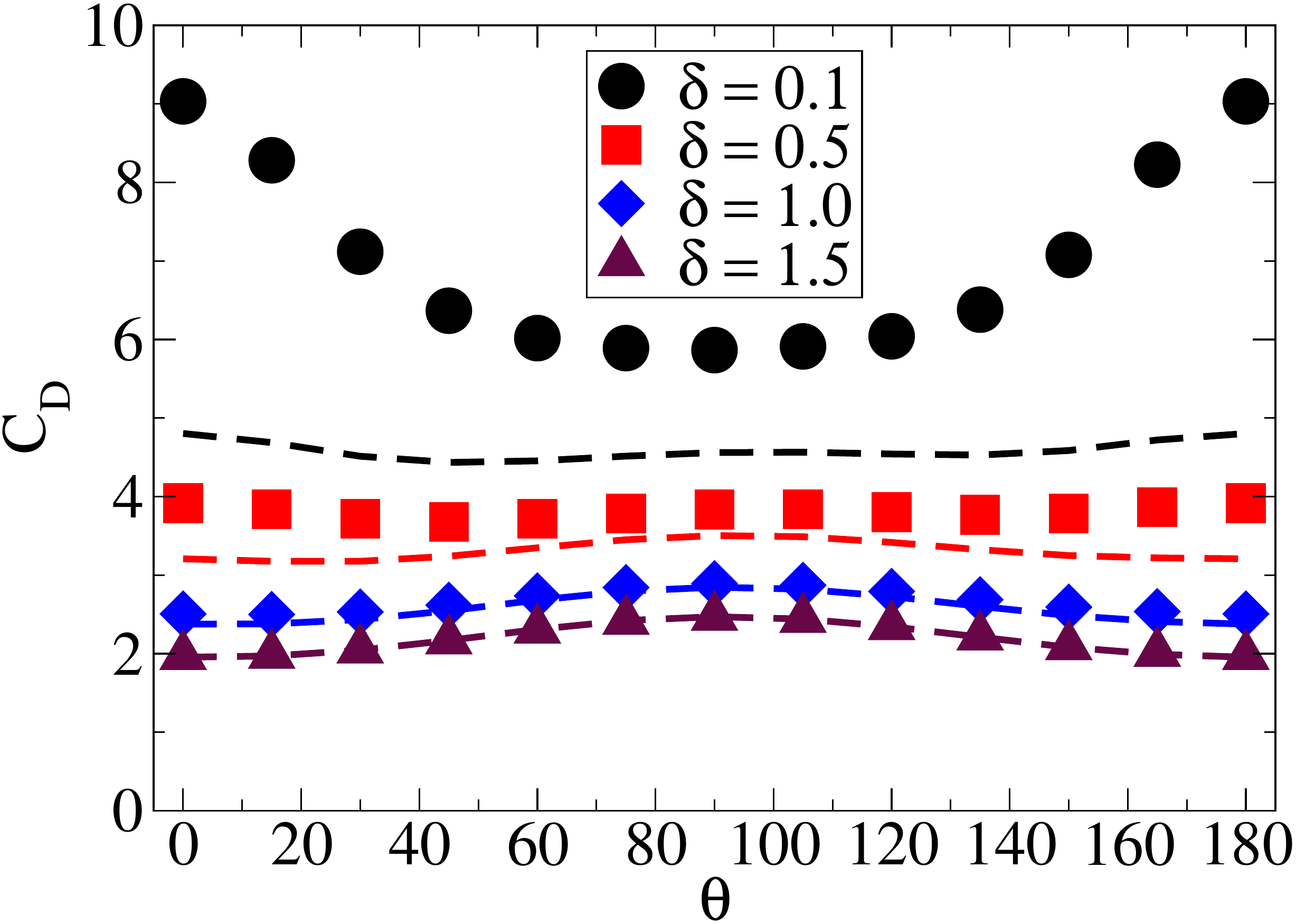}
		\caption{$Re_s = 10$}
		\label{fig:CdReShear10DeltaAll}
	\end{subfigure}
	\quad
	\begin{subfigure}[b]{0.45\textwidth}
		\includegraphics[width=\textwidth]{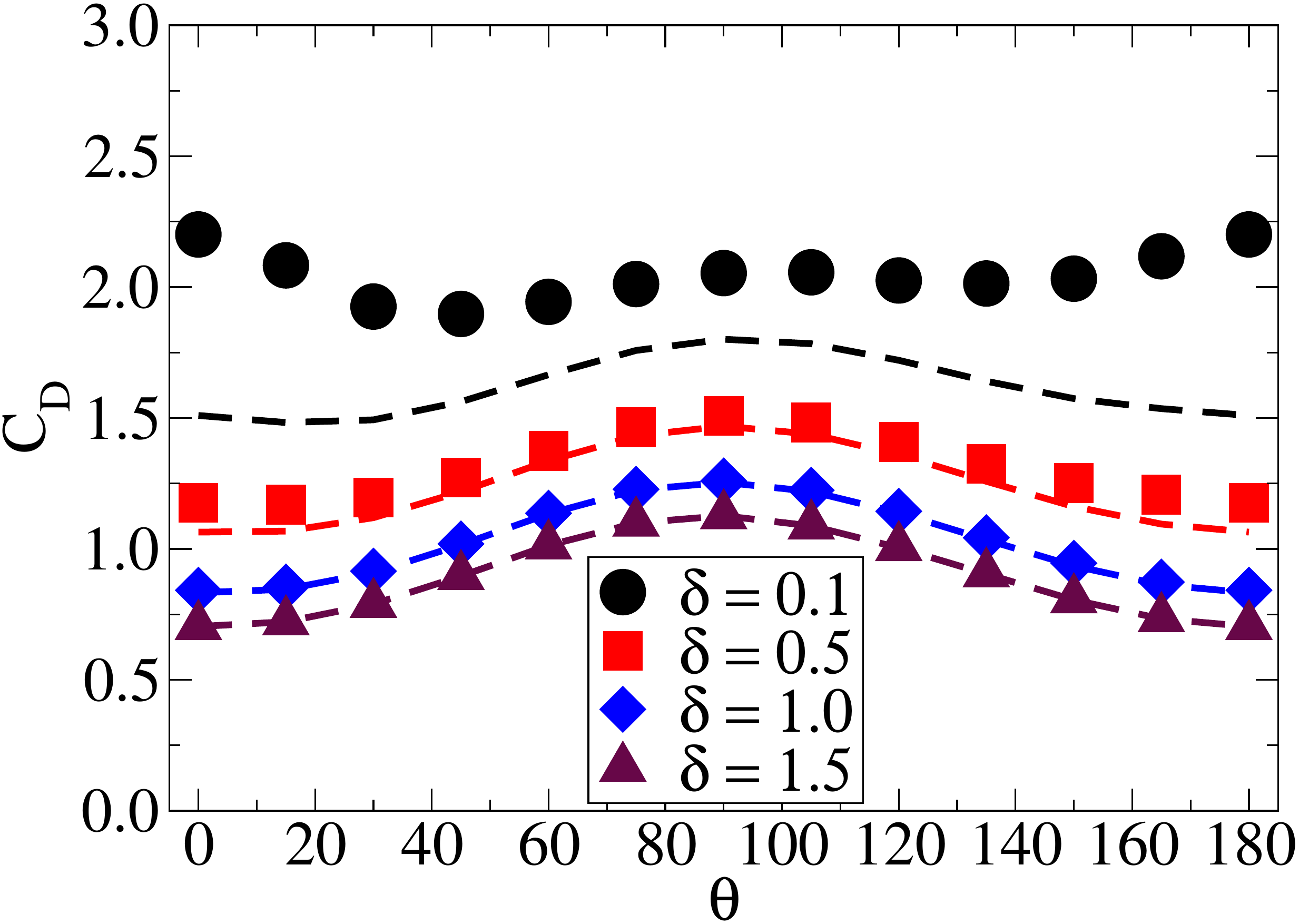}
		\caption{$Re_s = 50$}
		\label{fig:CdReShear50DeltaAll}
	\end{subfigure}
	\quad
	\begin{subfigure}[b]{0.45\textwidth}
		\includegraphics[width=\textwidth]{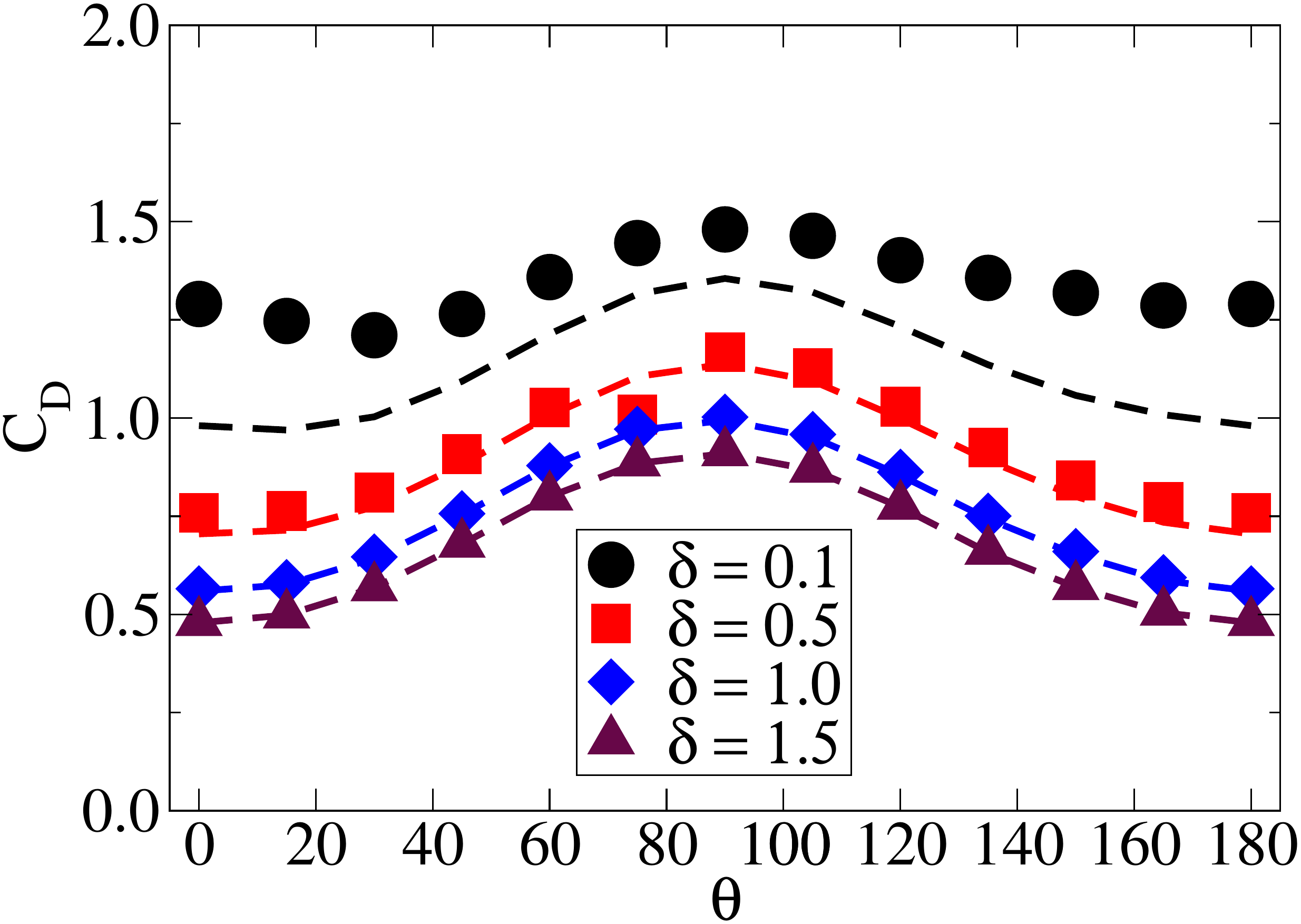}
		\caption{$Re_s = 100$}
		\label{fig:CdReShear100DeltaAll}
	\end{subfigure}
	\caption{Drag coefficient ($C_d$) as a function of orientation angle $\theta$ at several Reynolds number. A dashed line with same color represent the corresponding values of $C_d$ for smooth wall condition.}
	\label{fig:CdReShearallDeltaAll}
\end{figure}
\begin{figure}[htb]
	\centering
	\begin{subfigure}[b]{0.48\textwidth}
		\includegraphics[width=\textwidth]{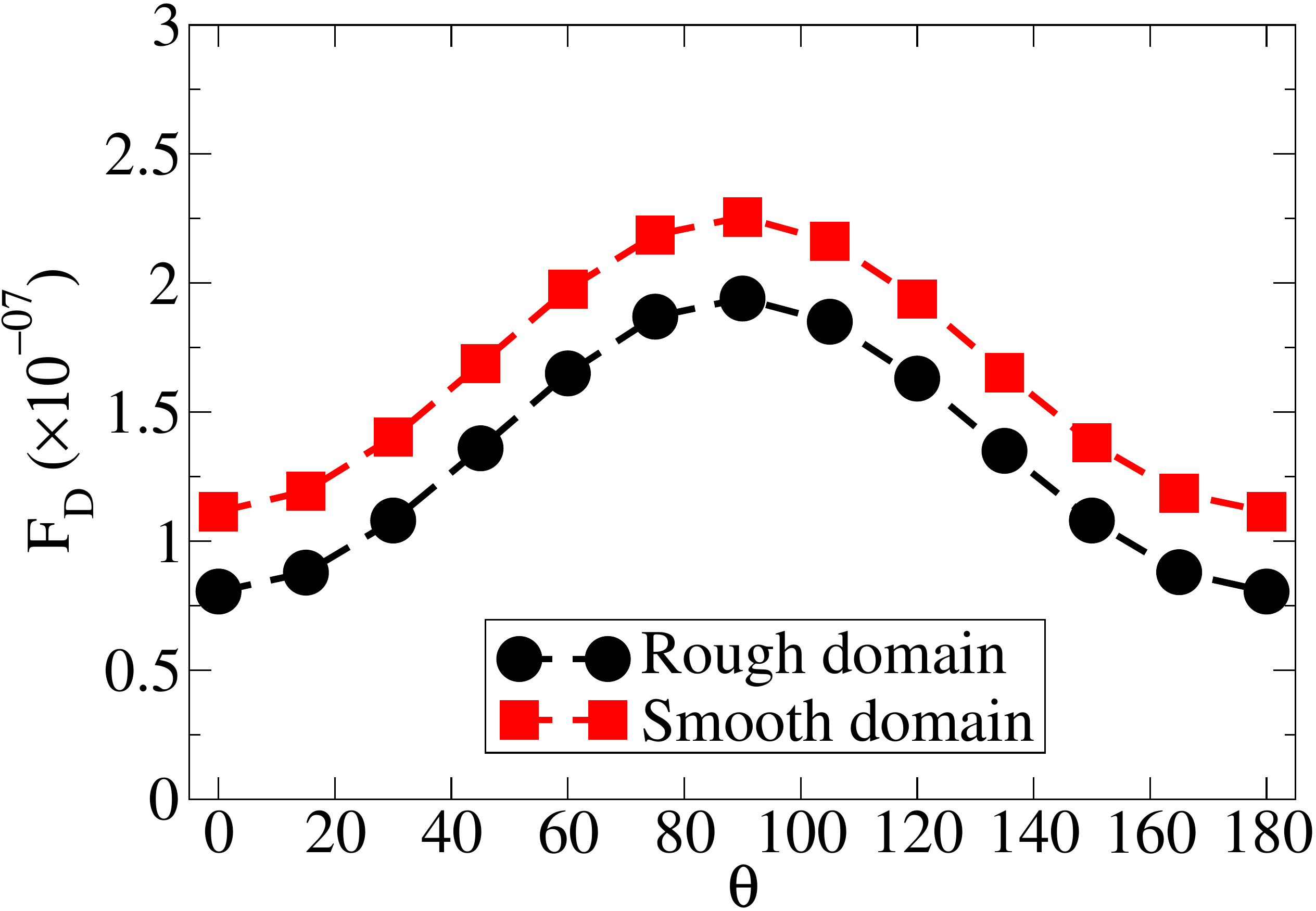}
		\caption{}
	\end{subfigure}
	\quad
	\begin{subfigure}[b]{0.48\textwidth}
		\includegraphics[width=\textwidth]{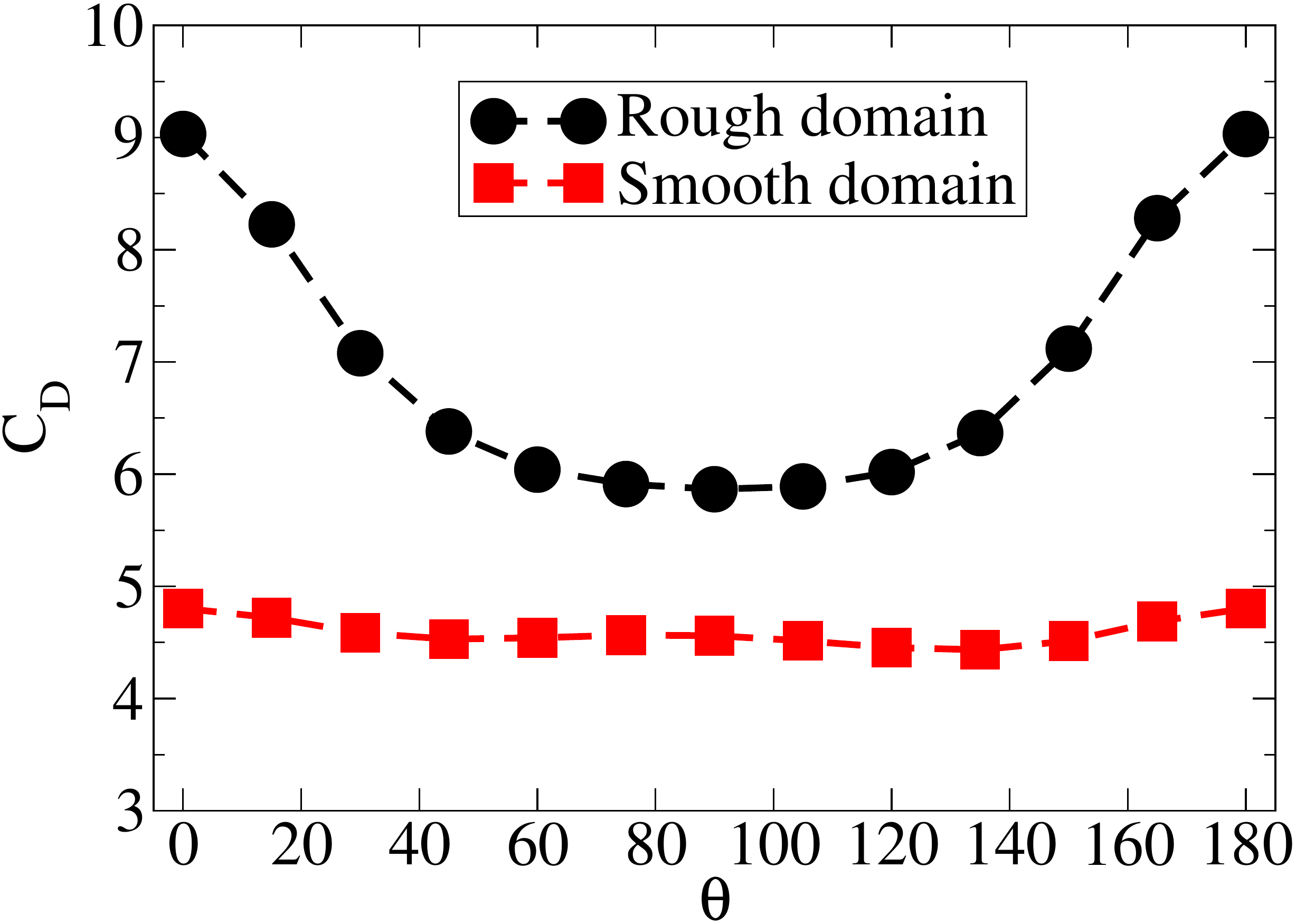}
		\caption{}
	\end{subfigure}
	\caption{Drag force $F_D$ and drag Coefficient $C_D$ as a function of $\theta$ for $Re_s$ = 10 at $\delta$ = 0.1}
	\label{fig:roughSmoothFdAndCdcomp}
\end{figure}

The second important parametric effect  to study is the orientation angle ($\theta$) of the particle with the flow direction. \autoref{fig:CdReShearallDeltaAll} shows the variation of drag coefficient with $\theta$ for different wall particle separation distance distance ($\theta$) and Reynolds number ($Re_s$). In the case of a smooth wall, the variation of $C_D$ with $\theta$ at higher separation distance follows a similar trend reported by \citet{HappelBook1983} for uniform flow over an ellipsoid particle. At low Reynolds number $Re_s = 10$, there is an insignificant variation of $C_D$ with angle of inclination, but with increase in  Reynolds number,  $C_D$ is maximum when $\theta=\ang{90}$. The effect of roughness becomes prominent when wall-particle separation ($\delta$) is less than 0.5. At low shear Reynolds number ($Re_s = 10$),  for the minimum separation distance reported here, drag coefficient is minimum when the particle is vertically aligned. For the horizontal alignment, the drag coefficient is almost 1.5 times higher than the vertical alignment. As shear Reynolds number increases, the above mentioned difference decreases. One important point to note here is that the undisturbed fluid velocity at the center of mass of the particle is higher in case of a vertically aligned particle than that of a horizontally aligned one. To address this ambiguity, we have reported the variation of drag force as well besides drag coefficient, which is shown in \autoref{fig:roughSmoothFdAndCdcomp}.

\begin{figure}[htb]
	\centering
	\begin{subfigure}[b]{0.48\textwidth}
		\includegraphics[width=\textwidth]{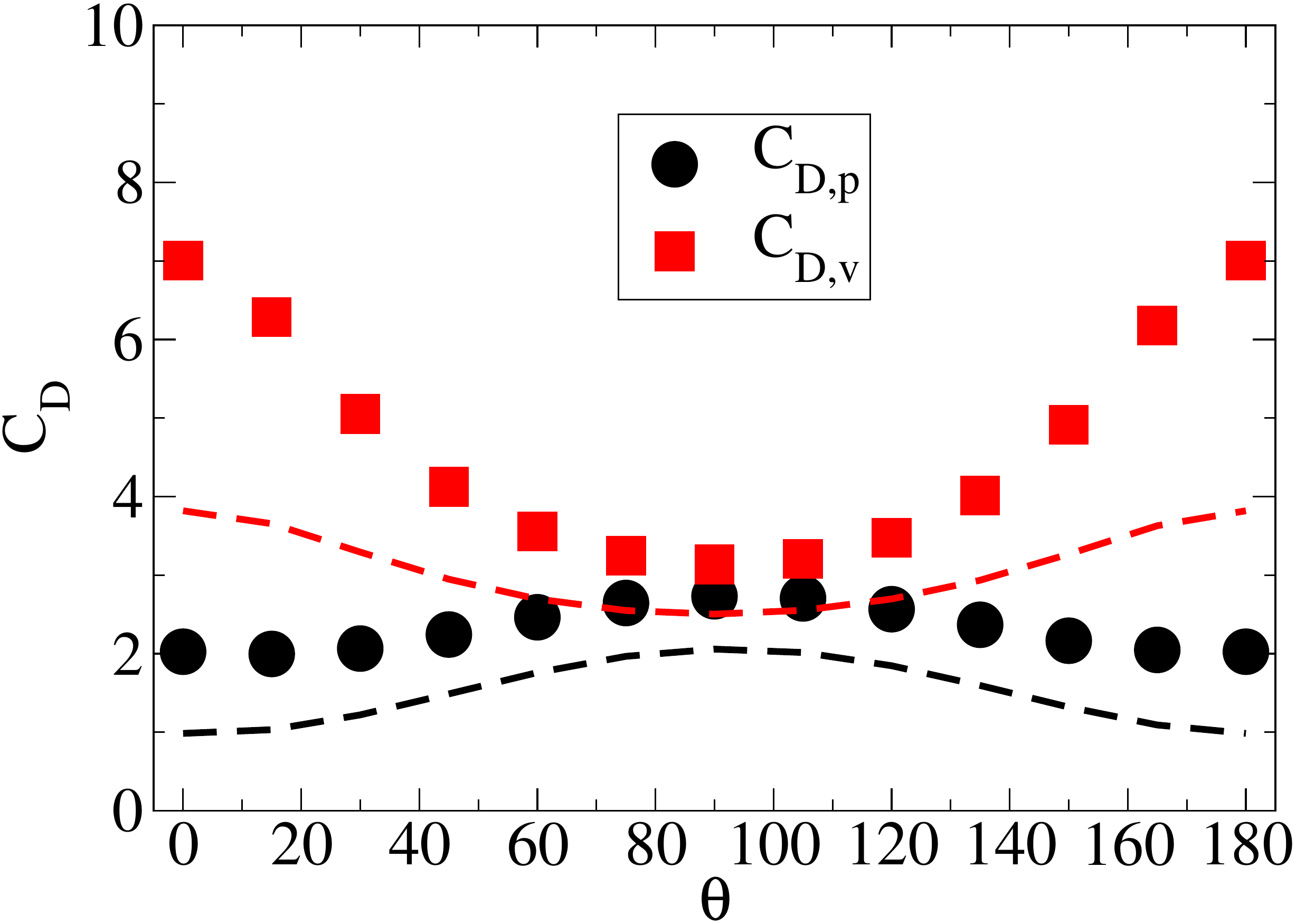}
		\caption{$Re_s = 10$}
		\label{fig:CdReShear10Delta0p1}
	\end{subfigure}
	\quad
	\begin{subfigure}[b]{0.48\textwidth}
		\includegraphics[width=\textwidth]{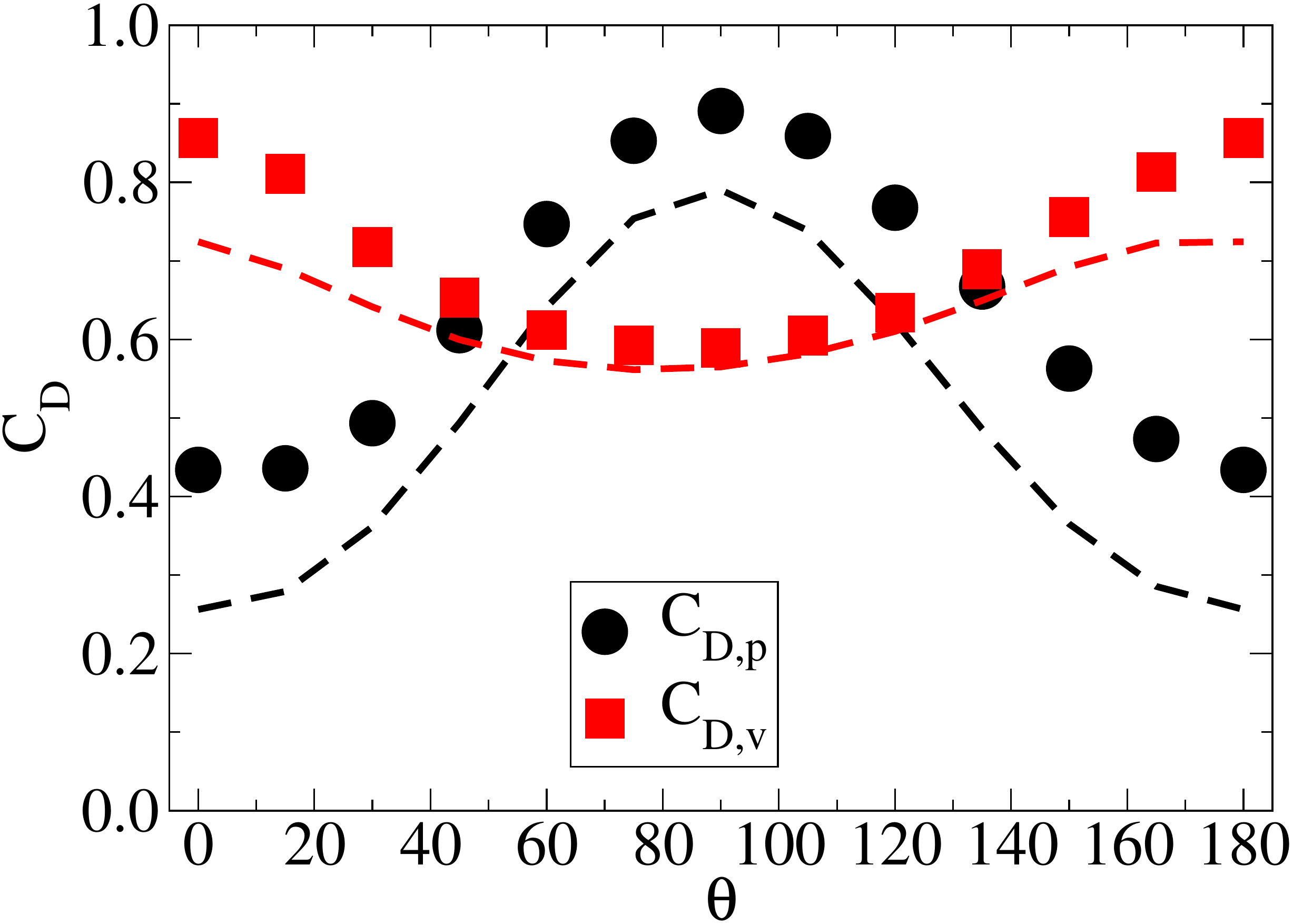}
		\caption{$Re_s = 100$}
		\label{fig:CdReShear100Delta0p1}
	\end{subfigure}
	\caption{Components of drop coefficient ($C_{d,p}$, $C_{d,v}$) as a function of orientation angle $\theta$ and, $\delta = 0.1$. A dashed line with same color represent the corresponding values of variable for smooth wall condition.}
	\label{fig:CdReShearAllDelta0p1}
\end{figure}

\autoref{fig:CdReShear10DeltaAll} shows that, away from the wall, there is a significant increase in the drag coefficient at $\theta=\ang{90}$. Similar to it, there is an increase in the undisturbed velocity, as mentioned earlier. To understand the contribution of viscous and pressure drag, we have computed these parameters separately and presented them in \autoref{fig:CdReShearAllDelta0p1}. We have restricted this exercise only for $\delta = 0.1$ and for two Reynolds numbers $Re_s = 10$ and $100$. It is observed that at low Reynolds number, a viscous component of drag always dominates over the pressure drag for both the smooth and rough walls. Effect of wall roughness plays a more prominent role in  case of the viscous drag for horizontally aligned particle. At high shear Reynolds number ($Re_s = 100$), the pressure drag coefficient is almost 1.5 times higher 
compared to the  viscous counterpart for vertically aligned particle. This could be related to the formation of a wake in the rear end.
One of the objectives of the present work is to express drag coefficient as a function of particle Reynolds number ($Re_p$), wall-normal distance, and  angle of inclination of the major axis with the flow direction. It is to be noted that in the present work, main emphasis is given to the effect of wall roughness on the hydrodynamic force coefficient. Here, we have computed all the coefficients using only one aspect ratio. To include the effect of aspect ratio, we have used the shape factor $k$ based upon theoretical formulation given in \cite{lamb1993} and relevant modifications suggested by \citet{ouchene2016}, which is applicable in the present scenario. It is observed from the simulations that the drag coefficients evolve as a function of $sin(\theta)$ between angle $\theta = \ang{0}$ and  $\ang{90}$ irrespective of the particle Reynolds numbers. Effect of particle Reynolds number is included in the drag coefficient of  $\theta = \ang{0}$ and  $\ang{90}$. Effect of aspect ratio has also been included in the expression for drag coefficients for $\theta = \ang{0}$ and  $\ang{90}$ \citep{loth2008, ouchene2016}. Separation distance from the wall ($\delta$), one of the important parameters of major focus in the present study, is found to contribute non-linearly to the drag coefficient.  A similar observation for the drag on a spherical particle has been reported by  \citet{ZengPOF2009} in an earlier study. With this described approach, the drag coefficient can be expressed by the following correlations.
	\begin{equation}
	C_D = C_{D,0} + (C_{D,90}-C_{D,0})sin^2\theta
	\end{equation}
	here,
	\begin{equation*}
	C_{D,0} = \big(\frac{24}{Re_p}\big)\big(k_0\alpha_0+\beta_0\omega^{-0.8}Re_p^{0.687}+\gamma_0(\omega-1)^{0.63}Re_p^{0.41}\big)
	\end{equation*}
	\begin{equation*}
	\alpha_0 = 1 + 0.001576exp(2\delta)+\frac{3.895}{1.611(1+2\delta)}
	\end{equation*}
	\begin{equation*}
	\beta_0 = 0.15-0.03(1+2.03\delta)exp(-0.392\delta)
	\end{equation*}
	\begin{equation*}
	\gamma_0 = 0.1497-0.01942(1+2.77\delta)exp(-0.2921\delta)
	\end{equation*}
	and,
	\begin{equation*}
	C_{D,90} = \big(\frac{24}{Re_p}\big)\big(k_{90}\alpha_{90}+\beta_{90}\omega^{-0.54}Re_p^{0.687}+\gamma_{90}\omega^{1.043}(\omega-1)^{-0.17}Re_p^{0.41}\big)
	\end{equation*}
	\begin{equation*}
	\alpha_{90} = 1 + 1.288exp(0.05953\delta^{2.804})+\frac{6.137}{6.479(1+0.05953\delta)}
	\end{equation*}
	\begin{equation*}
	\beta_{90} = 0.15-0.3824(1+2.599\delta)exp(-0.344\delta)
	\end{equation*}
	\begin{equation*}
	\gamma_{90} = 0.2516-0.247(1-0.11\delta)exp(\delta^{0.371})
	\end{equation*}
	Here, the shape factor $k$ is  presented by \citet{ouchene2016} as ,
	\begin{equation}
	k=k_{0}=k_{90}=\frac{8}{3}\omega^{-1/3}\big[\frac{2\omega}{(\omega^2-1)}+\frac{2\omega^2-1}{(\omega^2-1)^{(3/2)}}ln\frac{\omega+\sqrt{\omega^2-1}}{\omega-\sqrt{\omega^2-1}}\big]^{-1}
	\end{equation}

\begin{figure}[htb]
	\centering
	\begin{subfigure}[b]{0.48\textwidth}
		\includegraphics[width=\textwidth]{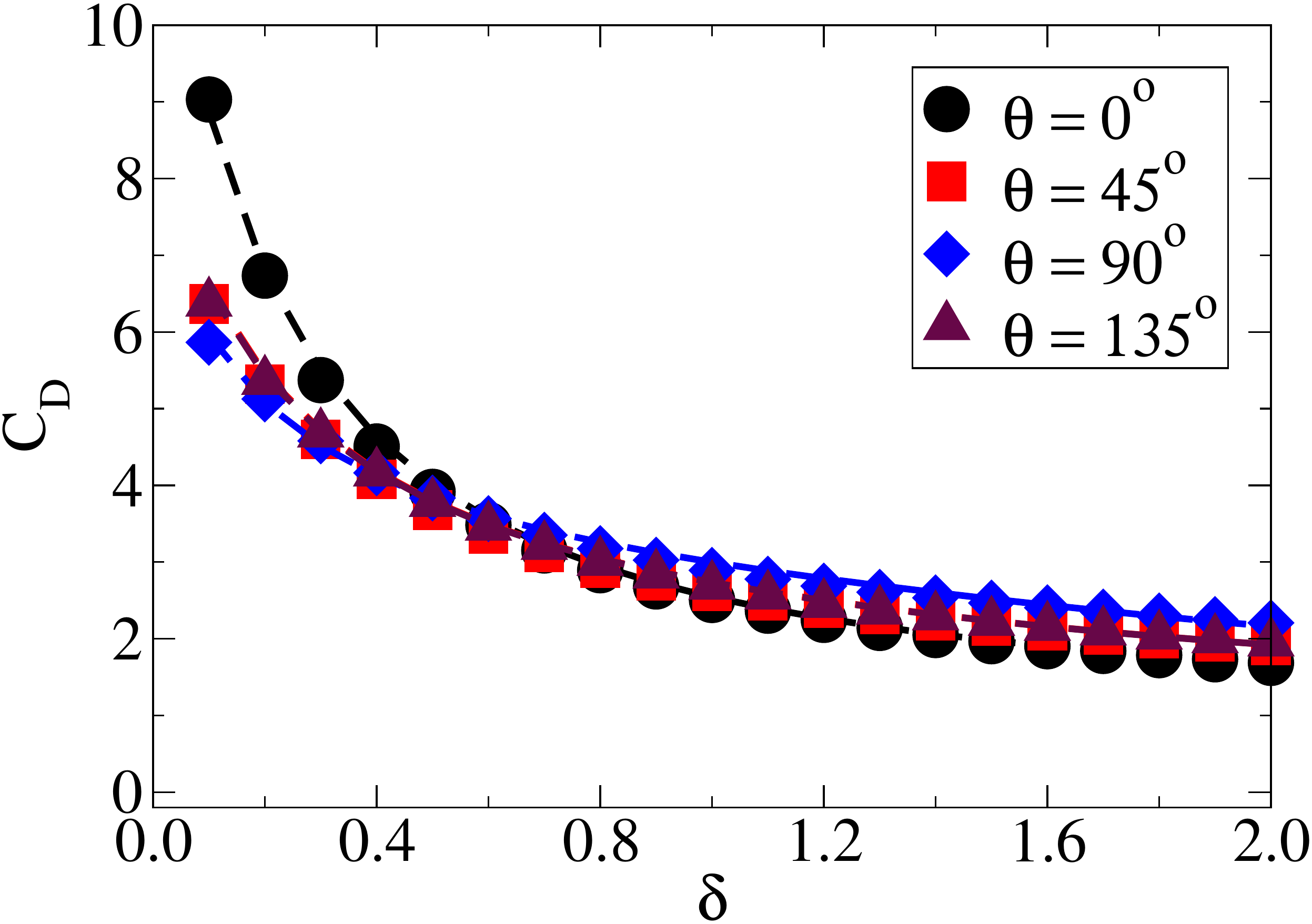}
		\caption{$Re_s = 10$}
	\end{subfigure}
	\quad
	\begin{subfigure}[b]{0.48\textwidth}
		\includegraphics[width=\textwidth]{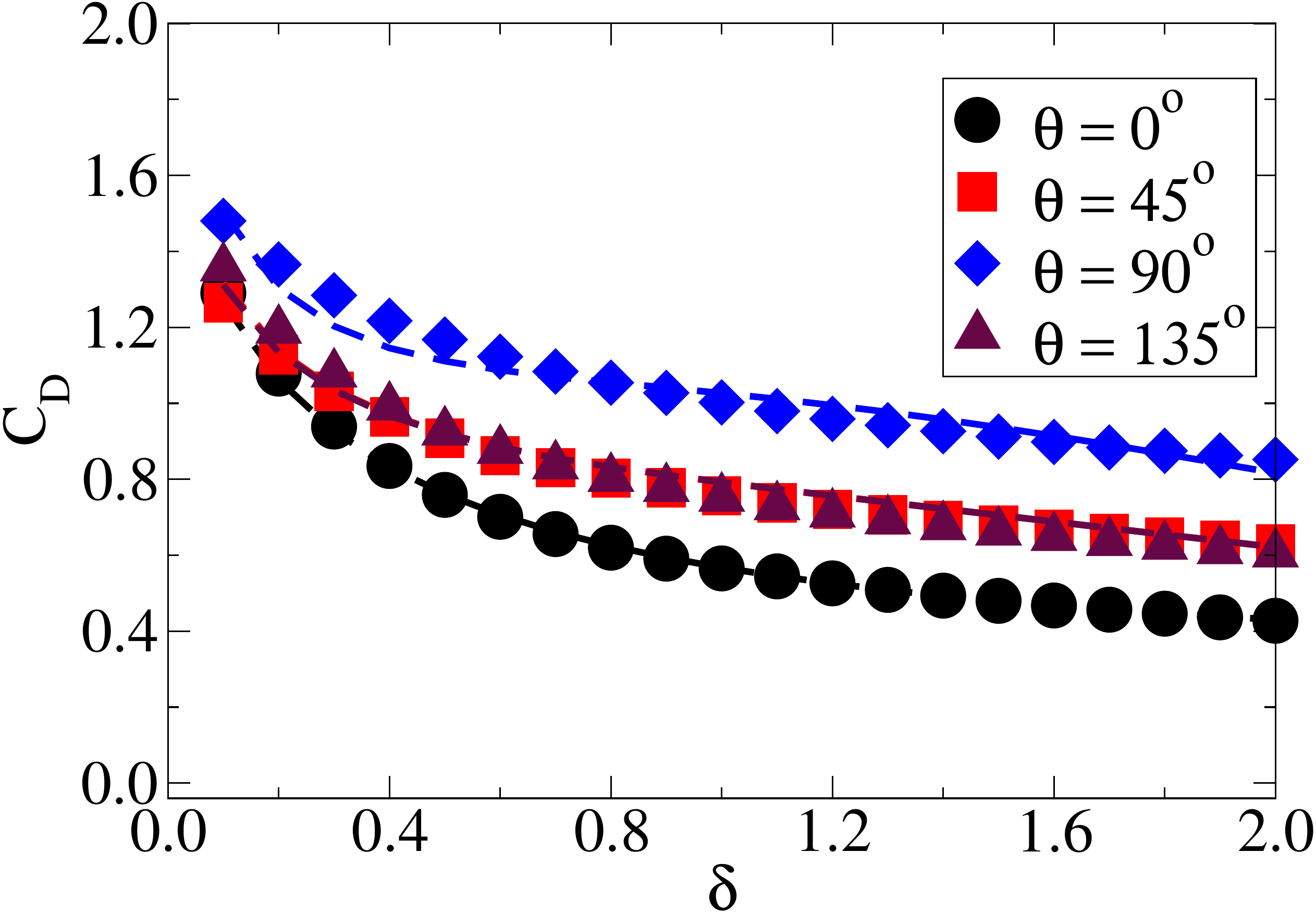}
		\caption{$Re_s = 100$}
	\end{subfigure}
	\caption{$C_D$ Vs $\delta$ for $Re_s$ = 10 and 100 at different orientation angle ($\theta$). The dashed line in corresponding color represents the fitted value from the correlation}
	\label{fig:CompResAllThetaAllCdVsDelta}
\end{figure}

\begin{figure}[htb]
	\centering
	\begin{subfigure}[b]{0.48\textwidth}
		\includegraphics[width=\textwidth]{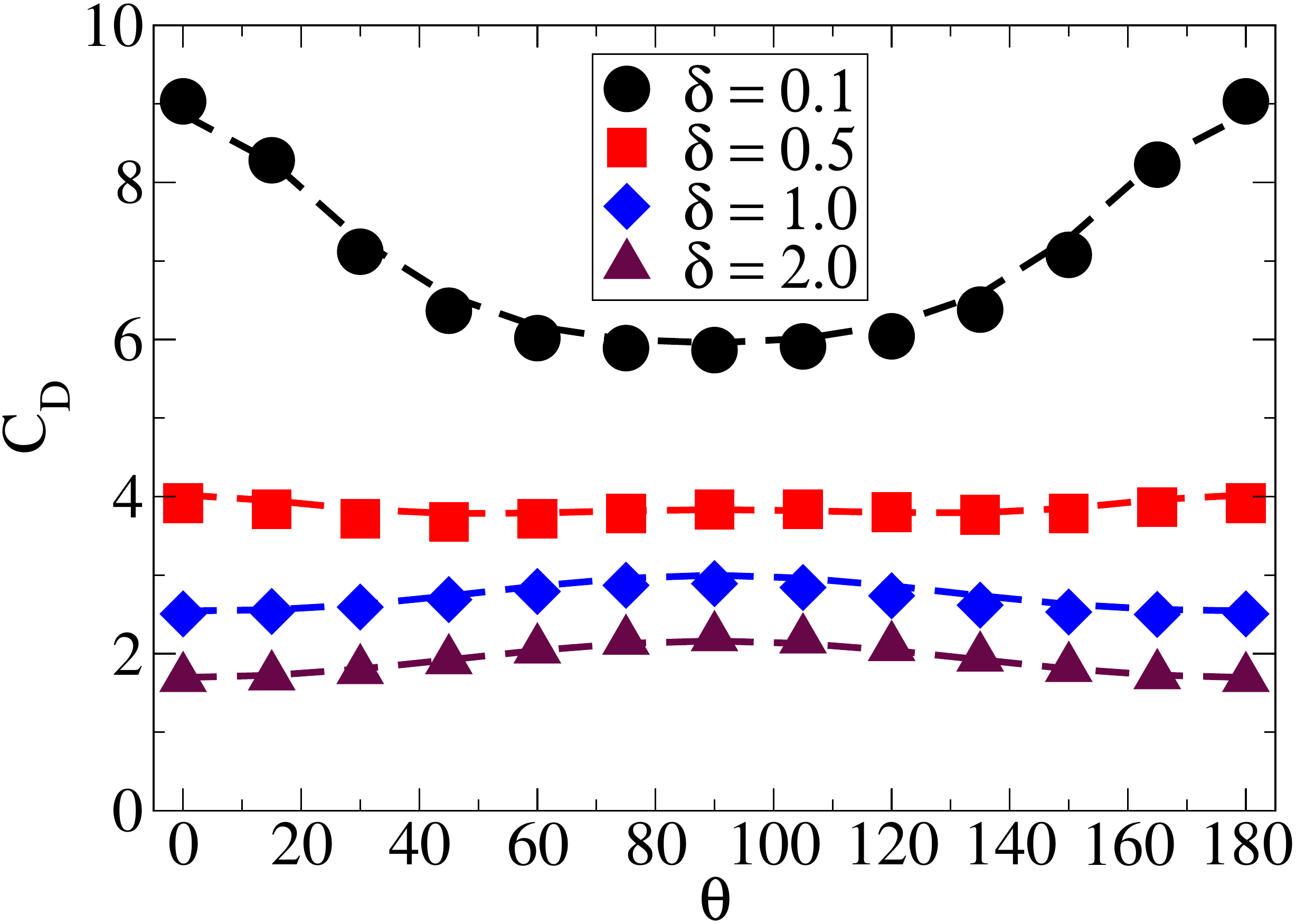}
		\caption{$Re_s = 10$}
	\end{subfigure}
	\quad
	\begin{subfigure}[b]{0.48\textwidth}
		\includegraphics[width=\textwidth]{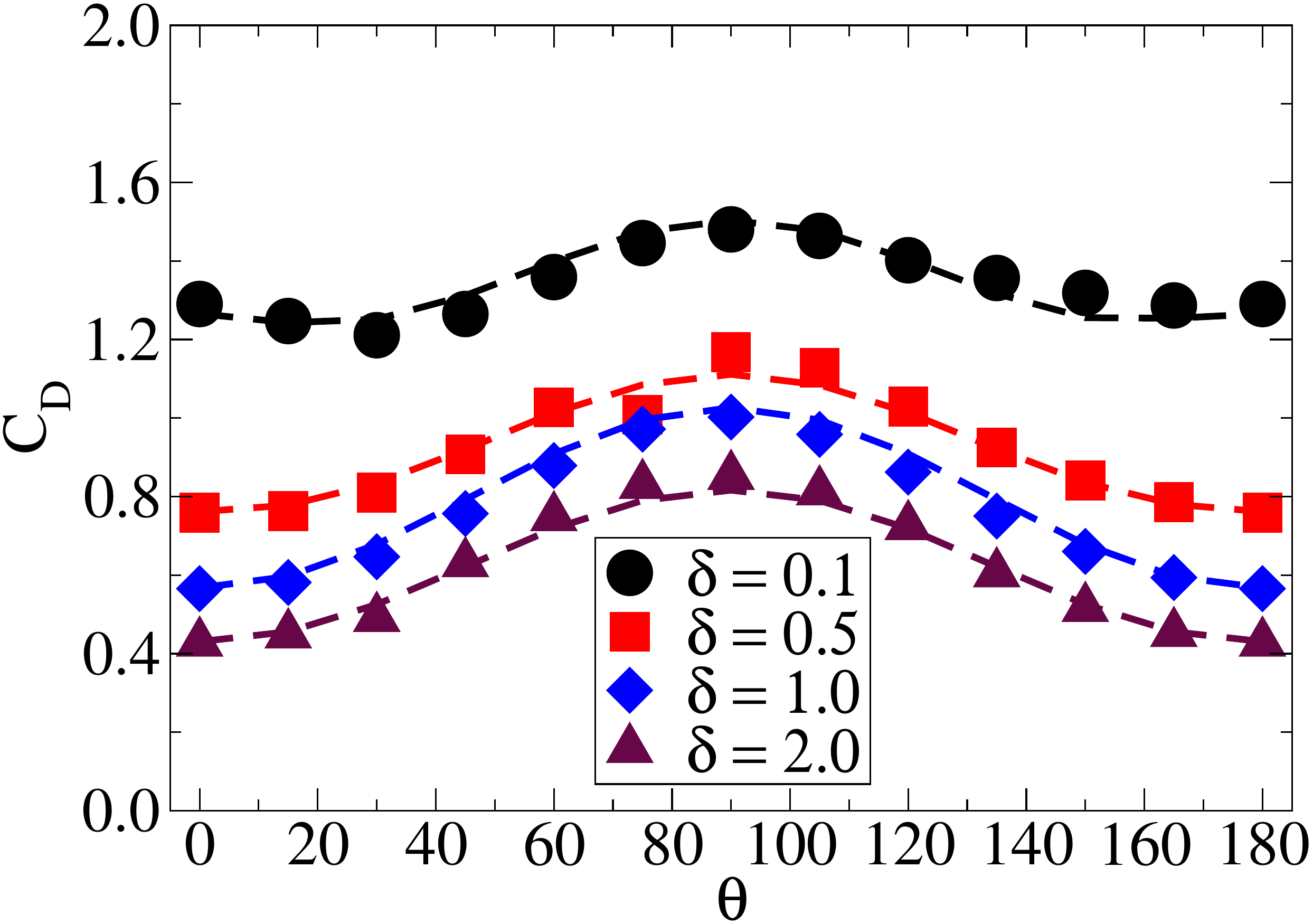}
		\caption{$Re_s = 100$}
	\end{subfigure}
	\caption{$C_D$ Vs $\theta$ for $Re_s$ = 10 and 100 at different wall separation distance ($\delta$). The dashed line in corresponding color represents the fitted value from the correlation}
	\label{fig:CompResAllDeltaAllCdVsTheta}
\end{figure}

\begin{figure}[htb]
		\centering
		\includegraphics[width=0.75\textwidth]{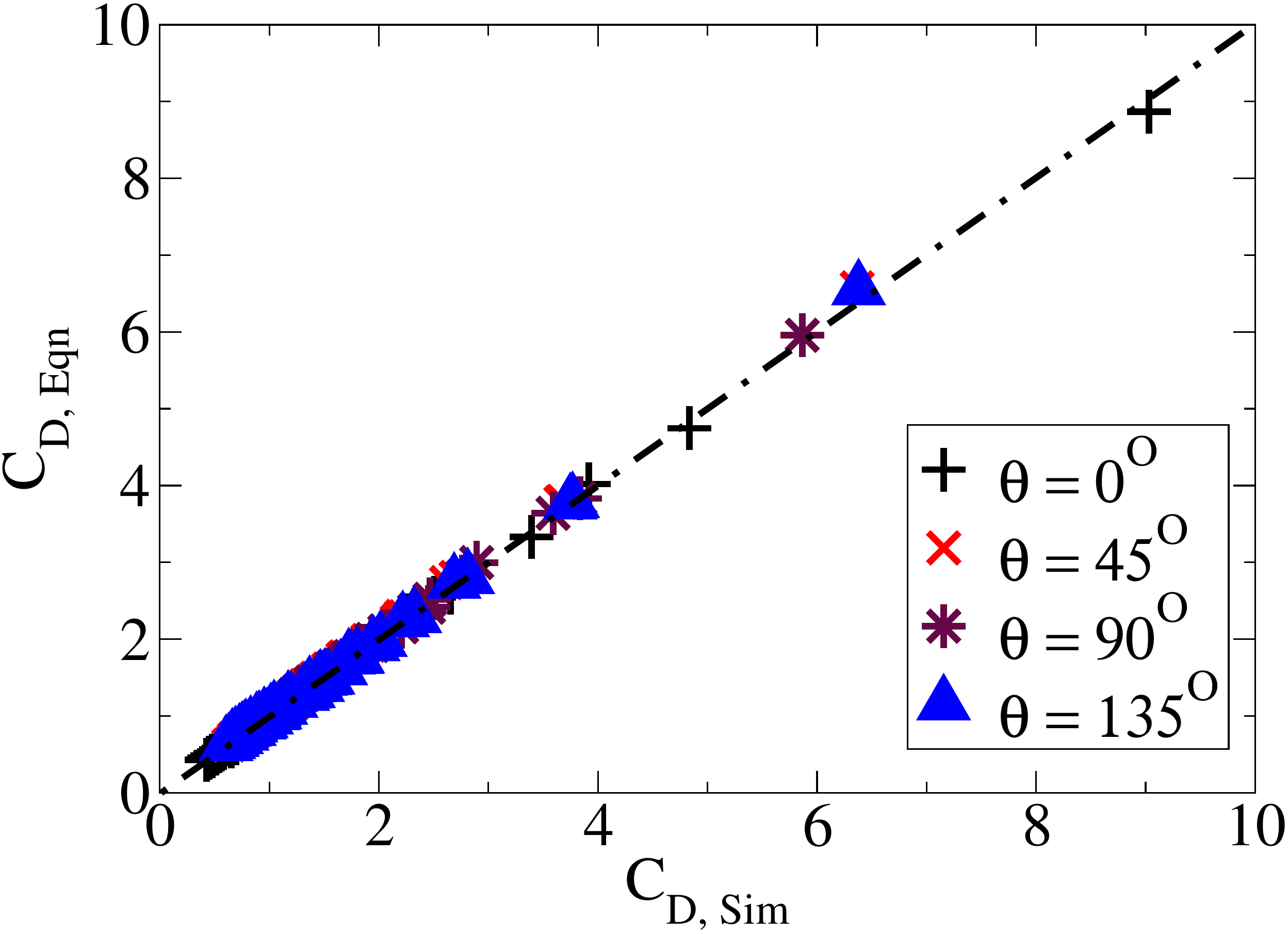}
		\caption{Predicted $C_D$ plotted against $C_D$ from simulation data at various $\theta$, and ($-\cdot-\cdot$) represent line of $y=x$}
		\label{fig:CdEqVsCdSimThetaAll}
	\end{figure}

Above equation  boils down to the correlation for the sphere reported by \citet{schiller1933} for $\omega = 1$ and  $\delta >> 1$. For $\omega = 1$ and in the limit of creeping flow and in the presence of a wall, the equation  follows the correlation of \citet{GOLDMAN1967} for all $\delta$, which was further simplified by \citet{ZengPOF2009}. Accuracy of the present non-trivial correlation is evaluated by comparing it with the simulation results for a wide range of parameters. \autoref{fig:CompResAllThetaAllCdVsDelta} shows comparison of drag coefficients obtained using correlation with that obtained from the simulations at two $Re_S = 10$ and $100$ and  for different  orientation angles $\theta$.  In the figure, the drag coefficients have been reported as a function of separation distance from the wall ($\delta$). It is observed from the figures that the prediction of the drag coefficients from the correlation matches well even for a very low separation distance $\delta<0.5$ for both the shear-Reynolds numbers. The maximum deviation of about $8\%$ between simulation and prediction by the correlation is observed at around $\delta \approx 1.0$ and $\theta = \ang{0}$ for all $Re_s$ considered in study. Whereas, for other separation distances, the deviation does not exceed 3\%.	A similar trend is also observed for other orientation angles ($\Theta$), considered in this study. At higher $\delta$ values, this deviation is reduced to  1\%. The overall mean deviation for all simulated cases lies within 3.5\%. Similarly, in \autoref{fig:CompResAllDeltaAllCdVsTheta}, comparison of correlation prediction with simulation results is shown for $C_D$ as function of $\theta$ at different $\delta$.  The figure shows that the correlation is fairly capable of producing accurate results from $\delta = 0.1$, where ellipsoid is placed very close to the wall to  several particle diameters away from the wall.
A goodness-of-fit between simulation results and correlation prediction is shown \autoref{fig:CdEqVsCdSimThetaAll} for all shear Reynolds numbers, separation distances from the rough wall $\delta$ (0.1, 0.5, 1.0, and 2.0) and orientation angles $\theta$ ($\ang{0},\ang{45},\ang{90},\ang{135}$). The figure shows that all the data collapse along the diagonal line with a maximum error of $8\%$; confirms the applicability of the above correlation for calculating drag on a ellipsoidal particle near a rough wall.

\subsection{Lift coefficient}
\begin{figure}[htb]
	\centering
	\begin{subfigure}[b]{0.45\textwidth}
		\includegraphics[width=\textwidth]{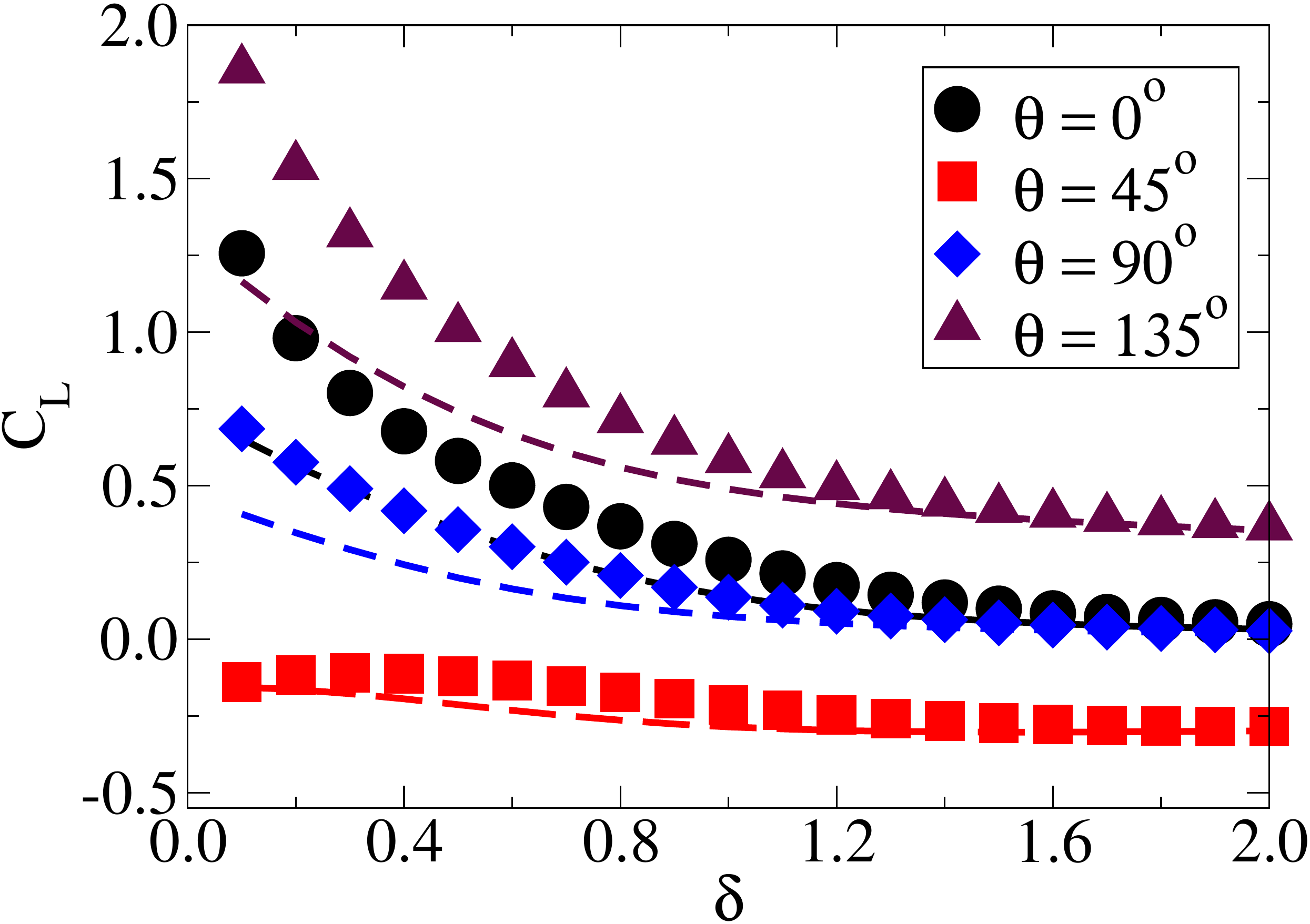}
		\caption{$Re_s = 10$}
		\label{fig:ClReShear10ThetaAll}
	\end{subfigure}
	\quad
	\begin{subfigure}[b]{0.45\textwidth}
		\includegraphics[width=\textwidth]{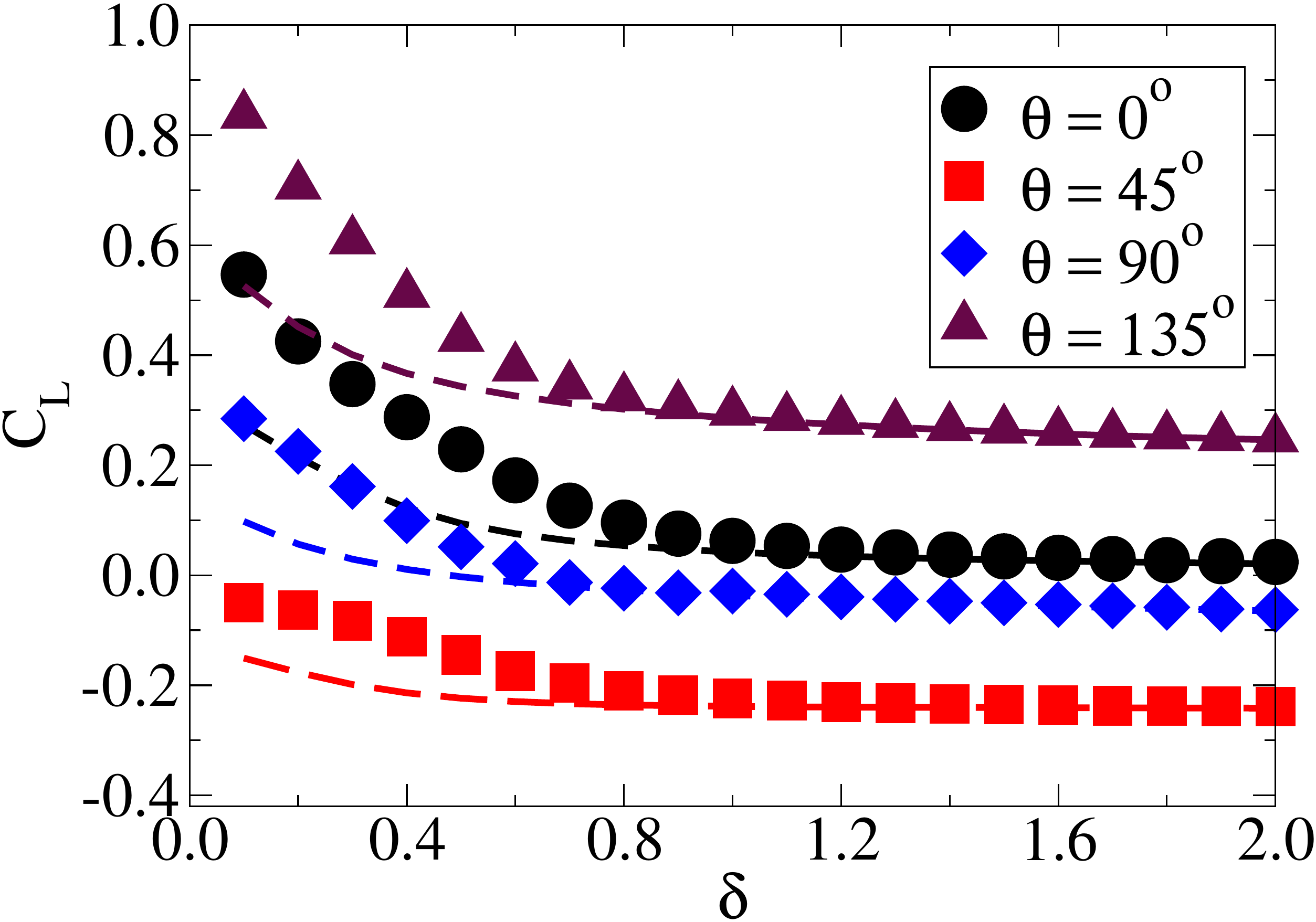}
		\caption{$Re_s = 50$}
		\label{fig:ClReShear50ThetaAll}
	\end{subfigure}
	\quad
	\begin{subfigure}[b]{0.45\textwidth}
		\includegraphics[width=\textwidth]{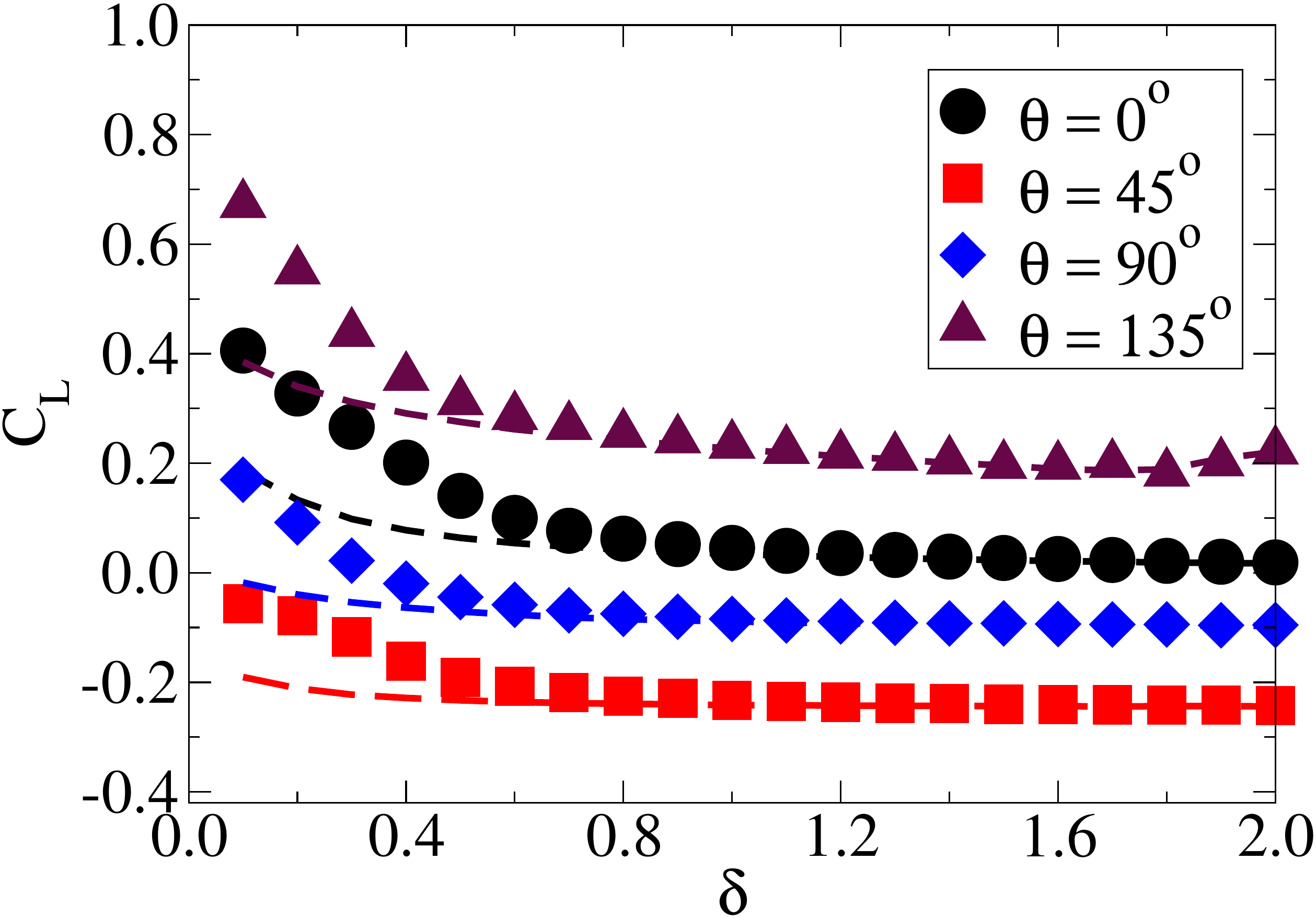}
		\caption{$Re_s = 100$}
		\label{fig:ClReShear100ThetaAll}
	\end{subfigure}
	\caption{Lift coefficient ($C_l$) as a function of wall normal distance $\delta$ at several Reynolds number. A dashed line with same color represent the corresponding values of $C_l$ for smooth wall condition.}
	\label{fig:ClReShearAllThetaAll}
\end{figure}

In this section, we report the lift coefficients for different particle-wall 
separation distances at three different shear-Reynolds numbers ($Re_s$).
\autoref{fig:ClReShearAllThetaAll}, shows the variation of lift coefficient ($C_L$) with $\delta$ at different inclination angles ($\ang{0}$ to $\ang{135}$). In addition to the results for rough wall, we have also presented the lift coefficients for a smooth wall. At lower shear-Reynolds number ($Re_s = 10$), the wall effects for both types of wall  are very prominent below $\delta = 1.5$. At higher separation, the lift coefficient becomes almost independent of the separation distance. It is observed from the \autoref{fig:ClReShear10ThetaAll} that for all angles, the lift force act away from the wall but for $\theta = \ang{45}$. At this angle, a particle is effectively pushed towards the wall. Effect of rough wall is significant for a separation distance up to $\delta = 1.0$ at low shear-Reynolds number and up to $\delta=0.5$ for higher shear Reynolds number as shown in \cref{fig:ClReShear50ThetaAll,fig:ClReShear100ThetaAll}. For a horizontally aligned ellipsoid very near the wall, lift induced by a rough wall is almost 1.5 to 2 times higher than that exerted by a smooth wall across all the Reynolds numbers.


\begin{figure}[htb]
	\centering
	\begin{subfigure}[b]{0.45\textwidth}
		\includegraphics[width=\textwidth]{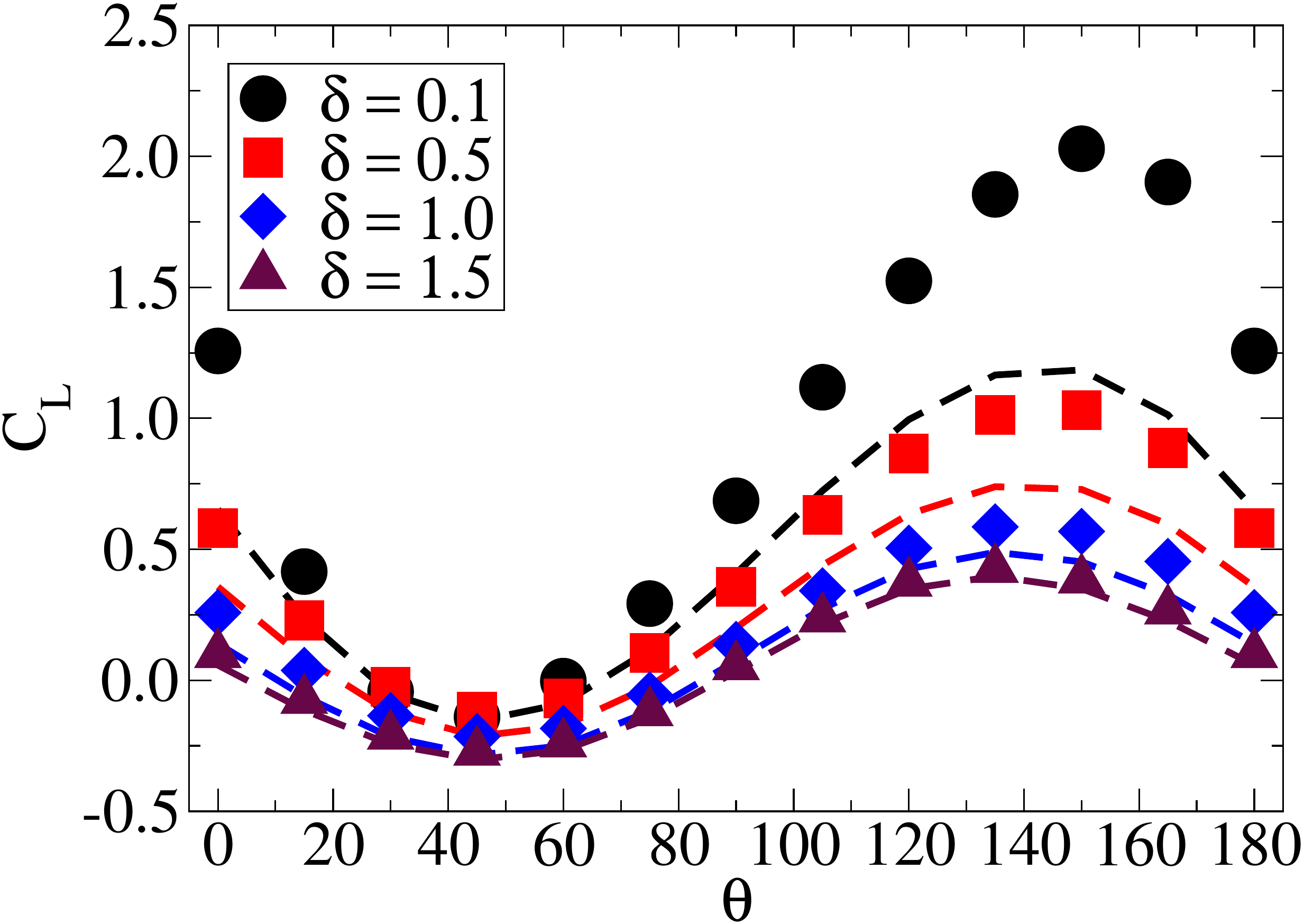}
		\caption{$Re_s = 10$}
		\label{fig:ClReShear10DeltaAll}
	\end{subfigure}
	\quad
	\begin{subfigure}[b]{0.45\textwidth}
		\includegraphics[width=\textwidth]{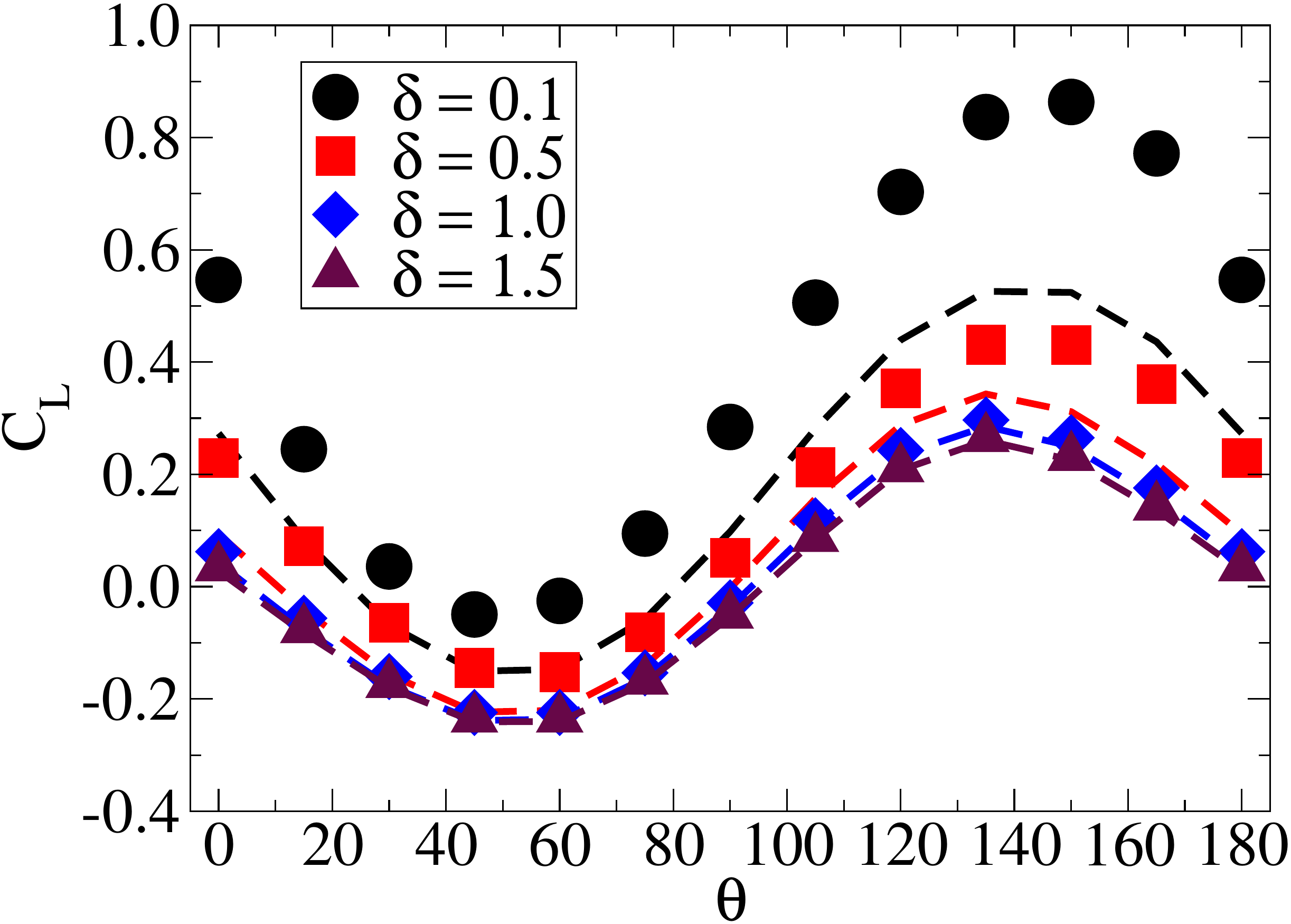}
		\caption{$Re_s = 50$}
		\label{fig:ClReShear50DeltaAll}
	\end{subfigure}
	\quad
	\begin{subfigure}[b]{0.45\textwidth}
		\includegraphics[width=\textwidth]{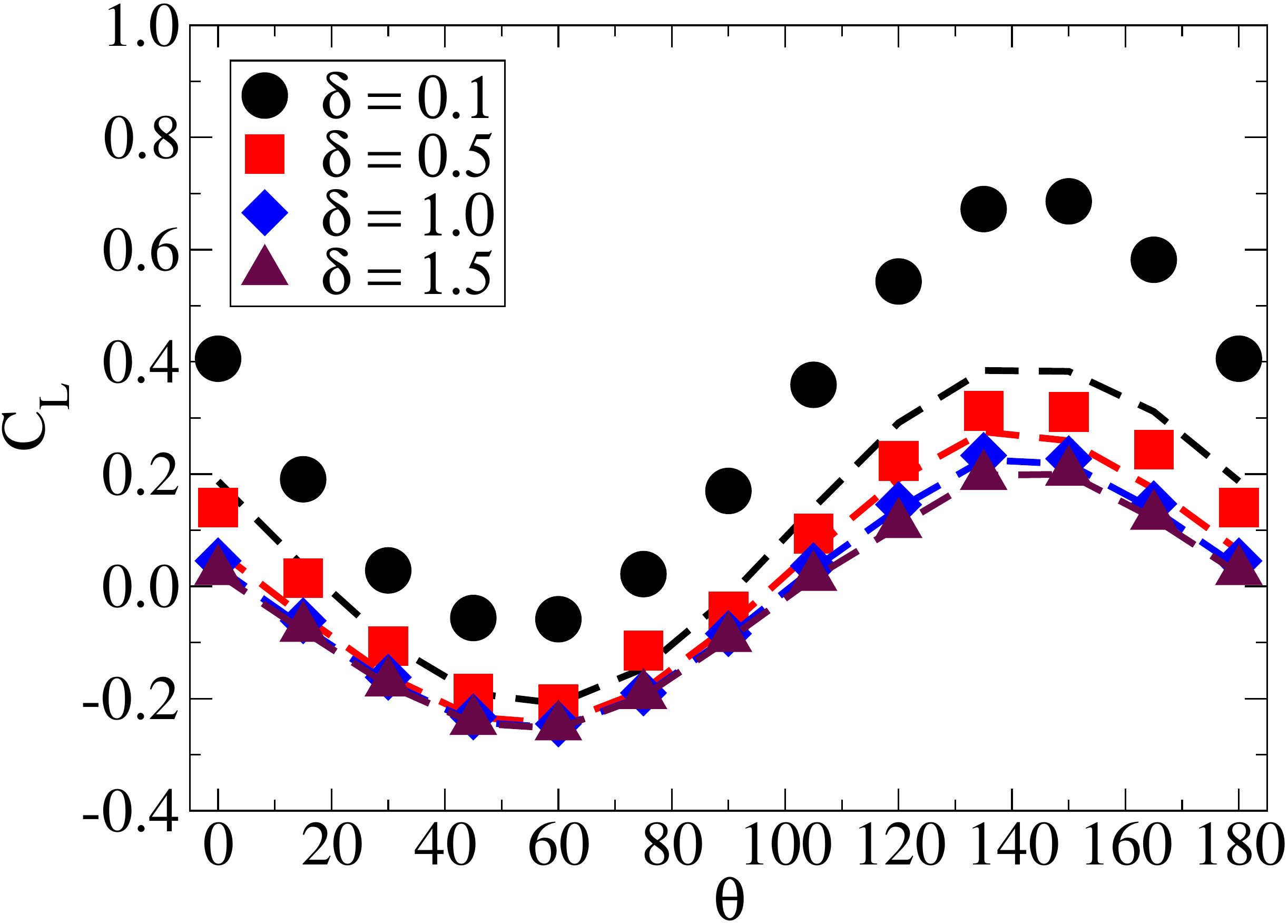}
		\caption{$Re_s = 100$}
		\label{fig:ClReShear100DeltaAll}
	\end{subfigure}
	\caption{Lift coefficient ($C_l$) as a function of orientation angle $\theta$ at several Reynolds number. A dashed line with same color represent the corresponding values of $C_l$ for smooth wall condition.}
	\label{fig:ClReShearAllDeltaAll}
\end{figure}

The variation of lift coefficients with  inclination angle ($\theta$) at different shear Reynolds numbers is  presented in \autoref{fig:ClReShearAllDeltaAll}. Here, the angle is measured between the wall (or the direction of mean flow) and the major axis of the ellipsoid in a counter clock-wise direction. The lift coefficients show asymmetric nature at intermediate angles. It is shown in  \autoref{fig:ClReShearAllDeltaAll} that the roughness plays a crucial role in deciding the asymmetry in lift coefficient  with change in angle at least when the particle is very close to the wall. It is also clear from the figures that with  angle up to $\theta = \ang{90}$, the difference in lift coefficients between the smooth and rough wall is less than that for angles higher than $\theta = \ang{90}$. Another important observation is that the particle encounters a lift force towards the wall for an angle between $\ang{30}$ and $\ang{50}$, which may be due to the creation of a low-pressure zone at the bottom rear end of the ellipsoid. An analysis of the contribution of viscous and pressure components towards the lift force  provides more insight into that, which is discussed below.

\begin{figure}[htb]
	\centering
	\begin{subfigure}[b]{0.48\textwidth}
		\includegraphics[width=\textwidth]{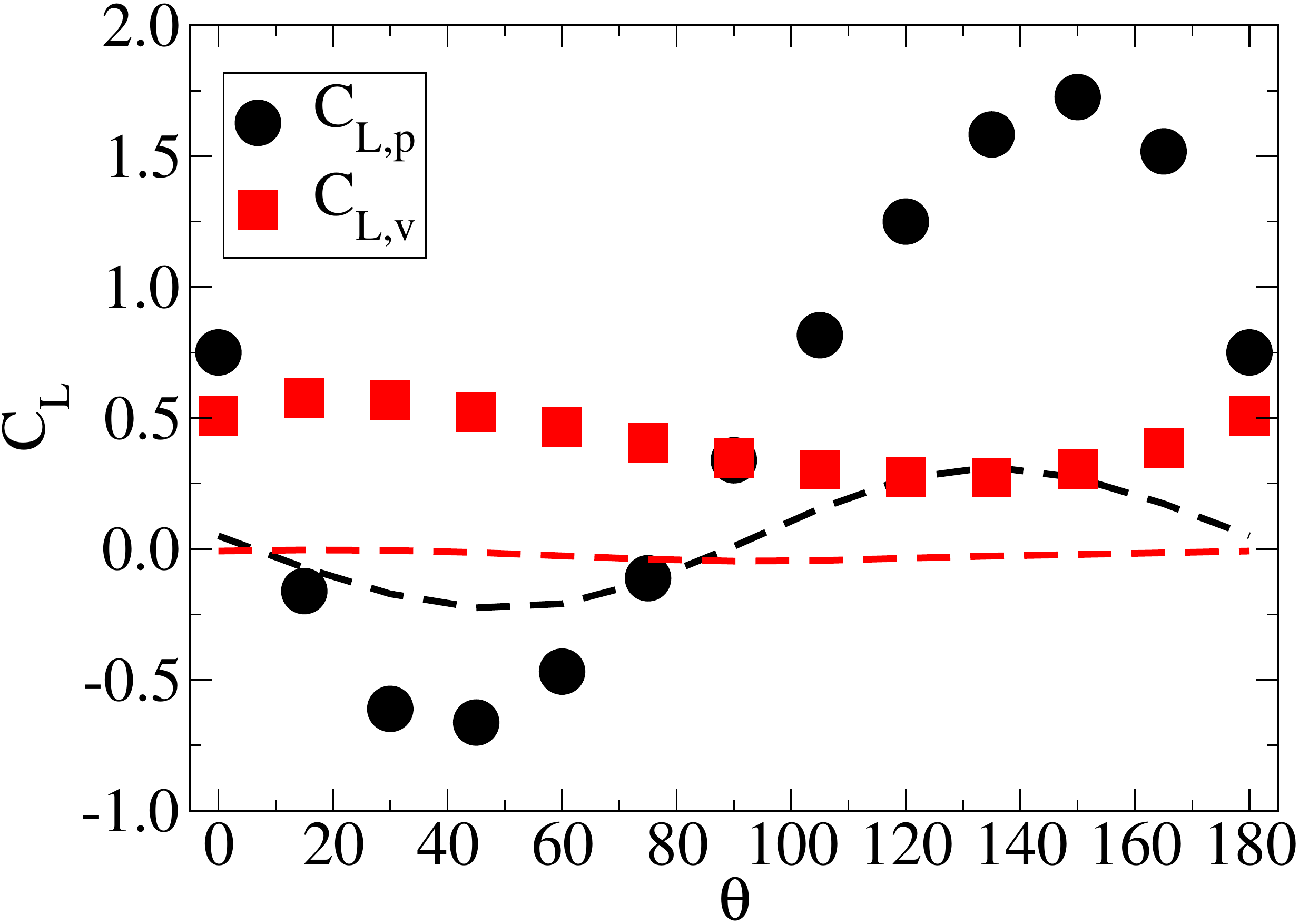}
		\caption{$Re_s = 10$}
		\label{fig:ClpClvRes10Delta0p1}
	\end{subfigure}
	\quad
	\begin{subfigure}[b]{0.48\textwidth}
		\includegraphics[width=\textwidth]{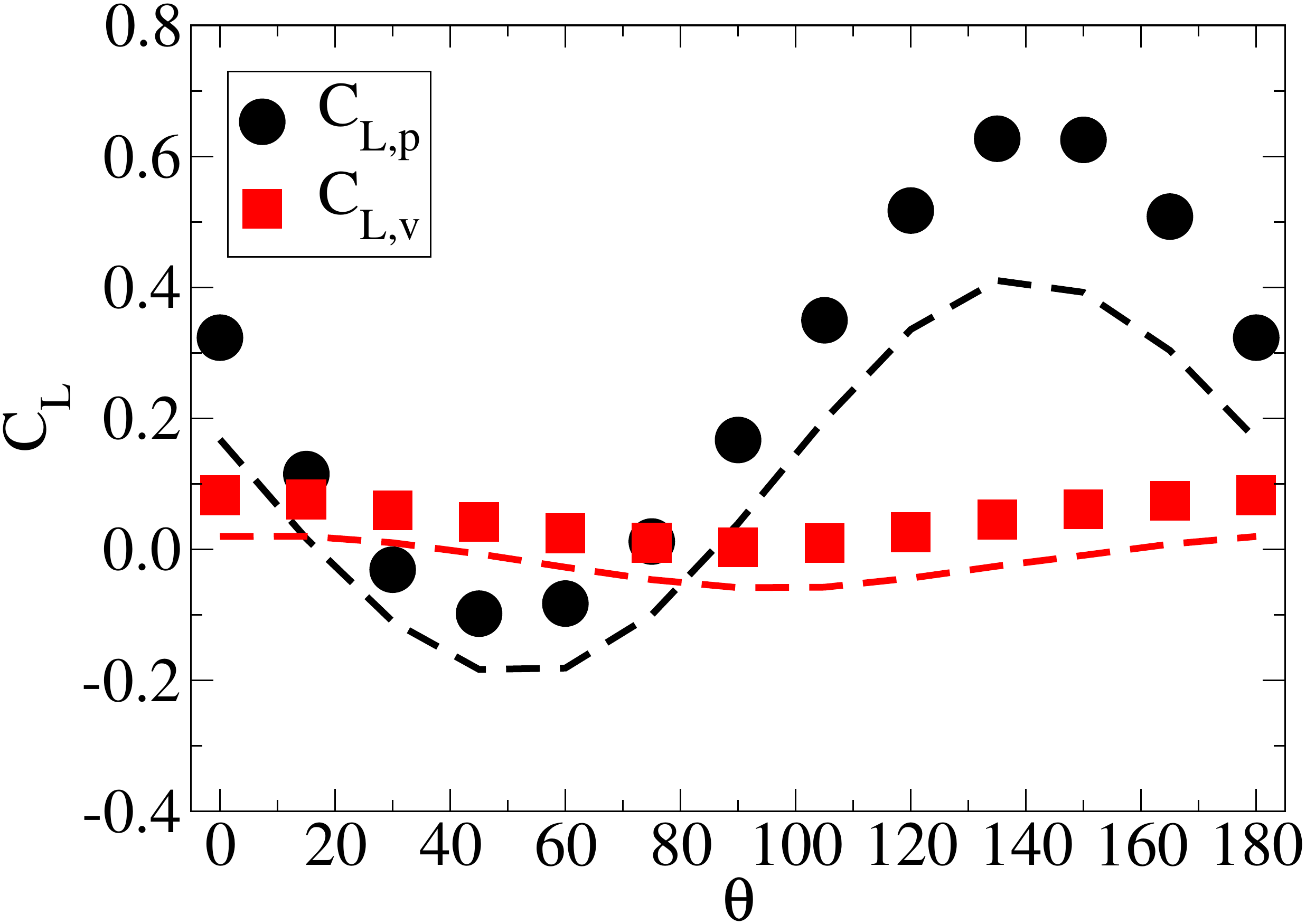}
		\caption{$Re_s = 100$}
		\label{fig:ClpClvRes100Delta0p1}
	\end{subfigure}
	\caption{components of lift coefficient ($C_{l,p}$, $C_{l,v}$) as a function of orientation angle $\theta$ and, $\delta = 0.1$. A dashed line with same color represent the corresponding values of variable for smooth wall condition.}
	\label{fig:ClpClvResAllDelta0p1}
\end{figure}

The contribution from the pressure variation and viscous effect towards the total lift is depicted in \autoref{fig:ClpClvResAllDelta0p1}. The figures show the variation of $C_{L,v}$ and $C_{L,p}$ Vs $\theta$ at $\delta = 0.1$ for $Re_s = 10$ and $100$. We have considered two different shear Reynolds numbers, $Re_s=10$ and $100$, for this analysis. Results are shown for the case when a particle is placed very near the wall ($\delta = 0.1$). It is observed from the \autoref{fig:ClpClvRes10Delta0p1} that pressure contribution towards the total lift is much higher compared to its viscous counterpart even at low shear Reynolds number. Pressure lift works towards the wall up to an angle of $\ang{80}$, but at a higher angle, net pressure force is exerted in the opposite direction. Maximum difference between the pressure to viscous contribution happens at an angle $\approx\ang{150}$; the ratio is about 6. A comparison with smooth wall at the lower shear Reynolds number (\autoref{fig:ClpClvRes10Delta0p1}) depicts that in the presence of a rough wall, both the viscous and pressure lift increases dramatically. The overall lift force acts away from the wall when a particle is very near the wall with an inclination angle of $>\ang{120}$. Such an observation has huge implications in the area of research that addresses the wall-deposition behaviour of nonspherical particles. It is shown in the figure that the pressure contribution to the lift for a smooth wall is much lower compared to the rough wall. The difference between those decreases at higher Reynolds number (\cref{fig:ClpClvRes10Delta0p1,fig:ClpClvRes100Delta0p1}). Another important outcome of the present analysis is that in the presence of roughness, there is a significant increase in viscous lift away from the wall when the particle is aligned horizontally. Accepting the point that the scenario may change for the  particle moves parallel to the wall, the present results can be a motivation for the wall modification to avoid particle deposition.


Turning our attention to develop the correlation for lift coefficient, 
we consider that the lift coefficient decreases exponentially with wall particle separation distance $\delta$, magnitude of lift coefficient is a function of shear Reynolds number $Re_s$, which is related to particle Reynolds number $Re_p$ as a function of $\delta$, $\theta$, and $\omega$ (aspect ratio of the ellipsoid). The effect of variation of angle is introduced through a skewed function of $\sin\ \theta \cos\ \theta$.


Using all the above criteria, we reach to the functional form of the correlation as, 
\begin{equation}
	C_L = C_{L,s}\big\{\boldsymbol{f}(\delta, Re_s)+\boldsymbol{g}(\delta, Re_s)\sin[\boldsymbol{h}(\delta, Re_s)\theta]\cdot\cos[\boldsymbol{h}(\delta, Re_s)\theta]\big\}
\end{equation}
Where,
$C_{L,s}$ represents the lift for ellipsoidal particle resting on wall as reported earlier by \citet{FILLINGHAM2021}.
\begin{equation}
	C_{L,s} = \frac{3.663}{(\omega^{2.021}Re_p^2+0.1173\omega^{1.559})^{0.22}}
\end{equation}
$\boldsymbol{f}(\delta, Re_s)$ takes care of the variation of lift coefficient 
including the change on direction of the lift.  
\begin{equation}
	\boldsymbol{f}(\delta, Re_s)=-0.043-0.00083Re_s+\frac{(0.93+0.0023Re_s)}{e^{(0.98\delta+0.016Re_s\delta)}}
\end{equation}
$\boldsymbol{g}(\delta, Re_s)$ is expressed as follows.
\begin{equation}
\boldsymbol{g}(\delta, Re_s) = \boldsymbol{g'}(\delta) + \boldsymbol{g''}(\delta, Re_s)
\end{equation} 

Here, the first part is an exclusive function of $\delta$,  while the second part contains effect of variation of both the $\delta$ and $Re_s$. 
\begin{equation}
	\boldsymbol{g'}(\delta) = 1.15e^{(-4.27\delta)}+0.8e^{(0.17\delta)}
\end{equation}
\begin{equation}
	\boldsymbol{g''}(\delta, Re_s)  = -\Big[0.285\delta+(0.001056Re_s\delta - 3.2)Re_s\Big]\cdot e^{-(6.91-\delta^{(0.16-0.0065Re_s)})}
\end{equation}
$\boldsymbol{h}(\delta, Re_s)$ contributes to capture the  skewed variation of $C_L$ with $\theta$. 
\begin{equation}
	\boldsymbol{h}(\delta, Re_s) = 0.018Re_s+0.96
\end{equation}
where $Re_s$ is related to $Re_p$ as,
\begin{equation}
	Re_s = \frac{Re_p}{(0.3165+\delta)+\frac{1}{2\sqrt{\omega}}+\frac{\sqrt{\omega}}{2}(1-\frac{1}{\omega})\sin^2 \theta}
\end{equation}

\begin{figure}[htb]
	\centering
	\begin{subfigure}[b]{0.48\textwidth}
		\includegraphics[width=\textwidth]{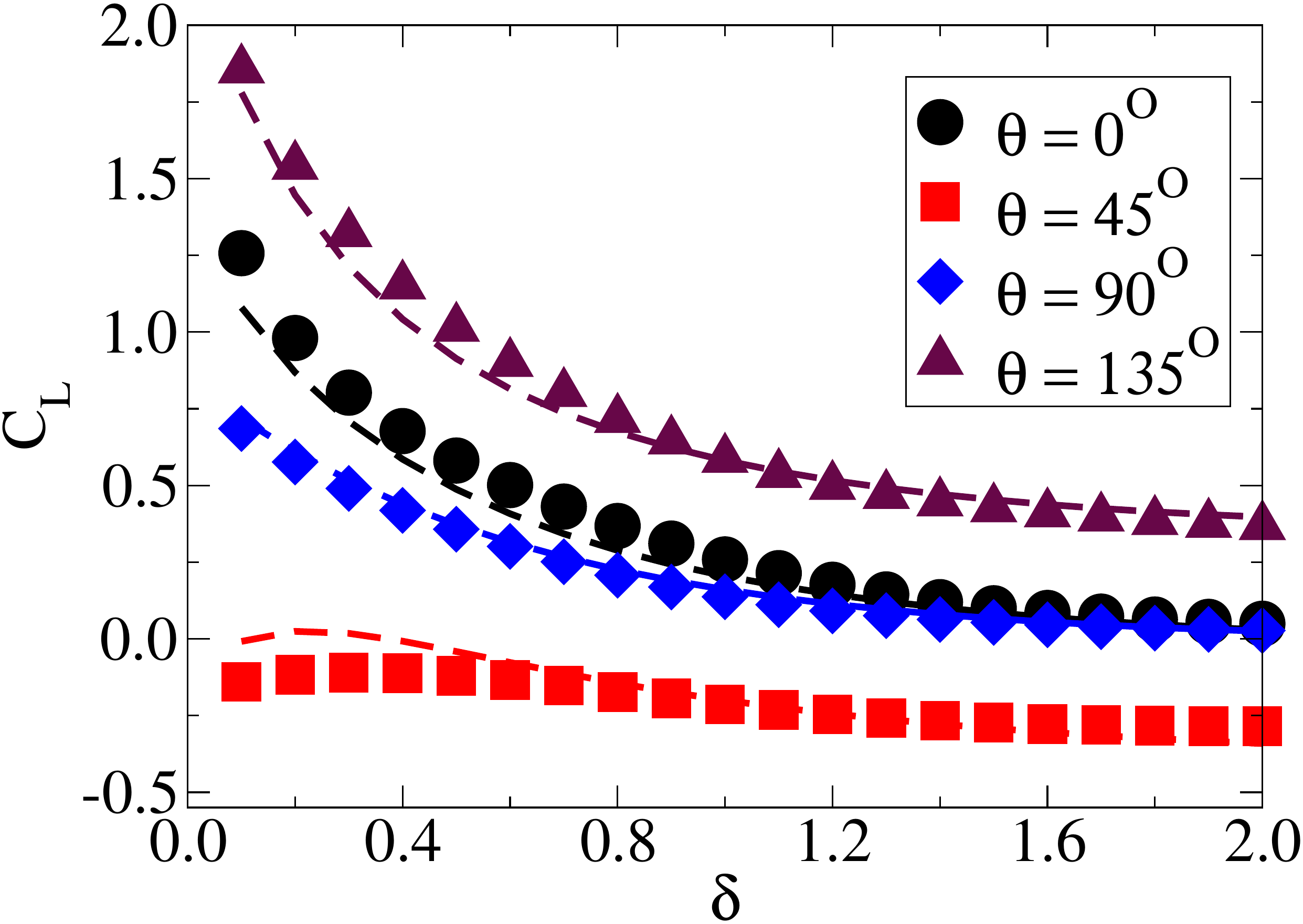}
		\caption{$Re_s = 10$}
		\label{fig:clVsDeltaAtRes10Theta04590135}
	\end{subfigure}
	\quad
	\begin{subfigure}[b]{0.48\textwidth}
		\includegraphics[width=\textwidth]{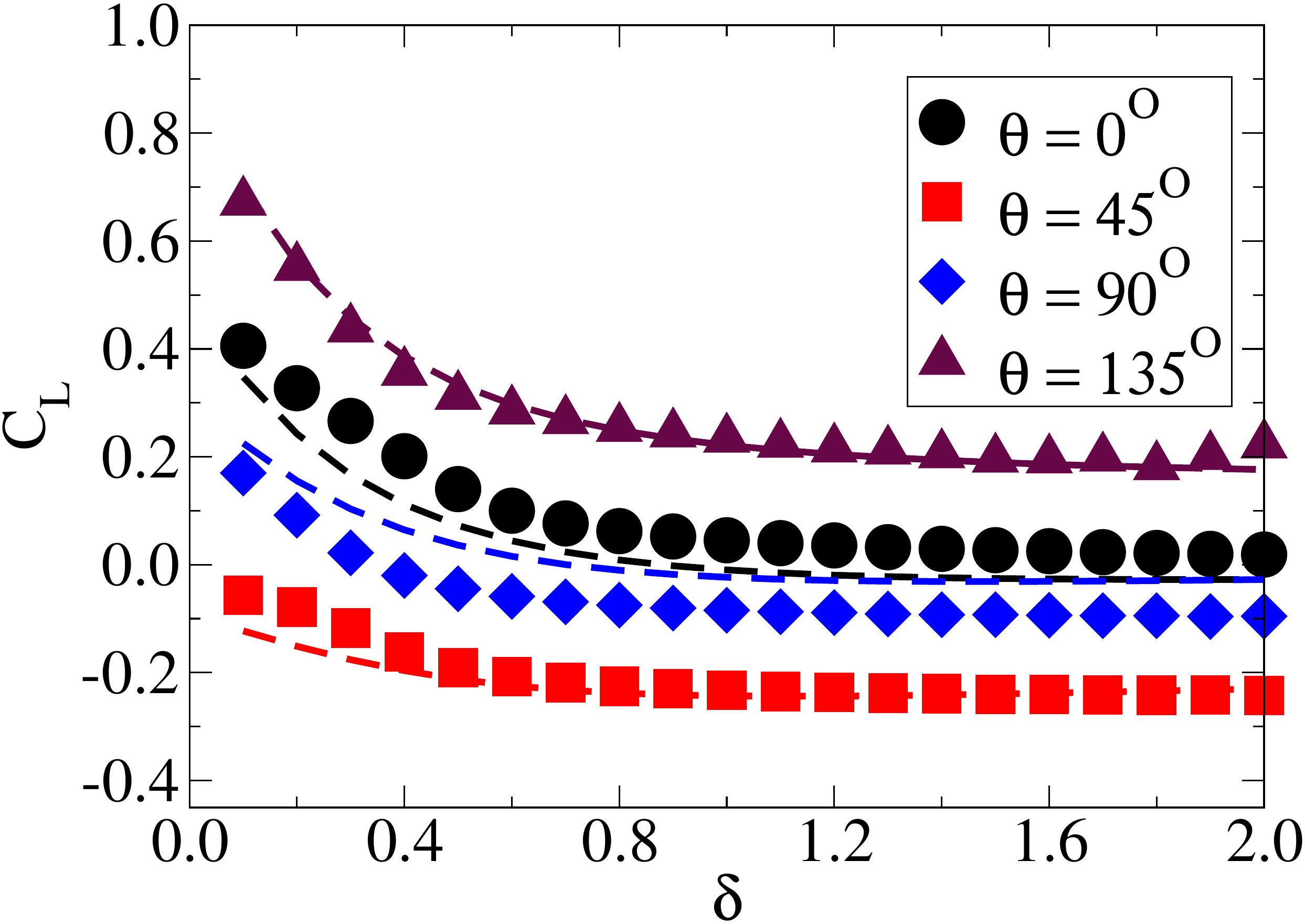}
		\caption{$Re_s = 100$}
		\label{fig:clVsDeltaAtRes100Theta04590135}
	\end{subfigure}
	\caption{$C_L$ Vs $\delta$ for $Re_s$ = 10 and 100 at different orientation angle ($\theta$). The dashed line in corresponding color represents the fitted value from the correlation}
	\label{fig:clVsDeltaAtResAllTheta04590135}
\end{figure}

\begin{figure}[htb]
	\centering
	\begin{subfigure}[b]{0.48\textwidth}
		\includegraphics[width=\textwidth]{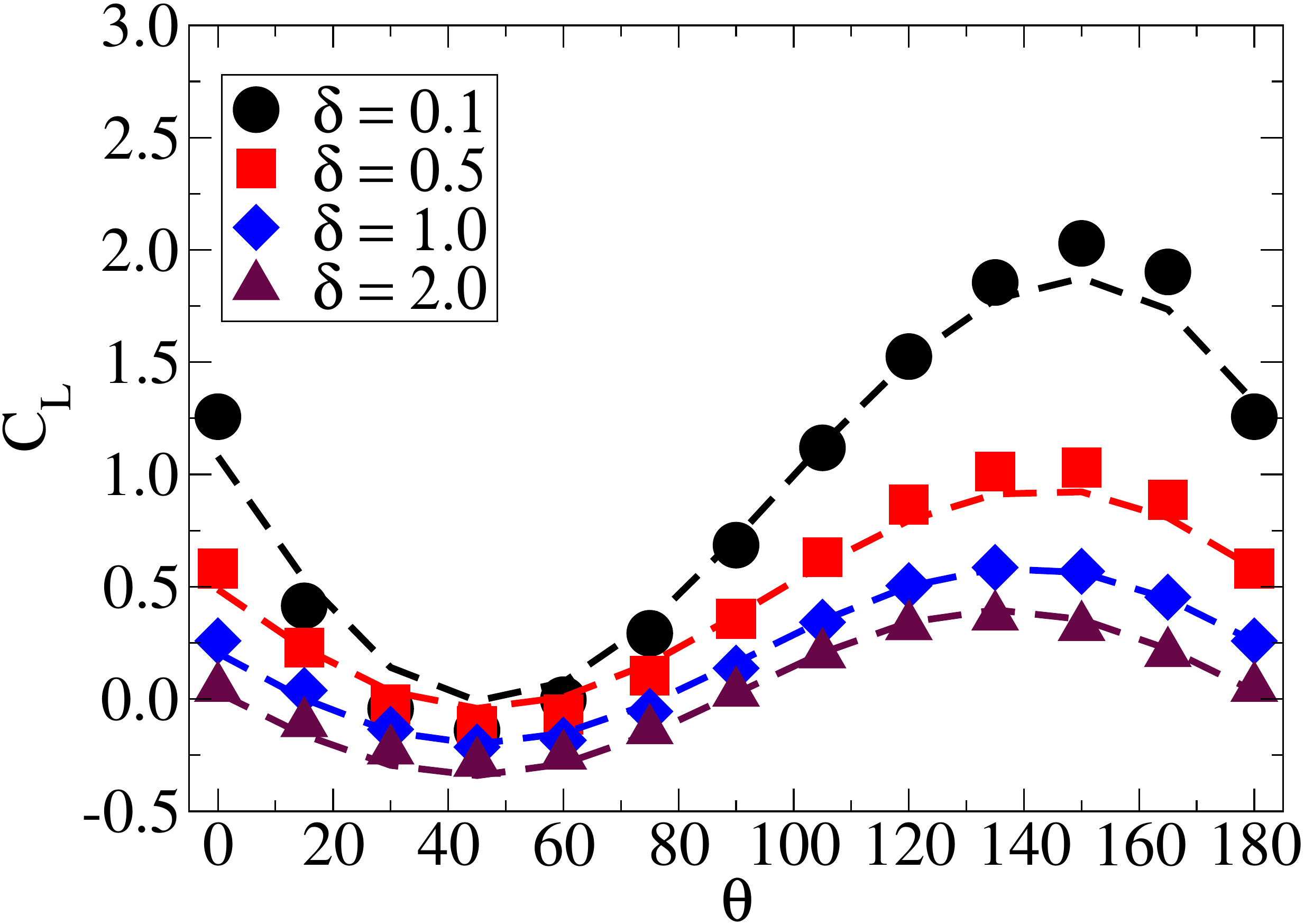}
		\caption{$Re_s = 10$}
	\end{subfigure}
	\quad
	\begin{subfigure}[b]{0.48\textwidth}
		\includegraphics[width=\textwidth]{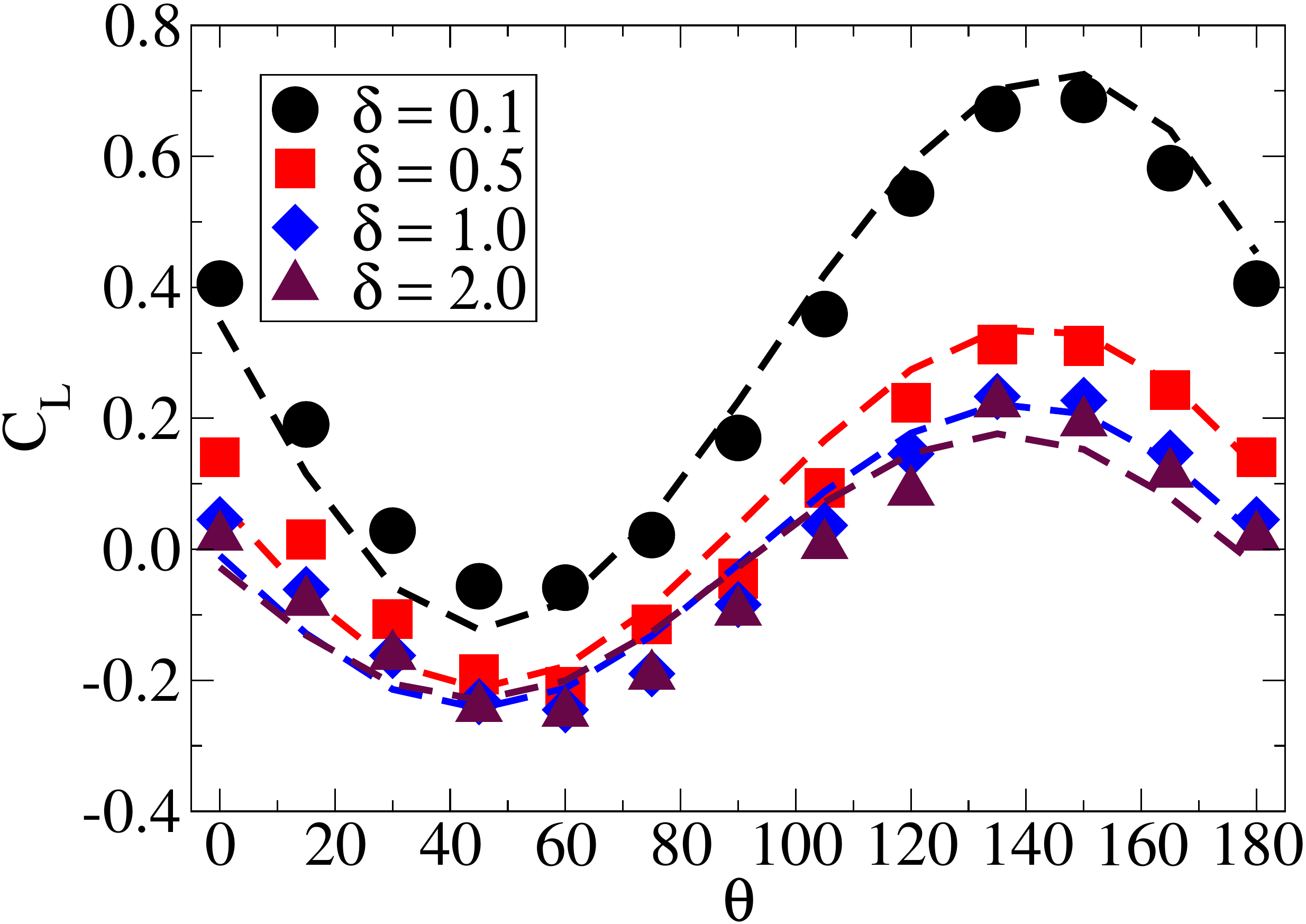}
		\caption{$Re_s = 100$}
	\end{subfigure}
	\caption{$C_L$ Vs $\theta$ for $Re_s$ = 10 and 100 at different wall separation distance ($\delta$). The dashed line in corresponding color represents the fitted value from the correlation}
	\label{fig:CompResAllDeltaAllClVsTheta}
\end{figure}

\begin{figure}[htb]
	\centering
	\includegraphics[width=0.75\textwidth]{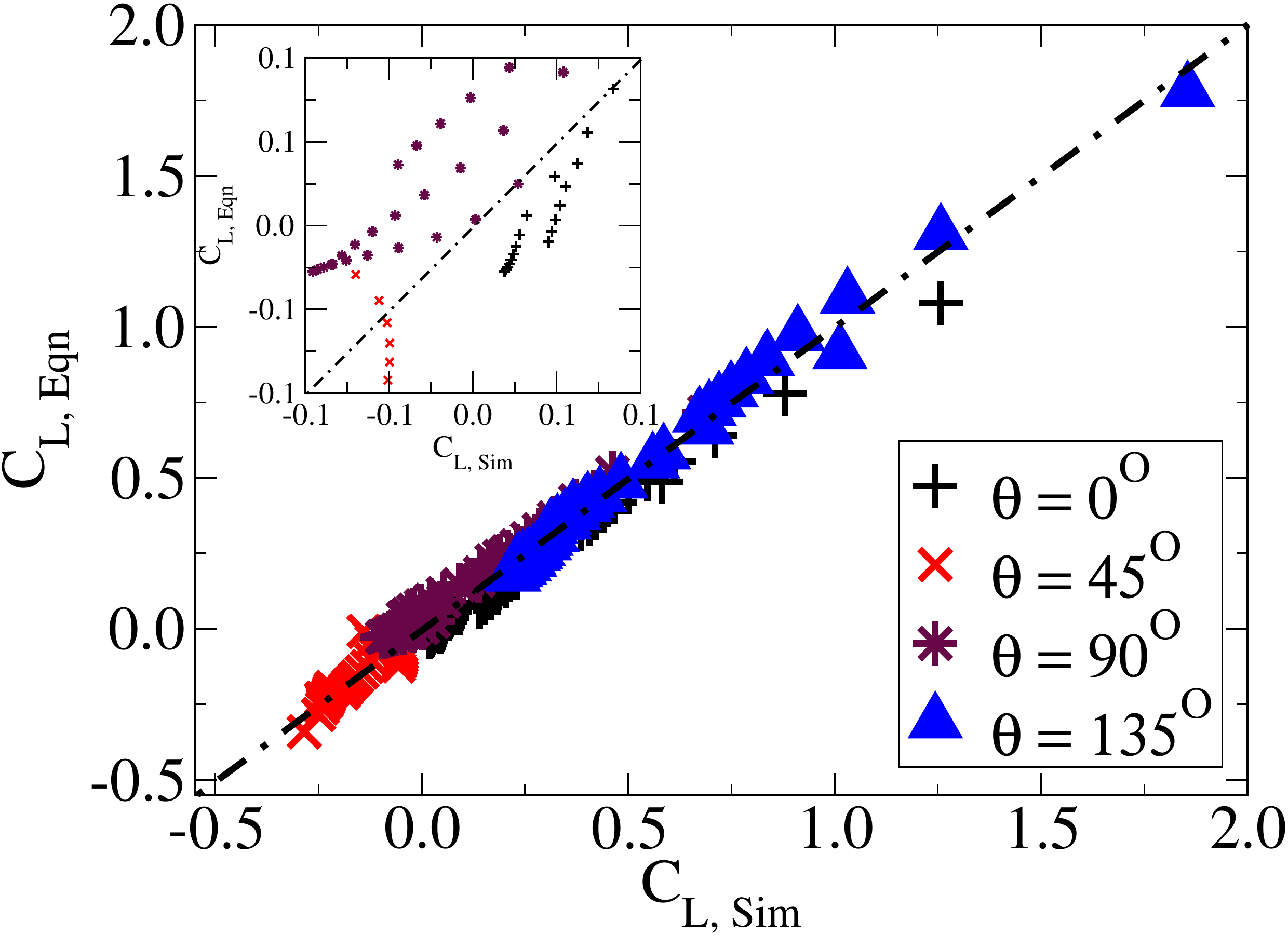}
	\caption{Predicted $C_L$ plotted against $C_D$ from simulation data at various $\theta$, and ($-\cdot-\cdot$) represent line of $y=x$}
	\label{fig:ClEqVsClSimThetaAll}
\end{figure}

Performance of the developed correlation in predicting the lift coefficient is assessed for different separation distance, angle of inclination, and for different Reynolds numbers. \autoref{fig:clVsDeltaAtResAllTheta04590135} shows the comparison between lift coefficient calculated using simulations and that predicted from the correlation as function of $\delta$ at $Re_s = 10$ (\autoref{fig:clVsDeltaAtRes10Theta04590135}) and $Re_s = 100$ (\autoref{fig:clVsDeltaAtRes100Theta04590135}). It is observed that the predictions from the correlation match well with the simulations results in majority of the cases. At $Re_s=10$, when the particle is placed very near the wall at an  $\ang{45}$, a correlation over-predicts the lift coefficient. In case of $Re_s=100$, at angle  $\ang{90}$, correlation over-predicts the lift coefficients for all the separation distances (\autoref{fig:clVsDeltaAtRes100Theta04590135}). However, the expression can successfully reports the trends of lift variation with wall separation distance at all shear Reynolds numbers.
\\Lift coefficients calculated from simulation and predicted by correlation are plotted as function of $\theta$ at several wall separation distance $\delta$, as shown in \autoref{fig:CompResAllDeltaAllClVsTheta}. A good agreement is observed for both the shear-Reynolds numbers. Most importantly, the asymmetry that occurs at a lower separation distance is well captured by the correlation. It is observed that, at high shear-Reynolds number ($Re_s=100$) when the wall separation distance ($\delta$) is $0.5$, correlation under-predicts the lift coefficients. To obtain a comprehensive assessment of the performance of correlation, developed in the present study, we have plotted  the lift coefficient predicted by correlation with that obtained from simulations for $\delta$ = 0.1, 0.5, 1.0, and 2.0, and $\theta$ = $\ang{0},\ang{45},\ang{90}$, and $\ang{135}$ in \autoref{fig:ClEqVsClSimThetaAll}. Even though majority of the data  collapses along a diagonal line in \autoref{fig:ClEqVsClSimThetaAll}, there is 
significant  deviation when the lift coefficient is less than $0.1$. 
  At a lower shear Reynolds number, the  proposed correlation can predict the simulated
 results fairly well with a maximum deviation of $20\%$ for all the orientation angles but  for $\theta = \ang{0}$. The error is maximum at those points where there is a 
 change is direction of the lift forces. 
  At a higher shear Reynolds number the lift becomes comparatively small and changes
 its direction as the particle moves away from the wall at a particular orientation angle. The
 region (wall separation distance) where lift is significantly small ($O(10^{-2}$))  and changes sign, the  proposed equation gives a large error as shown in the inset of \autoref{fig:ClEqVsClSimThetaAll}. 
 

\subsection{Torque coefficient}
\begin{figure}[htb]
	\centering
	\begin{subfigure}[b]{0.45\textwidth}
		\includegraphics[width=\textwidth]{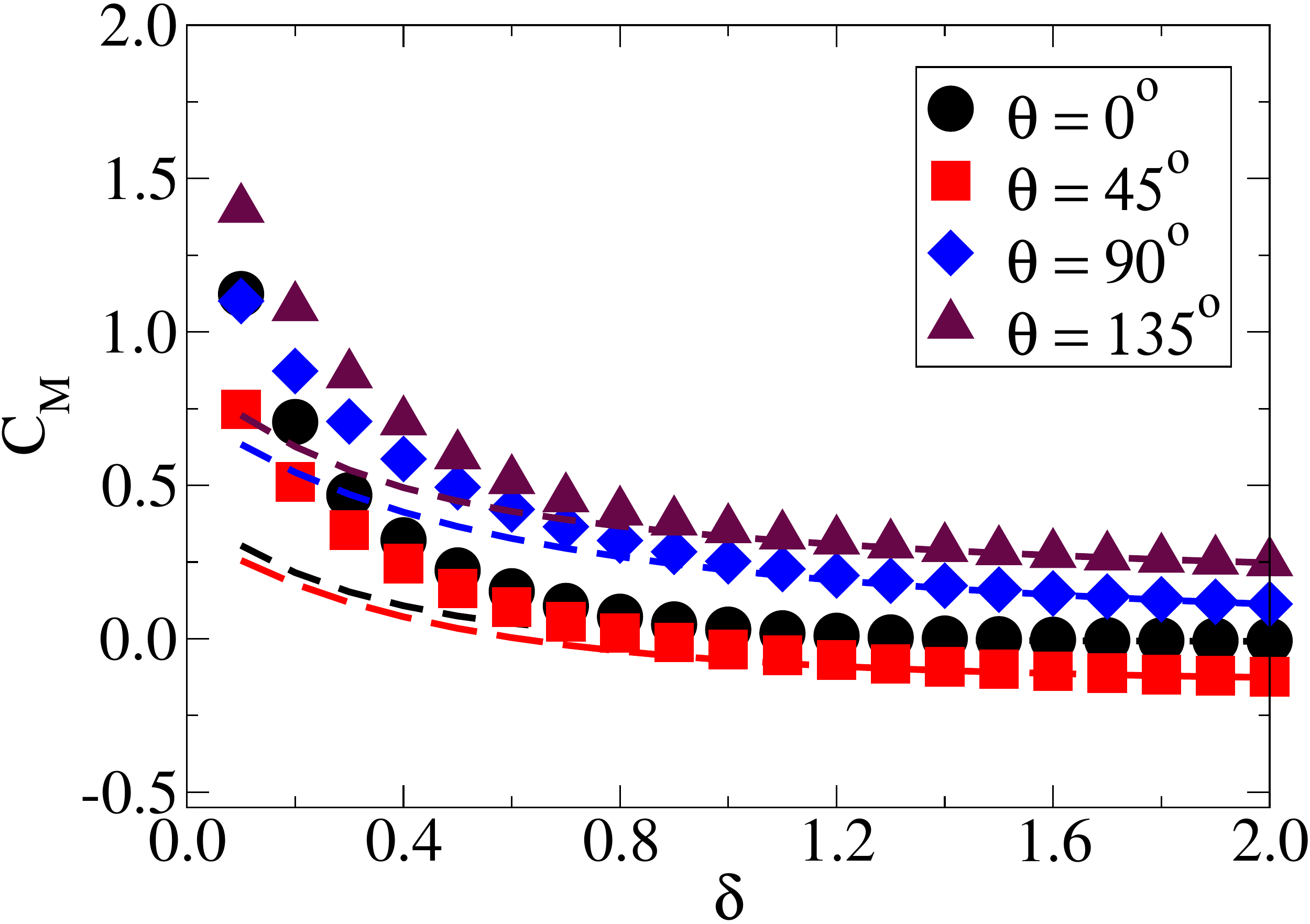}
		\caption{$Re_s = 10$}
		\label{fig:CmReShear10ThetaAll}
	\end{subfigure}
	\quad
	\begin{subfigure}[b]{0.45\textwidth}
		\includegraphics[width=\textwidth]{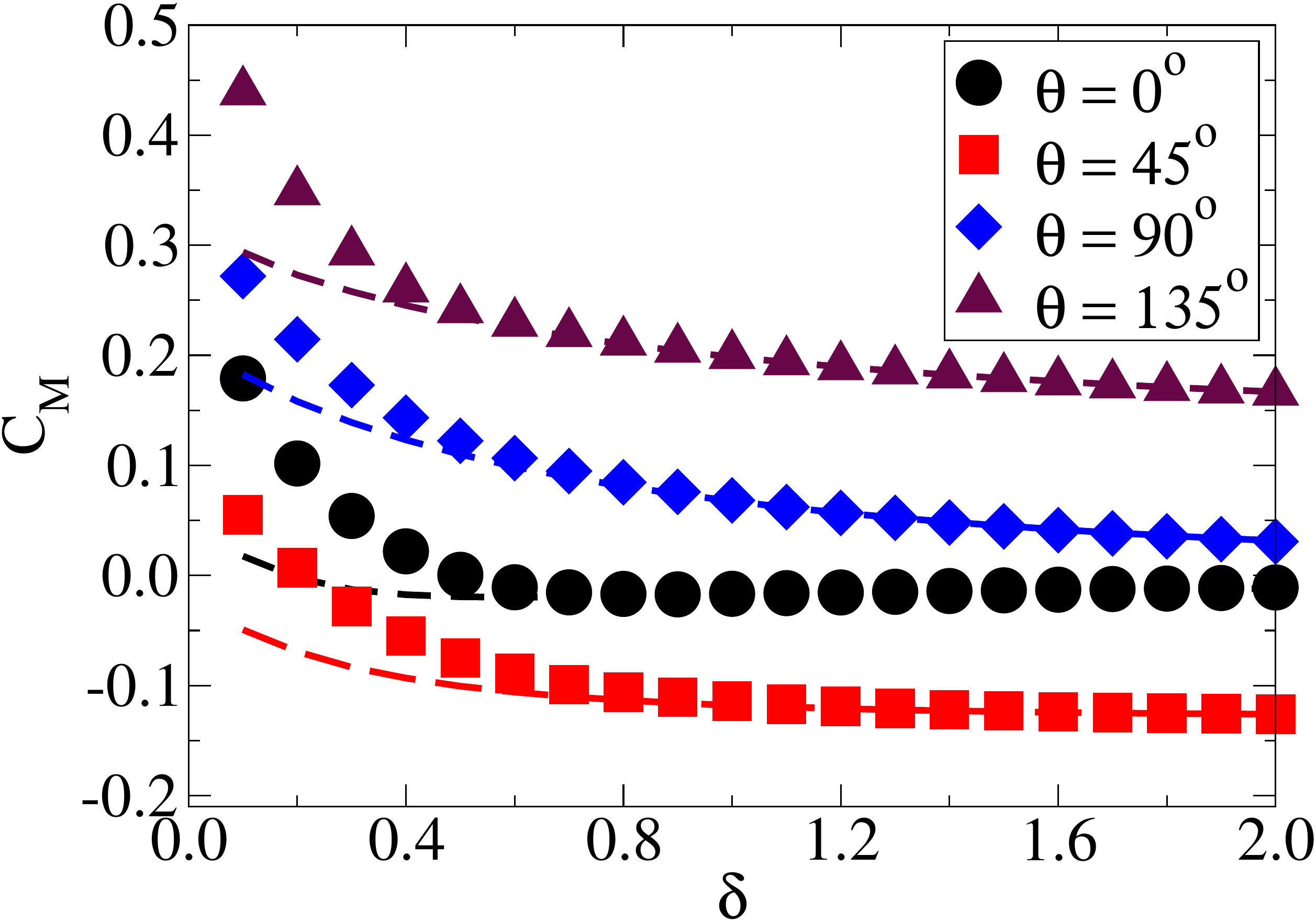}
		\caption{$Re_s = 50$}
		\label{fig:CmReShear50ThetaAll}
	\end{subfigure}
	\quad
	\begin{subfigure}[b]{0.45\textwidth}
		\includegraphics[width=\textwidth]{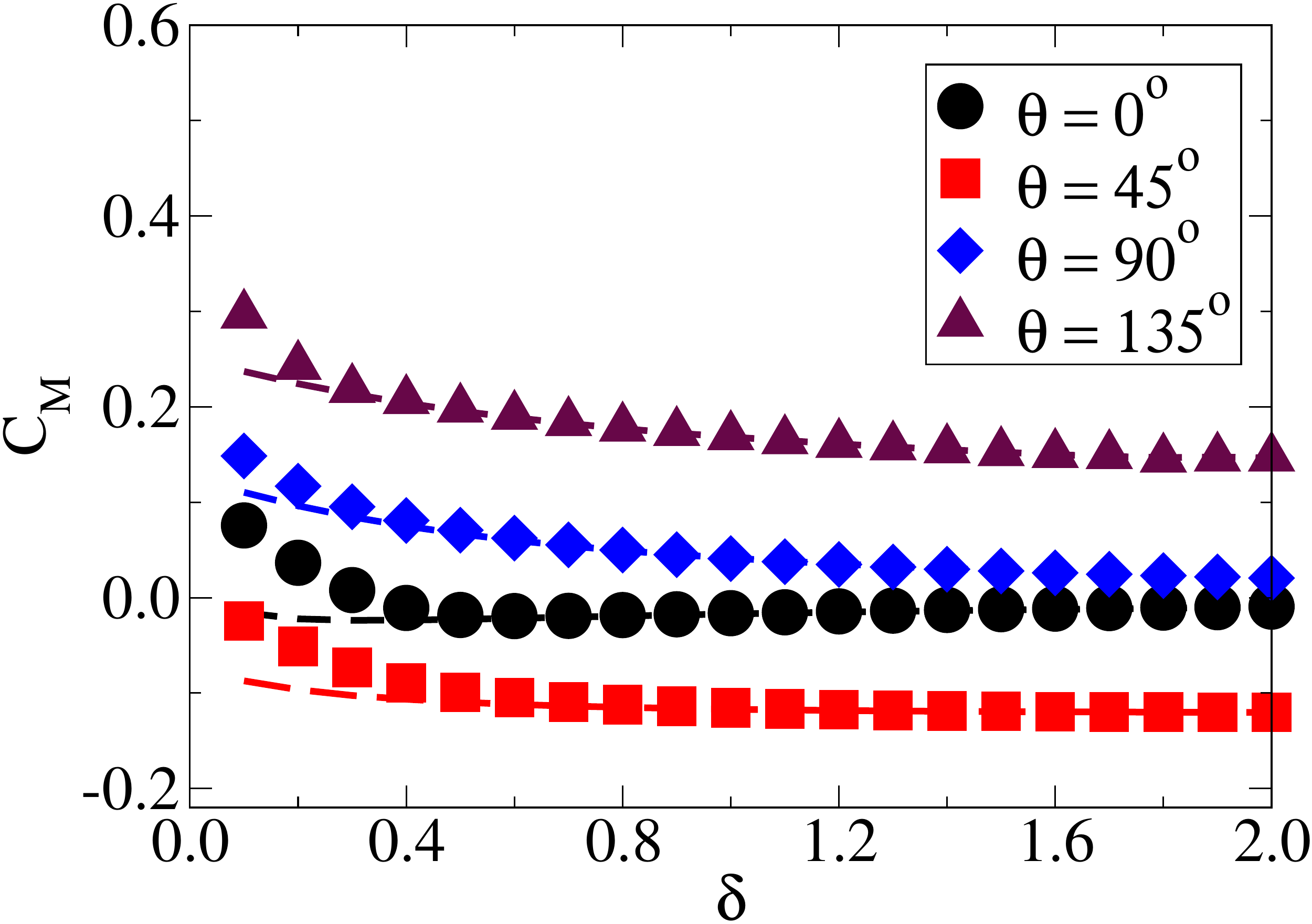}
		\caption{$Re_s = 100$}
		\label{fig:CmReShear100ThetaAll}
	\end{subfigure}
	\caption{Torque coefficient ($C_m$) as a function of wall normal distance $\delta$ at several Reynolds number. A dashed line with same color represent the corresponding values of $C_m$ for smooth wall condition.}
	\label{fig:CmReShearAllThetaAll}
\end{figure}

For a nonspherical particle immersed in a shear flow, besides calculating drag and lift, it is also essential to get an estimate of the torque.  This section analyses the  torque coefficient around  its center of mass point for three different shear  Reynolds numbers, different angles of orientation, and wall separation distances. We compare the results for both the rough and smooth walls. \autoref{fig:CmReShearAllThetaAll} shows a variation of torque coefficient $C_M$ with wall-normal distance $\delta$ for four different orientation angles ($\theta$) and at different Reynolds numbers.  We use the right-hand coordinate system to express the direction of torque. Legends used in the figures have same significances that have been stated earlier. It is observed in the figure that $C_M$ decreases exponentially as ellipsoid moves away from the wall, a similar trend for the drag and lift coefficients.

 
 At the lower shear Reynolds number, reported here ($Re_s = 10$)
 torque coefficient ($C_M$) is positive even for horizontally aligned particle placed near the wall.  For an angle of inclination $\ang{0}$, change in torque coefficient is observed up to wall separation distance, $\delta\approx1$. With an increase in Reynolds number, the distance of influence decreases gradually.  In case of $Re_s = 50$ (\autoref{fig:CmReShear50ThetaAll}), the $\delta$ after which $C_M$ does not show any variation is $\delta \approx 0.7$, whereas for $Re_s = 100$ that is $\delta < 0.5$ (\autoref{fig:CmReShear100ThetaAll}).  A similar trend is observed for other angles as well. 
 The figures illustrate that there is a three to five-fold decrease in torque coefficient for a ten times increase in Reynolds number for minimum wall separation distance.

\begin{figure}[htb]
	\centering
	\begin{subfigure}[b]{0.45\textwidth}
		\includegraphics[width=\textwidth]{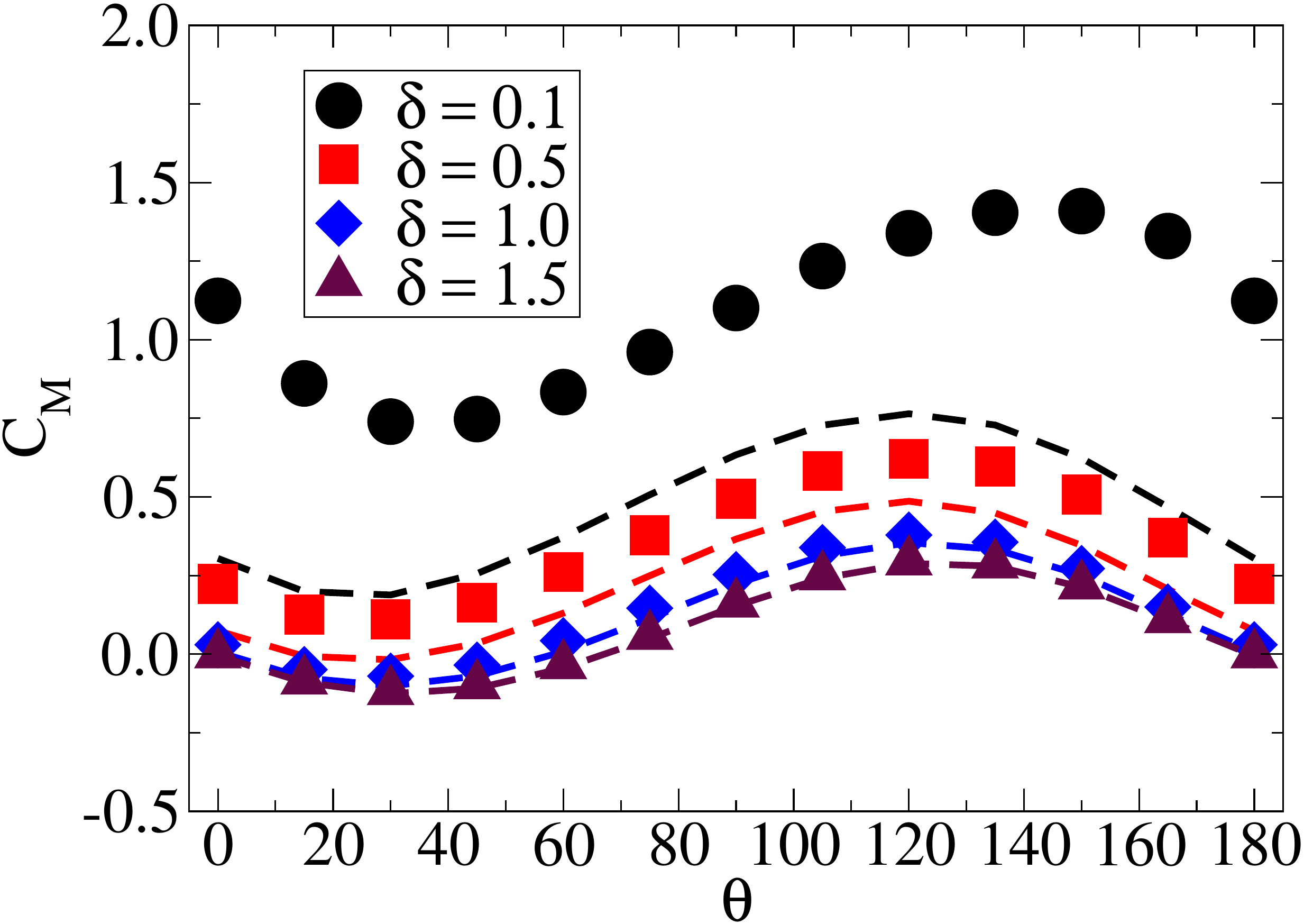}
		\caption{$Re_s = 10$}
		\label{fig:CmReShear10DeltaAll}
	\end{subfigure}
	\quad
	\begin{subfigure}[b]{0.45\textwidth}
		\includegraphics[width=\textwidth]{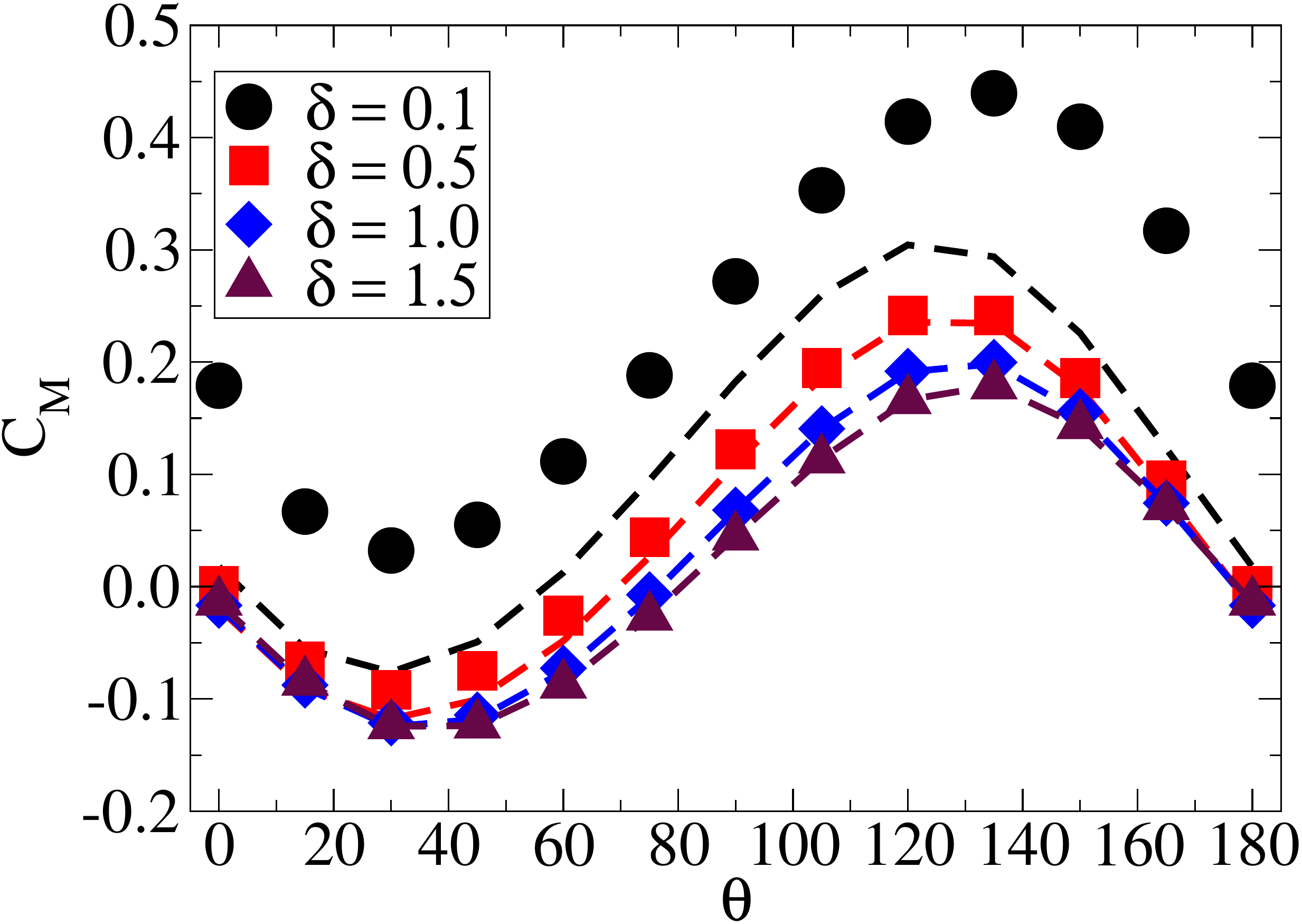}
		\caption{$Re_s = 50$}
		\label{fig:CmReShear50DeltaAll}
	\end{subfigure}
	\quad
	\begin{subfigure}[b]{0.45\textwidth}
		\includegraphics[width=\textwidth]{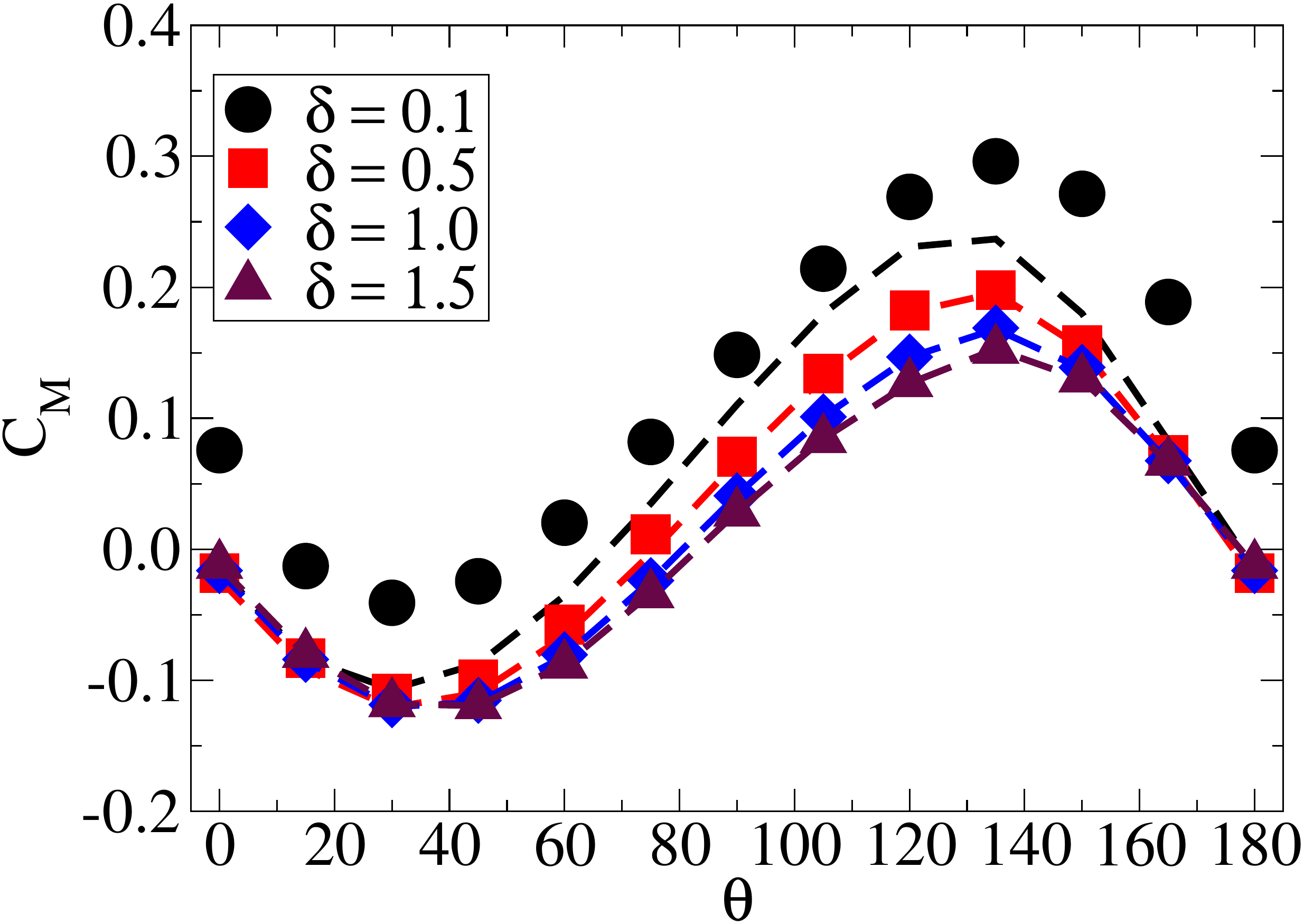}
		\caption{$Re_s = 100$}
		\label{fig:CmReShear100DeltaAll}
	\end{subfigure}
	\caption{Torque coefficient ($C_m$) as a function of orientation angle $\theta$ at several Reynolds number. A dashed line with same colour represent the corresponding values of $C_m$ for smooth wall condition.}
	\label{fig:CmReShearAllDeltaAll}
\end{figure}

\autoref{fig:CmReShearAllDeltaAll} show the variation of torque coefficient for different inclination angles and at three different shear Reynolds numbers. It is to be noted that the angle is changed in the plane of shear. It is shown in the \autoref{fig:CmReShear10DeltaAll} 
that the rough wall has  very prominent effect for the lowest wall-separation distance ($\delta=0.1$). With an increase in the distance from the wall, influence of the nature of the wall becomes insignificant. Extent of asymmetry in $C_M$ increases with increase in Shear Reynolds number (\autoref{fig:CmReShearAllDeltaAll}). At shear-Reynolds numbers, $Re_s=$ 50 and 100, if the particle-wall separation distance is $\ge0.5$, torque on the particle acts in the negative direction when the inclination angle is between $>\ang{0}$ and $\ang{80}$ 
(\cref{fig:CmReShear50DeltaAll,fig:CmReShear100DeltaAll}).

\section{Conclusion}\label{sec:conclusion}
Particle-laden flows with anisotropic particles find applications in industrial and natural processes. The interactions between anisotropic particles with fluid flow are complex in nature because the orientation of particles changes the fluid flow behaviour. Presence of an uneven wall in the vicinity of the particle adds additional complexity.  Earlier efforts were reported to estimate the forces in the presence of rough wall but for spherical particles only. We have performed numerical simulations (with high spatial resolution) to estimate the forces on an ellipsoidal particle placed in the vicinity of a rough wall at different wall separation distances and inclination angles. To create the wall roughness, we have followed the approach already demonstrated by \citet{LeeIJMF2017}. Simulations  have been performed for a range of Reynolds numbers $Re_s$ ($10\le Re_s \le 100$), $\delta$ ($0.1\le\delta\le 2$), and $\theta$ ($\ang{0}\le\theta\le\ang{180}$).  Undisturbed (by the ellipsoidal) fluid velocity at particle center of mass location is obtained using an empty channel simulation with wall roughness only. This velocity is used to calculate the drag, lift, and torque coefficients. Simulations with a smooth wall for the similar flow  conditions have also been performed and compared to quantify the effect of wall roughness.



Streamline plots over the ellipsoid qualitatively depicts the appearance of the flow separation and recirculating zone at the top edge for $\theta < 90$ and at the bottom edge for $\theta > 90$. Such a flow separation plays an essential role in deciding the direction of the lift force on the particle.   If the particle is placed near the wall at low shear Reynolds number, $C_d$ is almost two times higher than the smooth wall. The effect of a wall-roughness becomes insignificant away from the wall ($\delta>1.0$).  The drag coefficient for a vertically placed particle at a higher Reynolds number ($Re_s\approx100$) is larger than the horizontally placed particle,  even though a higher undisturbed velocity occurs at the particle location. At low Re, when the particle is placed very near the wall, a viscous component of the drag dominates over the pressure component for both types of wall.  If the particle is vertically aligned at a higher Reynolds number, the pressure drag is almost 1.5 times higher than the viscous counterpart. In case of lift coefficient, the effect of nature of wall is prominent below separation distance, $\delta=1.5$. In all the cases, lift acts away from the wall except the angle between $\ang{30}$ to $\ang{50}$. Effect of the rough wall is prominent up to a separation distance ($\delta$) of 1.0. 
Irrespective of the nature of the wall, an asymmetry is observed in the variation of lift coefficient as a function of an angle of the particle's major axis with the wall.
At low wall separation distance, the asymmetry is augmented by the wall roughness.


From the analysis of forces on the ellipsoidal particles, we develop and present  the semi-empirical 
correlations for the drag and lift coefficients for different wall separation distances 
(including roughness) and different inclination angles. The correlations are applicable for particle Reynolds numbers up to 250.  The accuracies of these correlations are verified by comparing predictions by the correlations with the simulation results for a wide range of parameters.   Maximum relative deviation for drag coefficient is found to be 8\%. 
However, correlation of the lift suffers from the accuracy 
at lower values of lift coefficients when there is a change in direction of the lift. \\
One of the important limitations of the present study is that the simulations have been performed for a single aspect ratio. Therefore, the developed correlations are to be justified through further studies using ellipsoid with other aspect ratios. The second important limitation is 
that the effect of roughness has been demonstrated only for a single roughness pattern. A detailed study to establish the effect of different roughness patterns will be an interesting future scope.


	
\section*{Acknowledgement}
This work is supported by IRCC IIT Bombay (No: RD/0512-IRCCSH0-017).
\section*{References}

\bibliography{mybib}

\end{document}